\UseRawInputEncoding
\documentclass[reprint,nofootinbib,showpacs]{revtex4}
\usepackage{graphicx}  
\usepackage{dcolumn}   
\usepackage{bm}        
\usepackage{amssymb}
\usepackage{multirow}
\usepackage{amsmath}
\usepackage{xcolor}
\usepackage{hyperref}
\usepackage[figurename=Figure]{caption}
\usepackage{caption}
\usepackage{subcaption}

\begin{document} 
\title{Bottom-up approach to texture zeros in the neutrino mass matrix}
 \author{Iffat Ara Mazumder${}$}
 \email{iffat\_rs@phy.nits.ac.in}
 \author{Rupak~Dutta${}$}
 \email{rupak@phy.nits.ac.in}
 \affiliation{${}$ National Institute of Technology Silchar, Silchar 788010, India}

\begin{abstract}
We investigate one and two texture zeros in the neutrino mass matrix using the latest oscillation data through a bottom-up approach. In 
this context, we begin by estimating the detailed features of each matrix element by varying the CP violating phases within $(0,\,2\pi)$, 
the lightest neutrino mass within $(0,\,1)\,{\rm eV}$ and the neutrino oscillation parameters such as three mixing angles and the two mass 
squared differences within $3\sigma$ of their central values. We find that $ee$ element can vanish only for normal ordering, whereas, 
four elements --- $e\mu$, $e\tau$, $\mu\mu$ and $\tau\tau$ --- vanish for normal,
inverted and degenerate mass ordering. The $\mu\tau$ element can vanish for inverted and degenerate mass ordering.
In case of two texture zeros, $(ee$, $e\mu)$ and $(ee$, $e\tau)$ can vanish only for normal ordering, whereas, $(e\mu$, $\tau\tau)$, 
$(e\tau$, $\tau\tau)$, $(e\mu$, $\mu\mu)$, $(e\tau$, $\mu\mu)$ and $(\mu\mu$, $\tau\tau)$ can vanish for normal, inverted and degenerate
mass ordering. For a particular vanishing case, we also estimate the range of the lightest 
neutrino mass and the CP violating phases. In particular, very interesting correlation among the CP violating phases and the lightest neutrino 
mass is obtained for each vanishing cases.
\end{abstract}
\maketitle

\section{Introduction}
Compelling evidences of neutrino oscillation from solar, atmospheric, accelerator and reactor~\cite{Fogli:2001vr,Super-Kamiokande:1998kpq,
SNO:2003bmh,K2K:2006yov} neutrino experiments clearly indicate lepton flavor mixing and the existence of non zero neutrino mass. The neutrino 
mass matrix can be parametrized in terms of nine free parameters namely, the lightest neutrino mass~$(m_1\,\, \rm{or}\,\, m_3)$, three mixing angles~$(\theta_{12},
\,\theta_{23},\,\theta_{13})$, three CP violating phases~$(\delta,\,\alpha,\,\beta)$ and two mass squared differences~$(\Delta m^2_{21},
\Delta m^2_{31})$. Out of these, the mixing angles and the absolute mass-squared differences are 
well-measured. However, the absolute mass scale of neutrino, three CP-violating phases and the sign of $\Delta m^2_{31}$ remain undetermined. 
The sign of $\Delta m^2_{31}$ distinguishes between normal ordering~$(\Delta m_{31}^2 > 0)$ and inverted ordering~$(\Delta m_{31}^2 < 0)$. 
Several experiments are trying to measure leptonic CP violation $\delta$ and the sign of $\Delta m^2_{31}$. The Tokai-to-Kamioka~(T2K) 
experiment supports normal ordering and suggests maximal CP violation with $\delta \thicksim \frac{3\pi}{2}$, ruling out $\delta=0$ and 
$\pi$ at $95\%$ confidence level~\cite{T2K:2019bcf}. In contrast, the NO$\nu$A experiment favors $\delta \thicksim \frac{\pi}{2}$ and rules 
out $\delta \thicksim \frac{3\pi}{2}$ at $2\sigma$ confidence level~\cite{NOvA:2021nfi} for normal ordering. The Super-Kamiokande~(Super-K) 
experiment slightly favors normal ordering and rules out inverted ordering with $81.9\%$ to $96.7\%$ confidence level. When combined with 
T2K data, the confidence level for disfavoring inverted ordering increases to $91.9\%$ to $94.5\%$~\cite{Super-Kamiokande:2017yvm}.

There exists several outstanding problems in neutrino physics today. The origin and form of neutrino mass matrix is one of the most sought 
after questions today not only in particle physics but also in cosmology. Moreover, we have no idea whether neutrinos are Dirac or Majorana 
particles. The Majorana neutrino mass matrix is a symmetric $3\times 3$ matrix having six independent complex elements. 
Indirect bounds on each matrix element can be established using neutrino oscillation data and the cosmological upper limit on the total 
neutrino mass. Incorporating the baryon 
acoustic oscillation~(BAO) data, the Planck experiment reported the upper limit on the total neutrino mass to be $\sum\,m_{i} < 0.12\,
{\rm eV}$ at $95\%$ confidence level~\cite{Zhang:2020mox} from cosmic microwave background~(CMB) anisotropies. This limit is further improved 
to $\sum\,m_{i} < 0.09\,{\rm eV}$ by combining the Planck data with Supernovae Ia luminosity distances, galaxy clustering and CMB lensing 
data~\cite{Palanque-Delabrouille:2019iyz}. Very recently, the most stringent upper bound of $\sum\,m_{i} < 0.072\,{\rm eV}$ on the total 
neutrino mass at $95\%$ confidence level was reported by DESI collaboration~\cite{DESI:2024mwx}. 
Direct bounds, on the other hand, comes from processes that directly depend on a specific matrix element. It is well known that Majorana
masses violate lepton number by two units, i.e, $\Delta L = 2$. Hence, the best way to address whether neutrinos are Majorana or Dirac 
particles is to use processes that violates lepton number by two units. One such process is neutrinoless double $\beta$ decay that can 
directly measure the effective Majorana mass term $M_{ee}$ which is identical to the $ee$ element of the neutrino mass matrix in the presence 
of light neutrinos. This neutrinoless double $\beta$ decay process is searched extensively in experiments. The KamLAND-Zen experiment searched 
for the $0\nu\beta\beta$ decay by using $^{136} Xe$ isotope and reported the upper bound on the effective Majorana mass term to be 
$M_{ee} < (0.06- 0.17)\,{\rm eV}$~\cite{KamLAND-Zen:2016pfg}. Similarly, using $^{136} Xe$ isotope, the EXO-200 experiment reported the upper 
bound of $M_{ee}$ to be $M_{ee} < (0.09 - 0.29)\,{\rm eV}$~\cite{EXO-200:2019rkq}. Moreover, the CUORE experiment searched for the 
$0\nu\beta\beta$ decay using the $^{130} Te$ isotope and reported the upper bound to be $M_{ee} < (0.09 - 0.31)\,{\rm eV}$~
\cite{CUORE:2021mvw}. There are several other lepton number violating processes by which one can measure and reconstruct other elements of 
the neutrino mass matrix~\cite{Zuber:2000ca,Rodejohann,Michele,Anupama}. The $e\mu$ element can be measured in $\mu - e$ conversion on nuclei.
Similarly, $\Delta L = 2$ process such as 
rare $K^+ \to \pi^-\mu^+\mu^+$ can provide information on the $\mu\mu$ entry of the mass matrix. To have better sensitivity to $\mu\mu$ 
element, instead of $K$ meson one can use $B$ and $D$ meson as well. Another possibility to probe the $\mu\mu$ entry is to use trimuon 
production in neutrino–nucleon scattering process. Moreover, a more general neutrino-nucleon scattering process such as 
$e^+\,p \to \bar{\nu}_e\,l_1^+\,l_2^+\,X$ can be used to provide information on $e\tau$, $\mu\tau$, $\mu\mu$ and $\tau\tau$ elements of the 
neutrino mass matrix. One can also use $B^+ \to X^-\tau^+\,l^+$ with $l = e,\,\mu,\,\tau$ and $X = \pi,\,K,\,D$ to get information on $e\tau$,
$\mu\tau$ and $\tau\tau$ elements of the mass matrix. Currently, the sensitivities of all direct searches for these $\Delta L = 2$ processes 
are significantly lower than that for neutrinoless double beta decay. 

There exists several top down schemes in literature such as 
texture zero~\cite{Lashin:2011dn,Singh:2018tqu,Gautam:2018izb,Frampton:2002yf,Lavoura:2004tu,Fritzsch:2011qv,Gautam:2016qyw,Channey:2018cfj,
Singh:2019baq,Denton:2023hkx,Dev:2006qe,Verma:2020gpl}, vanishing minor~\cite{Lashin:2009yd,Mazumder:2022ywa,Mazumder:2023ate,Lashin:2007dm,Dev:2010if,
Araki:2012ip}, and cofactor 
zero~\cite{Liao:2013saa, Whisnant:2015ovx,Dev:2015lya,Wang:2016tkm} which not only reduces the number of free parameters but also constrain 
the structure of the neutrino mass matrix. In Ref.~\cite{Lashin:2011dn} and Ref.~\cite{Gautam:2018izb}, the authors have studied one texture 
zero in the neutrino mass matrix using Pontecorvo~Maki~Nakagawa~Sakata~(PMNS) and Trimaximal~(TM) mixing matrix and found that all the six 
patterns of one texture zero are compatible with the data. In Ref.~\cite{Lashin:2011dn}, the authors reported that the $ee$ element of the 
mass matrix can vanish only for normal mass ordering, whereas, the $\mu\tau$ element can vanish for inverted and degenerate mass orderings. 
The $e\mu$, $e\tau$, $\mu\mu$ and $\tau\tau$ elements can vanish for normal, inverted, and degenerate mass orderings. With TM mixing, in 
Ref.~\cite{Gautam:2018izb}, the authors reported that the $ee$ element can vanish only for the normal mass ordering and all the other elements
$e\mu$, $e\tau$, $\mu\mu$, $\mu\tau$ and $\tau\tau$ can vanish for normal, inverted and degenerate mass orderings. In case of two texture 
zeros with the PMNS mixing matrix, however, only seven 
patterns out of fifteen possible patterns are compatible with the experimental data~\cite{Frampton:2002yf}. Similarly, with the TM mixing 
matrix, only two patterns are found to be compatible with the neutrino oscillation data~\cite{Gautam:2016qyw}. 
Very recently, in Ref.~\cite{Denton:2023hkx}, it was shown that only two patterns of two texture zeros, namely $(ee = 0,\,e\mu =0)$ and 
$(ee = 0,\,e\tau =0)$ patterns are allowed by the current oscillation
data. The phenomenological implications of one and two vanishing minors in the neutrino mass matrix were explored in Ref.~\cite{Lashin:2009yd}
and Ref.~\cite{Frampton:2002yf} with PMNS mixing 
matrix. It was found that all six possible one vanishing minors are compatible with the data, whereas, only seven out of fifteen possible two 
minor zeros are found to be allowed. Very recently, in Refs.~\cite{Mazumder:2022ywa,Mazumder:2023ate}, one and two vanishing minors in the 
neutrino mass matrix were explored with the TM mixing matrix. It was shown that only two patterns out of fifteen possible two minor zeros are 
viable and all the six possible one vanishing minors are compatible with the data. Moreover, in Ref.~\cite{Merle:2006du}, the authors 
explored the allowed ranges of the mass matrix elements and implication of texture zeros using the neutrino oscillation data.

This paper aims to explore the detailed features of all elements in the neutrino mass matrix and investigate the implications of texture 
zeros using the latest neutrino oscillation data through a bottom-up approach.
To this end, we first estimate the allowed range of each individual matrix elements by varying the CP 
violating phases~$(\alpha,\,\beta,\,\delta)$ within their full range $(0,\,2\pi)$, the lightest neutrino mass within $(0,\,1)\,{\rm eV}$ and 
all the neutrino oscillation parameters such as the mixing angles~$(\theta_{12},\,\theta_{13},\,\theta_{23})$ and the two mass squared 
differences~$(\Delta m^2_{21},\Delta m^2_{31})$ within $3\sigma$ of their central values. We have explored the possibility of one and 
two texture 
zeros in the neutrino mass matrix and by assuming any one or two of the entries of the mass matrix to be zero, we obtain phenomenological 
implications and correlations among various neutrino observables. Moreover, for each vanishing cases, we also estimate the allowed range of 
the lightest neutrino mass and the CP violating phases.

The paper is organized as follows. In Section.~\ref{section:2}, we introduce the PMNS mixing matrix and list all six independent elements of 
the neutrino mass matrix. Section.~\ref{section:3} provides a detailed analysis of these elements for both normal and inverted ordering, 
including plots of each element as functions of the lightest neutrino mass and CP-violating phases. We identify regions where matrix elements 
vanish and plot the correlations of unknown parameters for different values of the Dirac CP-violating phase $\delta$. We conclude with a brief
summary of our results in Section.~\ref{section:4}.

\section{Formalism}
\label{section:2}
The PMNS mixing matrix is parameterized by three mixing angles and one CP-violating phase and can be expressed as
\begin{equation} 
\label{eq:1}
  U_{\rm PMNS}=\begin{pmatrix}
 c_{12}c_{13} & s_{12}c_{13} & s_{13}e^{-i\delta}\\
 -s_{12}c_{23}-c_{12}s_{23}s_{13}e^{i\delta}& c_{12}c_{23}-s_{12}s_{23}s_{13}e^{i\delta}&s_{23}c_{13}\\
 s_{12}s_{23}-c_{12}c_{23}s_{13}e^{i\delta}& -c_{12}s_{23}-s_{12}c_{23}s_{13}e^{i\delta}&c_{23}c_{13}
\end{pmatrix}\,,
 \end{equation}
where $s_{ij}=\sin\theta_{ij}$ and $c_{ij}=\cos\theta_{ij}$ for $i,j=1,2,3$.  
In the charged lepton basis, the symmetric neutrino mass matrix $M$ can be expressed as
\begin{equation} 
\label{eq:2}
 M_{ab}=(V\,M_{\rm diag}\,V^{T})_{ab}\,\, {\rm with}\,\, a\,, b = e\,,\mu\,,\tau\,,
\end{equation}
where $M_{\rm diag} = {\rm diag}(m_1, m_2, m_3)$ is the diagonal matrix containing three mass states and $V=U_{\rm PMNS}\,P$ is the lepton flavor mixing 
matrix. The diagonal phase matrix $P$ can be written as
\begin{equation} 
\label{eq:3}
 P=\begin{pmatrix}
  1& 0 & 0\\
 0 & e^{i\alpha} &0\\
 0 & 0 & e^{i\beta}\
\end{pmatrix}\,,
\end{equation}
where $\alpha$ and $\beta$ are two CP violating Majorana phases, respectively.

Using Eq.~\ref{eq:2}, the elements of the neutrino mass matrix can be written as 
\begin{eqnarray} 
\label{eq:4}
  &&M_{ee}=(c^{2}_{12}c^{2}_{13})\,m_1+(s^{2}_{12}c^{2}_{13})\,m_2\,e^{2i\alpha}+(s^{2}_{13}e^{-2i\delta})\,m_3\,e^{2i\beta},  \nonumber \\ 
  &&M_{ e\mu}=c_{12}c_{13}(-s_{12}c_{23}-c_{12}s_{23}s_{13}e^{i\delta})\,m_1+s_{12}c_{13}(c_{12}c_{23}-s_{12}s_{23}s_{13}e^{i\delta})\,
m_2 e^{2i\alpha}+
(c_{13}s_{23}s_{13}e^{-i\delta})\,m_3 e^{2i\beta},  \nonumber \\
  &&M_{ e\tau}=c_{12}c_{13}(s_{12}s_{23}-c_{12}c_{23}s_{13}e^{i\delta})\,m_1+s_{12}c_{13}(-c_{12}s_{23}-s_{12}c_{23}s_{13}e^{i\delta})\,
m_2 e^{2i\alpha}+
(c_{13}c_{23}s_{13}e^{-i\delta})\,m_3 e^{2i\beta},  \nonumber \\
  &&M_{ \mu\mu}=(-s_{12}c_{23}-c_{12}s_{23}s_{13}e^{i\delta})^2\,m_1+(c_{12}c_{23}-s_{12}s_{23}s_{13}e^{i\delta})^2\,m_2 e^{2i\alpha}+
(c^{2}_{13}s^{2}_{23})\,m_3 e^{2i\beta},  \nonumber \\
  &&M_{ \mu\tau}=(-s_{12}c_{23}-c_{12}s_{23}s_{13}e^{i\delta})(s_{12}s_{23}-c_{12}c_{23}s_{13}e^{i\delta})\,m_1+(c_{12}c_{23}-s_{12}s_{23}
s_{13}e^{i\delta})(-c_{12}s_{23}-s_{12}c_{23}s_{13}e^{i\delta})  \nonumber \\
  &&\hspace{1.2cm}m_2 e^{2i\alpha}+(c^{2}_{13}s_{23}c_{23})\,m_3 e^{2i\beta},\nonumber \\
  &&M_{ \tau\tau}=(s_{12}s_{23}-c_{12}c_{23}s_{13}e^{i\delta})^2\,m_1+(-c_{12}s_{23}-s_{12}c_{23}s_{13}e^{i\delta})^2\,m_2 e^{2i\alpha}+
(c^{2}_{13}c^{2}_{23})\,m_3 e^{2i\beta}.  
 \end{eqnarray}

The elements of the neutrino mass matrix depend on nine parameters namely, three mass states $(m_{1}, m_{2}, m_{3})$, three mixing angles 
$(\theta_{12}, \theta_{13}, \theta_{23})$ and three CP violating phases $(\alpha, \beta, \delta)$. 
The mass states are related to the squared mass differences. In case of normal ordering~$(m_1 < m_2 < m_3)$, we have
\begin{eqnarray}
&& m_2 = \sqrt{m^{2}_{1}+\Delta m^2_{21}}\,,\qquad\qquad
m_3 = \sqrt{m^{2}_{1}+\Delta m^2_{31}}\,,
\end{eqnarray}
and in case of inverted ordering~$(m_3 < m_1 < m_2)$, we have
\begin{eqnarray}
&& m_1 = \sqrt{m^{2}_{3}+\Delta m^2_{13}}\,,\qquad\qquad
m_2 = \sqrt{m^{2}_{3}+\Delta m^2_{21}+\Delta m^2_{13}}\,.
\end{eqnarray}

Out of these nine parameters, the mixing angles~$(\theta_{12}, \theta_{13}, \theta_{23})$ and the mass squared 
differences~($\Delta m^2_{21},\, |\Delta m^2_{31}|$) are measured to a very good precision. The experimentally
measured values of all the mixing angles and the mass squared differences are listed in Table~\ref{tab:2}.  
\begin{table}[htbp]
\begin{tabular}{|c|c|c|c|c|}
 \hline
 \hline
Parameter & Normal ordering(best fit) & Inverted ordering ($\Delta \chi^2=7.1)$\\
&bfp$\pm 1\sigma$ \hspace{1.5cm} $3\sigma$ ranges&bfp$\pm 1\sigma$ \hspace{1.5cm} $3\sigma$ ranges\\
\hline
$\theta_{12}^\circ
$&$33.44^{+0.77}_{-0.74}$ \hspace{1.5cm} 31.27$\rightarrow $ 35.86&$33.45^{+0.77}_{-0.74}$\hspace{1.5cm} 31.27$\rightarrow $ 35.87\\
\hline
$\theta_{23}^\circ$&$49.2^{+1.0}_{-1.3}$ \hspace{1.5cm} 39.5$\rightarrow $ 52.0&$49.5^{+1.0}_{-1.2}$\hspace{1.5cm} 39.8$\rightarrow $ 52.1\\
\hline
$\theta_{13}^\circ$&$8.57^{+0.13}_{-0.12}$ \hspace{1.5cm} 8.20$\rightarrow $ 8.97&$8.60^{+0.12}_{-0.12}$ \hspace{1.5cm} 8.24$\rightarrow 
$ 8.98\\
\hline
$\delta^\circ$&$194^{+52}_{-25}$ \hspace{1.5cm} 105$\rightarrow $ 405&$287^{+27}_{-32}$ \hspace{1.5cm} 192$\rightarrow $ 361\\
\hline
$\frac{\Delta m^2_{21}}{10^{-5}eV^2}$&$7.42^{+0.21}_{-0.20}$ \hspace{1.5cm} 6.82$\rightarrow $ 8.04&$7.42^{+0.21}_{-0.20}$ \hspace{1.5cm} 
6.82$\rightarrow $ 8.04\\
\hline
$\frac{\Delta m^2_{3l}}{10^{-3}eV^2}$&$+2.515^{+0.028}_{-0.028}$ \hspace{1.5cm} +2.431$\rightarrow $ +2.599&$-2.498^{+0.028}_{-0.029}$ 
\hspace{1.5cm} -2.584$\rightarrow $ -2.413\\
\hline
\hline
\end{tabular}
\caption{Neutrino oscillation parameters from NuFIT.~\cite{Esteban:2020cvm}. }
\label{tab:2}
\end{table}

We now proceed to analyse all the six elements of the neutrino mass matrix one by one and explore the implications of one and two texture 
zeros in the neutrino mass matrix.

\section{\textbf{One Texture Zero}}
\label{section:3}
In this section, we wish to explore the implications of one texture zeros in the neutrino mass matrix through a bottom up approach using the 
latest oscillation data. 
In our analysis, we vary the lightest neutrino mass within 
$(0,\,1)\,{\rm eV}$, the neutrino oscillation parameters $\theta_{12}$, $\theta_{13}$, $\theta_{23}$, 
$\Delta  m^2_{21}$ and $\Delta m^2_{3l}$ within $3\sigma$ of their experimental central values, and the CP violating phases $\alpha$, $\beta$ 
and $\delta$ across the full range $(0,\,2\pi)$.

\subsection{$M_{ee}$: the $ee$ element of the mass matrix}
\subsubsection{Normal ordering:}
\begin{figure}[htbp!]
\begin{subfigure}{0.35\textwidth}
\includegraphics[width=\textwidth]{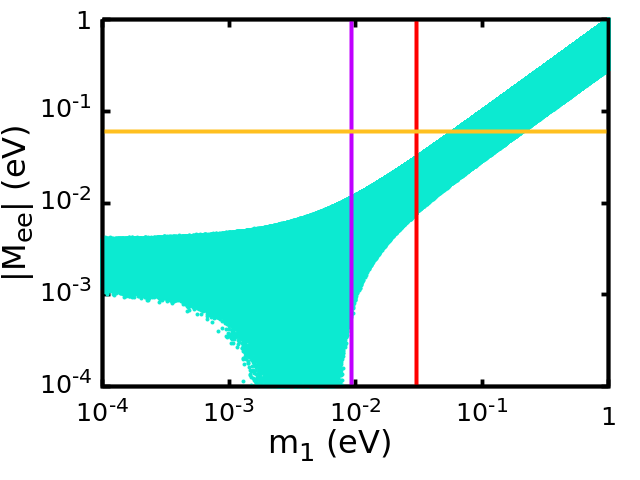}
\end{subfigure}
\captionsetup{justification=raggedright,singlelinecheck=false}
\caption{Variation of $|M_{ee}|$ with the lightest neutrino mass $m_1$ in case of normal ordering. The red and the magenta lines represent 
total neutrino mass $\sum m_{i} = 0.12\,{\rm  eV}$~\cite{Zhang:2020mox} and $\sum m_{i} = 0.072\,{\rm  eV}$~\cite{DESI:2024mwx}, respectively.
The yellow line represents $|M_{ee}|= 0.06\,{\rm eV}$~\cite{KamLAND-Zen:2016pfg}.}
\label{fig:1}
\end{figure}

The variation of $|M_{ee}|$ as functions of the lightest neutrino mass $m_1$ is shown in Fig.~\ref{fig:1}. We show the total neutrino mass 
$\sum m_{i} = 0.12\,{\rm  eV}$ and $\sum m_{i} = 0.072\,{\rm  eV}$ in Fig.~\ref{fig:1} with red and magenta vertical lines, respectively.  
It is observed that upper bound on the total neutrino mass imposes constraints on the lightest neutrino mass $m_1$ and on $|M_{ee}|$. Notably, 
for a vanishing $|M_{ee}|$, the allowed values of $m_1$ with the total neutrino mass constraint are exactly same as those without the 
constraint. 

\begin{figure}[htbp!]
\begin{subfigure}{0.30\textwidth}
\includegraphics[width=\textwidth]{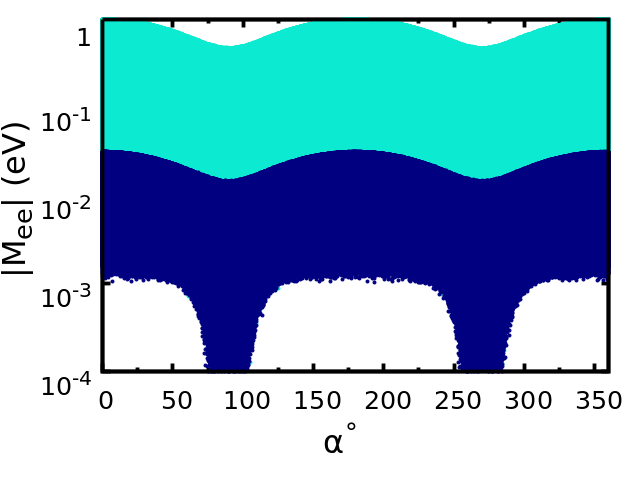}
\caption{}
\label{fig:1b}
\end{subfigure}
\begin{subfigure}{0.30\textwidth}
\includegraphics[width=\textwidth]{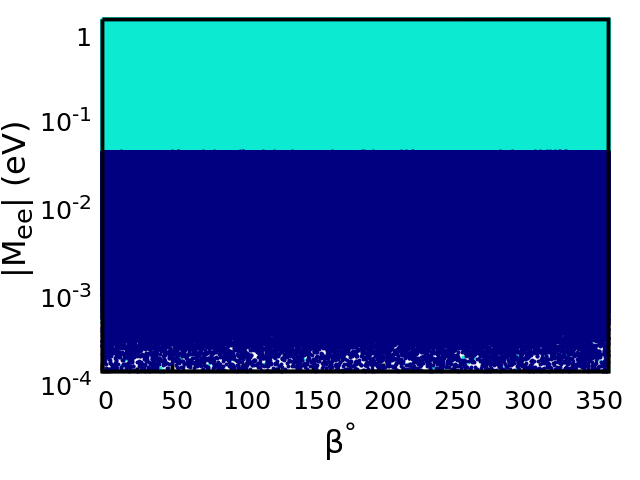}
\caption{}
\label{fig:1c}
\end{subfigure}
\begin{subfigure}{0.30\textwidth}
\includegraphics[width=\textwidth]{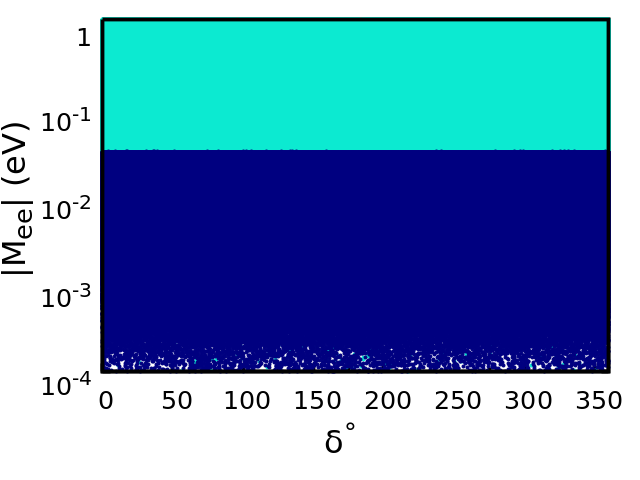}
\caption{}
\label{fig:1d}
\end{subfigure}
\caption{Variation of $|M_{ee}|$ with the three CP violating phases $\alpha$, $\beta$ and $\delta$ in case of normal ordering.}
\label{fig:1cp_phases}
\end{figure}

We also show the variation of $|M_{ee}|$ with the CP violating phases $\alpha$, $\beta$ and $\delta$ in  Fig.~\ref{fig:1b}, Fig.~\ref{fig:1c}
and Fig.~\ref{fig:1d} with~(blue band) and without~(cyan band) the total neutrino mass $\sum m_{i} < 0.12\,{\rm  eV}$ constraint,
respectively.
  
Our main observations are as follows:  
\begin{itemize}
 \item It is clear from Fig.~\ref{fig:1} that $|M_{ee}|$ can vanish for a very limited parameter range of the lightest neutrino mass 
$m_1$, i.e, $|M_{ee}| \to 0$ is only possible if $m_1$ lies between $(2.0,\,8.0)\times 10^{-3}\,{\rm eV}$ with and without the
total neutrino mass constraint.
 \item  In order to have vanishing $|M_{ee}|$, the value of the Majorana CP violating phase $\alpha$ has to lie either between 
$(75^{\circ},\,110^{\circ})$ or between $(250^{\circ},\,285^{\circ})$, irrespective of the total neutrino mass.
 \item From Fig.~\ref{fig:1c} and Fig.~\ref{fig:1d}, it is clear that there is no constraint on the values of $\beta$ and $\delta$ for 
vanishing $|M_{ee}|$.
 \end{itemize}
Now, let us examine vanishing of $|M_{ee}|$ analytically for different values of the lightest neutrino mass $m_1$. In case of normal ordering,
the effective Majorana mass term $M_{ee}$ can be written as
\begin{eqnarray}
 M_{ee}=c^{2}_{12}c^{2}_{13}\,m_1+s^{2}_{12}c^{2}_{13}\,\sqrt{m_1^2 + \Delta m^2_{21}}\,e^{2i\alpha}+s^{2}_{13}e^{2i(\beta - \delta)}\,
 \sqrt{m_1^2 + \Delta m^2_{31}}\,.
\end{eqnarray}

\begin{itemize}
\item {\bf Case I}: $m_{1}<<\sqrt{\Delta m^2_{21}}$ : \\
In this case, $|M_{ee}|$ can be expressed as
  \begin{eqnarray}
  \label{eq:5}
  |M_{ee}|\simeq \Big|s^{2}_{12}c^{2}_{13}\sqrt{\Delta m^2_{21}}e^{2i\alpha}+s^{2}_{13}\sqrt{\Delta m^2_{31}}e^{2i(\beta-\delta)}\Big|. 
  \end{eqnarray}
We find the allowed range of $|M_{ee}|$ to be $(10^{-3},\,10^{-2})\,{\rm eV}$.
It is clear that $|M_{ee}|$ can not vanish under this condition. From Eq.~\ref{eq:5}, it is clear that we get the minimum value
of $|M_{ee}|$ if there is a maximal cancellation between the two terms and that is only possible if $\alpha = 90^{\circ},\,270^{\circ}$ and 
$\beta - \delta = 0^{\circ},\,180^{\circ}$. Under such condition, the minimum value of $|M_{ee}|$ will be
\begin{eqnarray}
|M_{ee}| = \Big|s^{2}_{12}c^{2}_{13}\sqrt{\Delta m^2_{21}} - s^{2}_{13}\sqrt{\Delta m^2_{31}}\Big|
\end{eqnarray}
In this case, $|M_{ee}|$ will vanish only if
\begin{eqnarray}
\sin^2\theta_{12} = \tan^2\theta_{13}\sqrt{\frac{\Delta m^2_{31}}{\Delta m^2_{21}}}
\end{eqnarray}
Using the latest best fit value of $\theta_{13}$, $\Delta m^2_{21}$ and $\Delta m^2_{31}$, we get $\theta_{12} \equiv 21^{\circ}$ which is 
quite different from the $3\sigma$ experimental range of $\theta_{12} = (31.27^{\circ},\, 35.86^{\circ})$. Hence, in the $m_1 \to 0$ limit, 
$|M_{ee}|$ does not vanish.

 \item {\bf Case II}: $m_{1} >>\sqrt{\Delta m^2_{31}}$ : \\
 In this case, $|M_{ee}|$ can be expressed as
  \begin{eqnarray}
\label{meebigm1}
  |M_{ee}| = \Big|m_1\Big(c^{2}_{12}c^{2}_{13}+s^{2}_{12}c^{2}_{13}\,e^{2i\alpha}+s^{2}_{13}e^{2i(\beta - \delta)}\Big)\Big|
  \end{eqnarray}
  Minimum value of
  $|M_{ee}|$ corresponds to $\alpha = 90^{\circ},\,270^{\circ}$ and $\beta - \delta = 90^{\circ},\,270^{\circ}$. Under these condition,
  $|M_{ee}|$ can be written as
  \begin{eqnarray}
  |M_{ee}| = \Big|m_1\Big(c^{2}_{12}c^{2}_{13} - s^{2}_{12}c^{2}_{13} - s^{2}_{13}\Big)\Big| = m_1\,\Big|2\,c_{12}^2\,c_{13}^2 - 1\Big|
  \end{eqnarray}
If we consider $c_{13} \simeq 1$, then vanishing of $|M_{ee}|$ implies $\theta_{12} \simeq 45^{\circ}$.
Hence, when $m_1 >> \sqrt{\Delta m^2_{31}}$, $|M_{ee}|$ does not vanish.
  \item {\bf Case III}: $m_{1} \simeq \sqrt{\Delta m^2_{21}}$ : \\
  In this case, $|M_{ee}|$ can be written as
  \begin{eqnarray}
 |M_{ee}|=\Big|c^{2}_{12}c^{2}_{13}\,\sqrt{\Delta m^2_{21}}+s^{2}_{12}c^{2}_{13}\,\sqrt{2\,\Delta m^2_{21}}\,e^{2i\alpha}+s^{2}_{13}
e^{2i(\beta - \delta)}\,\sqrt{\Delta m^2_{31}}\Big|
\end{eqnarray}
Minimum value of $|M_{ee}|$ will be obtained for $\alpha = 90^{\circ},\,270^{\circ}$ and $\beta - \delta = 90^{\circ},\,270^{\circ}$. Under
these conditions, the Majorana mass term can be written as
\begin{eqnarray}
 |M_{ee}|=\Big|c^{2}_{12}c^{2}_{13}\,\sqrt{\Delta m^2_{21}} - s^{2}_{12}c^{2}_{13}\,\sqrt{2\,\Delta m^2_{21}} - s^{2}_{13}\,
\sqrt{\Delta m^2_{31}}\Big|
\end{eqnarray}
In this case, $|M_{ee}|$ will vanish if
\begin{eqnarray}
 \tan^2\theta_{13} = (\cos^2\theta_{12} - \sqrt{2}\sin^2\theta_{12})\sqrt{\frac{\Delta m_{21}^2}{\Delta m_{31}^2}}   
\end{eqnarray}
With the experimental value of $\theta_{12}$ and the mass squared differences, we obtain $\theta_{13} \simeq 9.5^{\circ}$ which is quite close 
to the experimental value of $\theta_{13}$. Hence, we can have vanishing $|M_{ee}|$ if $m_1$ is close to $\sqrt{\Delta m_{21}^2}$. This is
what we have obtained in Fig.~\ref{fig:1}.
  \end{itemize}

\begin{figure}[htbp!]
\begin{subfigure}{0.40\textwidth}
\includegraphics[width=5cm,height=4cm]{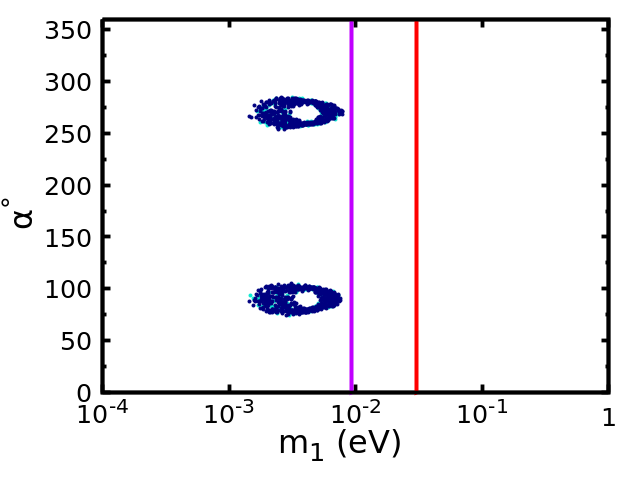}
\caption{}
\label{fig:1.1a}
\end{subfigure}
\begin{subfigure}{0.40\textwidth}
\includegraphics[width=5cm,height=4cm]{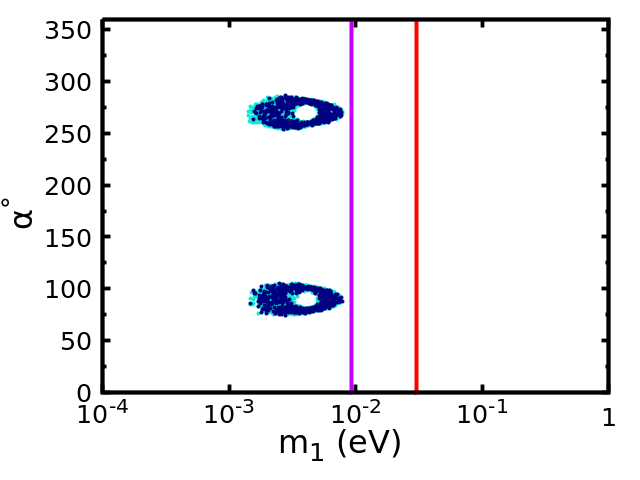}
\caption{}
\label{fig:1.1b}
\end{subfigure}
 \begin{subfigure}{0.40\textwidth}
\includegraphics[width=5cm,height=4cm]{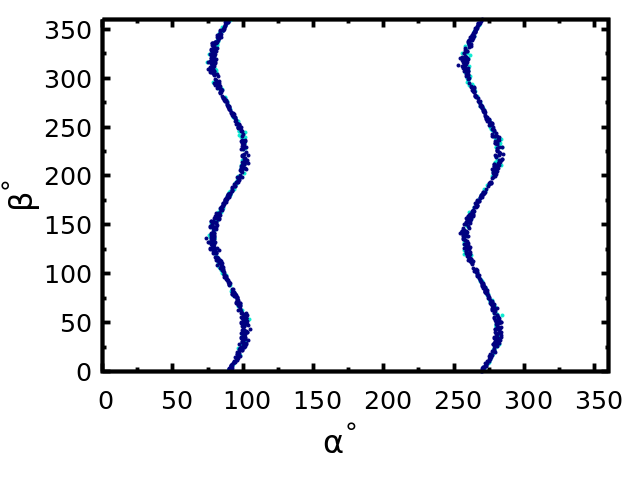}
\caption{}
\label{fig:1.1c}
\end{subfigure} 
\begin{subfigure}{0.40\textwidth}
\includegraphics[width=5cm,height=4cm]{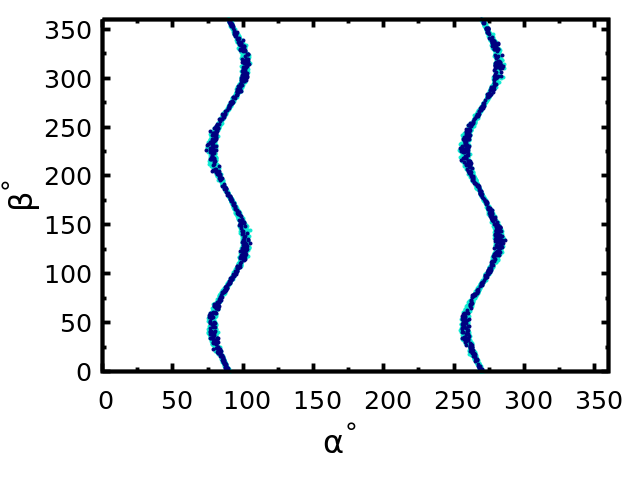}
\caption{}
\label{fig:1.1d}
\end{subfigure}
\captionsetup{justification=raggedright,singlelinecheck=false}
\caption{Correlation between $\beta$ and $\alpha$, and between $\alpha$ and $m_{1}$ for $\delta=0^{\circ}$~(Left panel) and 
$\delta=90^{\circ}$~(Right panel) with vanishing $|M_{ee}|$ in case of normal ordering. The red and magenta lines 
represent $\sum m_i = 0.12\,{\rm eV}$~\cite{Zhang:2020mox} and $\sum m_i = 0.072\,{\rm eV}$~\cite{DESI:2024mwx}, respectively.}
\label{fig:1.1}
\end{figure}

We now wish to obtain the correlation between the unknown parameters for vanishing $|M_{ee}|$ with and without the constraint coming from the 
upper bound of $\sum m_i < 0.12\,{\rm eV}$ on the total neutrino mass. In Fig.~\ref{fig:1.1}, we have shown 
the correlation between some of the parameters for $\delta = 0^{\circ}$ and $90^{\circ}$, respectively. It is evident that the blue region
that was obtained using the constraint coming from the upper bound of $\sum m_i < 0.12\,{\rm eV}$ overlaps with the cyan region that was
obtained without the total neutrino mass constraint. Fig.~\ref{fig:1.1a} and 
Fig.~\ref{fig:1.1b} are scatter plot of the lightest neutrino mass $m_1$ against the Majorana CP 
violating phase $\alpha$ for $\delta =0^{\circ}$ and $90^{\circ}$, respectively. It is observed that, irrespective of the $\delta$ value, 
value of $\alpha$ should be around $90^{\circ}$ or $270^{\circ}$ and the value of the lightest neutrino mass 
should be $m_1 \in (2.0,\,8.0)\times 10^{-3}\,{\rm eV}$ to have vanishing $|M_{ee}|$. Similarly, in Fig.~\ref{fig:1.1c} and 
Fig.~\ref{fig:1.1d}, we show the correlation between the 
two Majorana CP violating phases, namely between $\alpha$ and $\beta$ for $\delta = 0^{\circ}$ and $90^{\circ}$, respectively. As expected,
there is no constraint on the value of $\beta$, i.e, $|M_{ee}|$ will vanish for any value of $\beta$. It should be mentioned that the 
correlation between $\alpha$ and $\beta$ for $\delta = 0^{\circ}$ is different from that for $\delta =90^{\circ}$. We also show the correlation between the atmospheric mixing angle $\theta_{23}$ and Majorana CP violating phase $\alpha$ with vanishing $|M_{ee}|$ in Fig.~\ref{fig:Nmalpth23meezero_no} of Appendix~\ref{app1}.

\subsubsection{Inverted Ordering:}
\begin{figure}[h!]
\begin{subfigure}{0.32\textwidth}
\includegraphics[width=\textwidth]{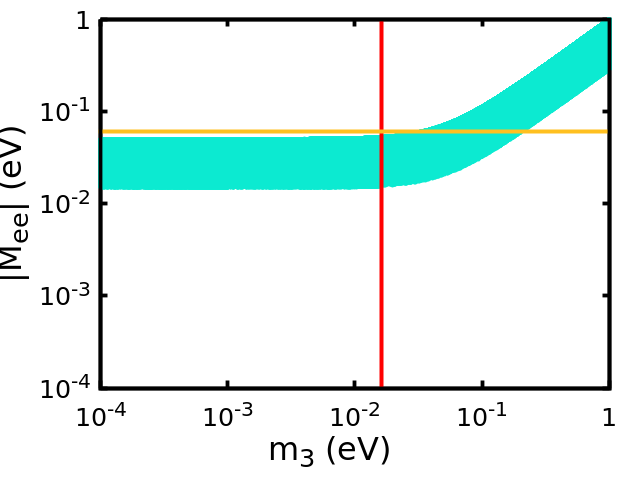}
\end{subfigure}
\captionsetup{justification=raggedright,singlelinecheck=false}
\caption{Variation of $|M_{ee}|$ with the lightest neutrino mass $m_{3}$ in case of inverted ordering. Total neutrino mass 
$\sum m_{i} = 0.12\,{\rm  eV}$~\cite{Zhang:2020mox} is shown with the red vertical line. The yellow line represents 
$|M_{ee}| = 0.06\,{\rm eV}$~\cite{KamLAND-Zen:2016pfg}.}
\label{fig:2}
\end{figure}

\begin{figure}[htbp!]
\begin{subfigure}{0.32\textwidth}
\includegraphics[width=\textwidth]{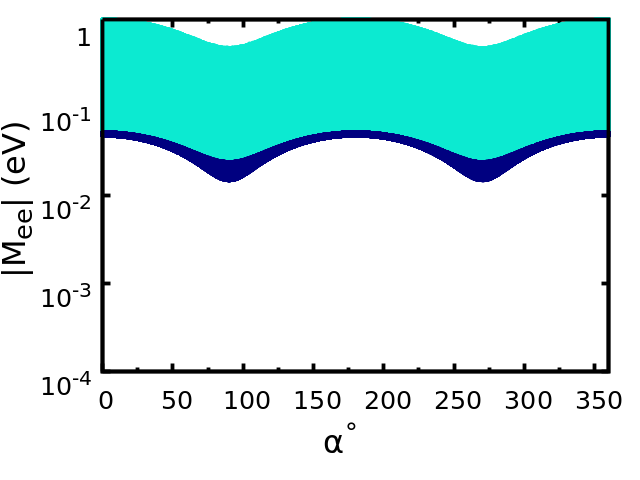}
\caption{}
\label{fig:2b}
\end{subfigure}
\begin{subfigure}{0.32\textwidth}
\includegraphics[width=\textwidth]{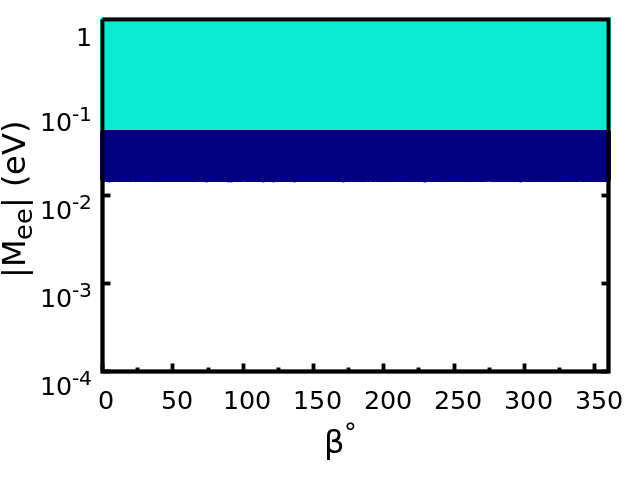}
\caption{}
\label{fig:2c}
\end{subfigure}
\begin{subfigure}{0.32\textwidth}
\includegraphics[width=\textwidth]{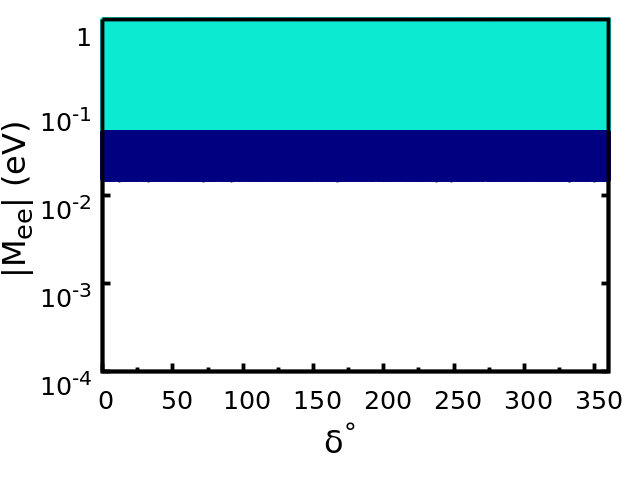}
\caption{}
\label{fig:2d}
\end{subfigure}
\captionsetup{justification=raggedright,singlelinecheck=false}
\caption{Variation of $|M_{ee}|$ with the CP violating phases $\alpha$, $\beta$ and $\delta$ in case of inverted ordering.}
\label{fig:2_cp_phase}
\end{figure}

We show the variation of $|M_{ee}|$ with the lightest neutrino mass~$(m_3)$ in Fig.~\ref{fig:2}. The red and the yellow lines represent
$\sum m_{i} = 0.12\,{\rm  eV}$~\cite{Zhang:2020mox} and $|M_{ee}| = 0.06\,{\rm eV}$~\cite{KamLAND-Zen:2016pfg}, respectively. It is evident 
that $|M_{ee}|$ does not vanish for any value of the lightest neutrino mass in case of inverted ordering. The lowest value of $|M_{ee}|$ is 
found to be of $\mathcal{O}(10^{-2})\,{\rm eV}$. 
We also show the variation of $|M_{ee}|$ with the CP violating phases $\alpha$, $\beta$ and $\delta$ in  Fig.~\ref{fig:2b}, Fig.~\ref{fig:2c}
and Fig.~\ref{fig:2d} with~(blue band) and without~(cyan band) the total neutrino mass $\sum m_{i} < 0.12\,{\rm  eV}$ constraint,
respectively.

We wish to analyse few extreme cases analytically. In case of inverted ordering, $M_{ee}$ can be expressed as
\begin{eqnarray}
 M_{ee}=c^{2}_{12}c^{2}_{13}\,\sqrt{m_3^2 + \Delta m^2_{13}}+s^{2}_{12}c^{2}_{13}\,\sqrt{m_3^2 + \Delta m^2_{13} + \Delta m^2_{21}}\,
e^{2i\alpha}+s^{2}_{13}e^{2i(\beta - \delta)}\,m_{3}\,.
\end{eqnarray}
\begin{itemize}
 \item {\bf Case I:} $m_{3}<<\sqrt{\Delta m^2_{21}}$:\\
 In this case, $|M_{ee}|$ can be written as
 \begin{eqnarray}
|M_{ ee}| \simeq \sqrt{\Delta m^2_{13}}\Big|c^{2}_{12}c^{2}_{13} + s^{2}_{12}c^{2}_{13} e^{2i\alpha}\Big|
\end{eqnarray}
Minimum value of $|M_{ee}|$ can be obtained for $\alpha = 90^{\circ}$ and $270^{\circ}$. Under this condition, $|M_{ee}|$ can be 
written as
\begin{equation}
 |M_{ee}|= c^{2}_{13}\cos2\theta_{12}\sqrt{\Delta m^{2}_{13}} =\mathcal{O}(10^{-2})
\end{equation}
Vanishing of $|M_{ee}|$ implies $\cos2\theta_{12} \approx 0$, i.e, $\theta_{12} \approx 45^{\circ}$. In this case $|M_{ee}|$ can not vanish.
 \item {\bf Case II}: $m_{3} >>\sqrt{\Delta m^2_{13}}$ : \\
 In this case, we have 
  \begin{eqnarray}
  |M_{ee}| = \Big|m_3\Big(c^{2}_{12}c^{2}_{13}+s^{2}_{12}c^{2}_{13}\,e^{2i\alpha}+s^{2}_{13}e^{2i(\beta - \delta)}\Big)\Big|
  \end{eqnarray}
This is exactly same as Eq.~\ref{meebigm1} of the normal ordering case. It is expected because, under this condition, we have 
$m_1 \approx m_2 \approx m_3$. Hence, in this case $|M_{ee}|$ will not vanish.
 \end{itemize} 

\subsection{$M_{e\mu}$: the $e\mu$ element of the mass matrix} 
\subsubsection{Normal Ordering:}
\begin{figure}[htbp!]
\begin{subfigure}{0.35\textwidth}
\includegraphics[width=\textwidth]{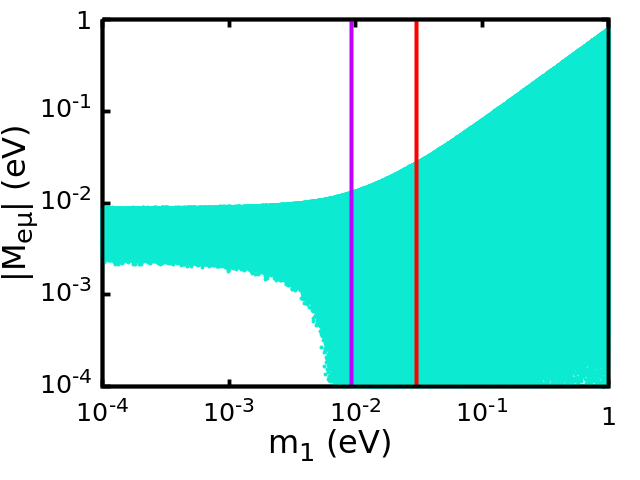}
\end{subfigure} 
\captionsetup{justification=raggedright,singlelinecheck=false}
\caption{Variation of $|M_{e\mu}|$ with the lightest neutrino mass $m_1$ in 
case of normal ordering. Total neutrino mass $\sum m_{i} = 0.12\,{\rm  eV}$~\cite{Zhang:2020mox} and 
$\sum m_{i} = 0.072\,{\rm  eV}$~\cite{DESI:2024mwx} are 
shown with red and magenta vertical lines, respectively.}
\label{fig:3}
\end{figure}

We plot $|M_{e\mu}|$ as functions of the unknown parameter $m_1$ in Fig.~\ref{fig:3}. 
We show the total neutrino mass $\sum m_{i} = 0.12\,{\rm  eV}$ and $\sum m_{i} = 0.072\,{\rm  eV}$ with red and magenta vertical lines, 
respectively.
The upper bound on the total neutrino mass imposes constraint on the lightest neutrino mass $m_1$ and the matrix element
$|M_{e\mu}|$. It is worth mentioning that we get vanishing $|M_{e\mu}|$ with and without the total neutrino mass constraint.

\begin{figure}[htbp!] 
\begin{subfigure}{0.30\textwidth}
\includegraphics[width=\textwidth]{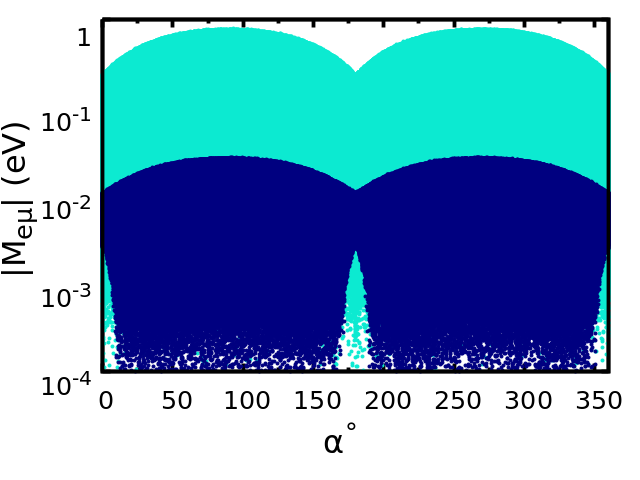}
\caption{}
\label{fig:3b}
\end{subfigure}
\begin{subfigure}{0.30\textwidth}
\includegraphics[width=\textwidth]{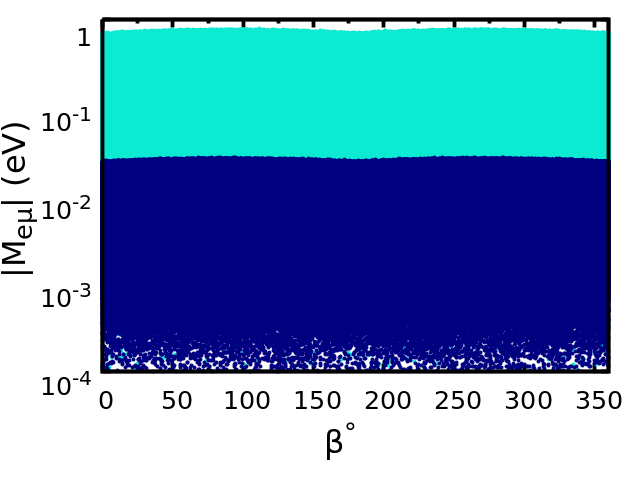}
\caption{}
\label{fig:3c}
\end{subfigure}
\begin{subfigure}{0.30\textwidth}
\includegraphics[width=\textwidth]{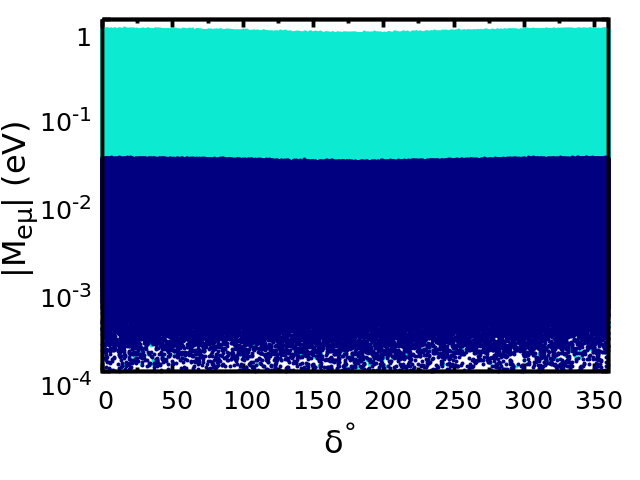}
\caption{}
\label{fig:3d}
\end{subfigure}
\caption{Variation of $|M_{e\mu}|$ with the CP violating phases $\alpha$, $\beta$ and $\delta$ in case of normal ordering.}
\label{fig:3_cp_phases}
\end{figure}

In Fig.~\ref{fig:3b}, Fig.~\ref{fig:3c} and Fig.~\ref{fig:3d}, we show the variation of $|M_{e\mu}|$ with the CP violating phases $\alpha$, 
$\beta$ and $\delta$, respectively. The blue region is obtained using the constraint coming from the upper bound of 
$\sum m_{i} < 0.12\,{\rm  eV}$ on the total neutrino mass, whereas, the cyan region is obtained without the total neutrino mass 
constraint.\\
 
Our main observations are as follows:
\begin{itemize} 
 \item We get vanishing $|M_{e\mu}|$ if the lightest neutrino mass $m_1 \geq 6.0\times 10^{-3}\,{\rm eV}$.
 \item It is evident that we get vanishing $|M_{e\mu}|$ if the Majorana CP violating phase $\alpha$ lies between either 
$(20^{\circ},\,160^{\circ})$ or 
$(200^{\circ},\,340^{\circ})$ if we impose the constraint coming from the upper bound of $\sum m_i < 0.12\,{\rm eV}$. However, without the
constraint, $|M_{e\mu}|$ can vanish for any value of $\alpha$.
 \item It is evident from Fig.~\ref{fig:3c} and Fig.~\ref{fig:3d} that $|M_{e\mu}|$ can vanish for any value of $\beta$ and $\delta$ 
within $(0,\,360^{\circ})$, irrespective of the value of the total neutrino mass.
 \end{itemize}

We now proceed to analyse vanishing of $M_{e\mu}$ for different values of the lightest neutrino mass $m_1$. In case of normal ordering, the $M_{e\mu}$ can be expressed as
\begin{eqnarray}
\label{memu}
M_{ e\mu} &=& c_{12}c_{13}(-s_{12}c_{23}-c_{12}s_{23}s_{13}e^{i\delta})\,m_1+s_{12}c_{13}(c_{12}c_{23}-s_{12}s_{23}s_{13}e^{i\delta})\,
\sqrt{m_1^2 + \Delta m_{21}^2}\,e^{2i\alpha} \nonumber \\
&&+ (c_{13}s_{23}s_{13}e^{-i\delta})\,\sqrt{m_1^2 + \Delta m_{31}^2}\,e^{2i\beta} 
\end{eqnarray}

\begin{itemize}
 \item {\bf Case I:} $m_{1}<<\sqrt{\Delta m^2_{21}}$:\\ 
 In this case, $|M_{e\mu}|$ can be written as
  \begin{equation}
   \label{eq:11}
   |M_{e\mu}|\simeq \Big|s_{12}c_{13}(c_{12}c_{23}-s_{12}s_{23}s_{13}e^{i\delta})\sqrt{\Delta m^2_{21}}e^{2i\alpha}+(c_{13}s_{23}s_{13}
e^{-i\delta})\sqrt{\Delta m^2_{31}}e^{2i\beta}\Big|.
  \end{equation}
The allowed range of $|M_{e\mu}|$ is found to be $(2.0,\, 9.0) \times 10^{-3}\,{\rm eV}$. 
Minimum value of $|M_{e\mu}|$ can be obtained if there is a maximal cancellation between the two terms. This can be achieved by assuming 
$\alpha = 90^{\circ},\,270^{\circ}$ and $\beta = \delta = 0^{\circ},\,180^{\circ}$. 
Moreover, if we assume $s_{23} \simeq c_{23}$, then $|M_{e\mu}|$ can be written as
\begin{equation}\label{eq:11}
  |M_{e\mu}| \simeq s_{23}c_{13}\Big|s_{12}(c_{12}-s_{12}s_{13})\sqrt{\Delta m^2_{21}} - s_{13}\sqrt{\Delta m^2_{31}}\Big|
    \simeq s_{23}c_{13}\Big|s_{12}c_{12}\sqrt{\Delta m^2_{21}} - s_{13}\sqrt{\Delta m^2_{31}}\Big|\,,
  \end{equation}
 where we have neglected the term containing $s_{12}s_{13}$, since $c_{12} >> s_{12}s_{13}$. 
Under this condition, $|M_{e\mu}|$ can vanish only if
 \begin{eqnarray}
  \sin2\theta_{12} = 2\,\sin\theta_{13}\sqrt{\frac{\Delta m^2_{31}}{\Delta m^2_{21}}}\,,   
 \end{eqnarray}
and it implies $\sin2\theta_{12} > 1$. Hence, we can not have vanishing $|M_{e\mu}|$ in the $m_1 \to 0$ limit. 
 \item {\bf Case II:} $m_{1} >> \sqrt{\Delta m^2_{31}}$:\\ 
 In this case $|M_{e\mu}|$ can be written as
 \begin{eqnarray}
\label{memuno}
&&|M_{ e\mu}|=m_1\Big|c_{12}s_{12}c_{13}c_{23}(e^{2i\alpha} - 1) + c_{13}s_{23}s_{13}\Big[-(c_{12}^2 + s_{12}^2\,e^{2i\alpha}) + 
e^{2i(\beta - \delta)}\Big]\,e^{i\delta}\Big| \,.
\end{eqnarray}
The $e\mu$ element $|M_{e\mu}|$, in this case, can be zero if we can have both the terms equal to zero. The first term can be zero if we have 
$e^{2i\alpha} = 1$ or $\alpha = 0,\,180^{\circ},\,360^{\circ}$. Similarly, the second term can be zero if we have $e^{2i(\beta - \delta)} = 1$, i.e, 
$(\beta - \delta) = 0,\,180^{\circ},\,360^{\circ}$. Hence, we can have vanishing $|M_{e\mu}|$ in this case.

 \item {\bf Case III:} $m_{1}\simeq \sqrt{\Delta m^2_{31}}$: \\
In this case, we can write $|M_{e\mu}|$ as
 \begin{eqnarray}
|M_{ e\mu}|\simeq\sqrt{\Delta m^2_{31}}\,c_{13}\,c_{23}\Big|c_{12}s_{12}(e^{2i\alpha} - 1) + s_{13}\Big[-(c_{12}^2 + s_{12}^2\,e^{2i\alpha}) +
\sqrt{2}\,e^{2i(\beta - \delta)}\Big]e^{i\delta}\Big| \,,
\end{eqnarray}
where we have assumed $s_{23} \simeq c_{23}$. It is observed that $|M_{e\mu}|$ can 
vanish in this case. Using $\beta = \delta = 0^{\circ}$, vanishing of $|M_{e\mu}|$ implies 
\begin{eqnarray}
\label{alpha}
\cos2\alpha\approx\frac{\Big[s_{12}c_{12}-s^{2}_{12}s_{13}\Big]^2+\Big[s_{12}c_{12}-(\sqrt{2}-c^{2}_{12})s_{13}\Big]^2}{2\Big[s_{12}c_{12}-
s^{2}_{12}s_{13}][s_{12}c_{12}-(\sqrt{2}-c^{2}_{12})s_{13}\Big]}\approx 1. 
\end{eqnarray}
If we neglect $\mathcal O(s_{13}^2)$ terms from Eq.~\ref{alpha}, then we get $\cos2\alpha = 1$, i.e, $\alpha$ value has to be either $0^{\circ}$ or 
$180^{\circ}$ to have vanishing $|M_{e\mu}|$. Hence, in this case, the Majorana CP violating phase $\alpha$ has to be either close to zero or $180^{\circ}$ 
to have vanishing $|M_{e\mu}|$.
\end{itemize}

\begin{figure}[htbp!]
\begin{subfigure}{0.30\textwidth}
\includegraphics[width=5cm,height=4cm]{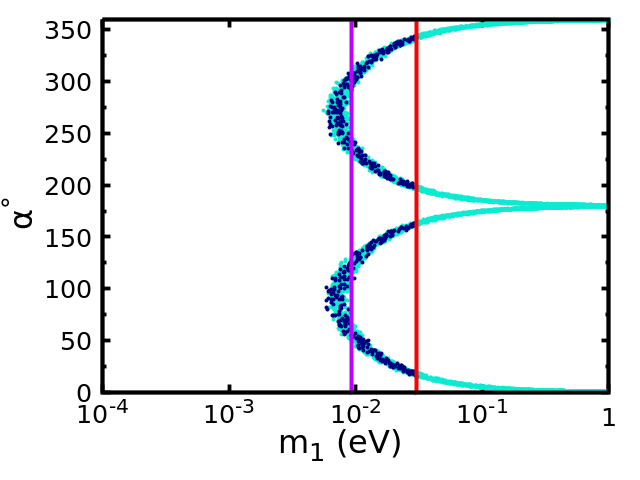}
\caption{}
\label{fig:3.1a}
\end{subfigure}
\begin{subfigure}{0.30\textwidth}
\includegraphics[width=5cm,height=4cm]{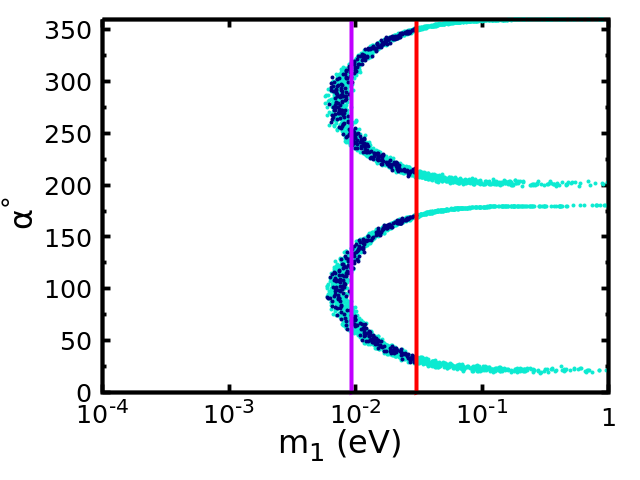}
\caption{}
\label{fig:3.1b}
\end{subfigure}
\begin{subfigure}{0.30\textwidth}
\includegraphics[width=5cm,height=4cm]{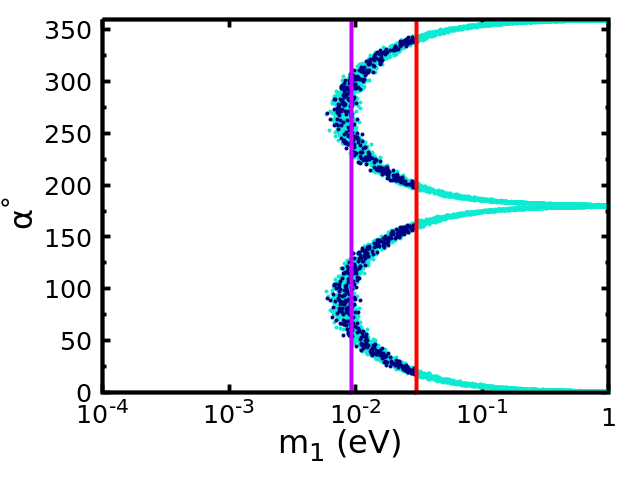}
\caption{}
\label{fig:3.1c}
\end{subfigure}
\begin{subfigure}{0.30\textwidth}
\includegraphics[width=5cm,height=4cm]{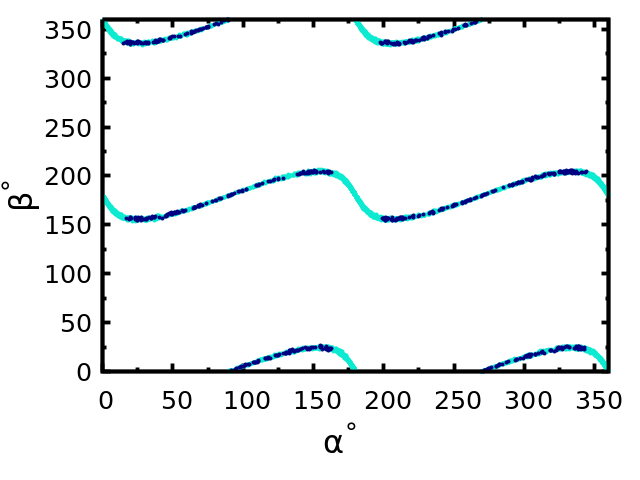}
\caption{}
\label{fig:3.1d}
\end{subfigure} 
\begin{subfigure}{0.30\textwidth}
\includegraphics[width=5cm,height=4cm]{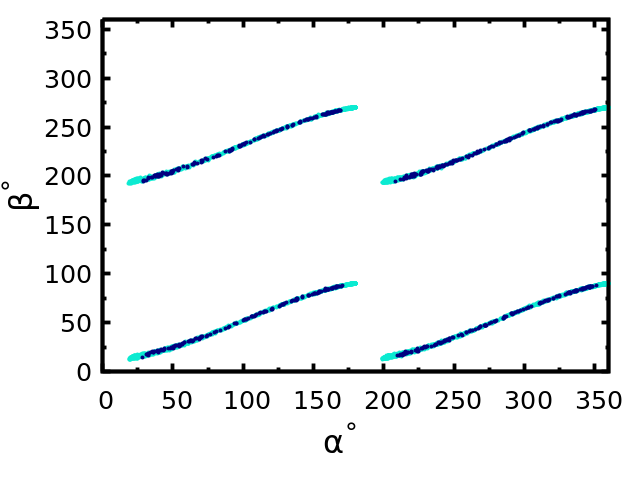}
\caption{}
\label{fig:3.1e}
\end{subfigure} 
\begin{subfigure}{0.30\textwidth}
\includegraphics[width=5cm,height=4cm]{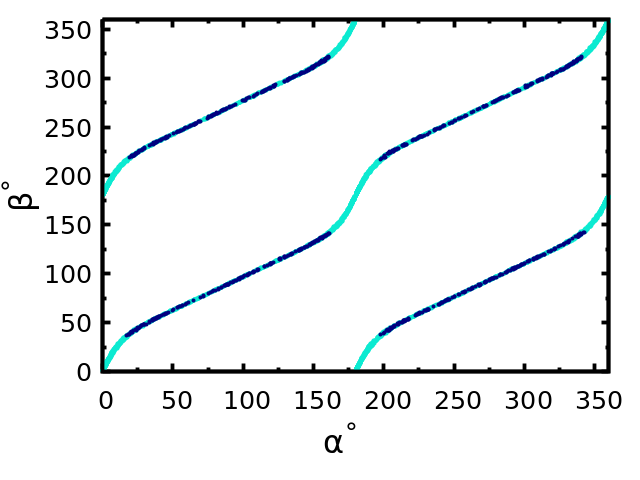}
\caption{}
\label{fig:3.1f}
\end{subfigure} 
\captionsetup{justification=raggedright,singlelinecheck=false}
\caption{Correlation between $\beta$ and $\alpha$, and between $\alpha$ and $m_{1}$ for $\delta=0^{\circ}$~(Left panel), 
$\delta=90^{\circ}$~(Middle panel) and $\delta=180^{\circ}$~(Right panel) with vanishing $|M_{e\mu}|$ in case of normal ordering. The red and magenta lines 
represent $\sum m_i = 0.12\,{\rm eV}$~\cite{Zhang:2020mox} and $\sum m_i = 0.072\,{\rm eV}$~\cite{DESI:2024mwx}, respectively.}
\label{fig:3.1}
\end{figure}

In Fig~\ref{fig:3.1}, we show correlation between the unknown parameters with vanishing $|M_{e\mu}|$ for several values of the Dirac CP 
violating phase $\delta$. The blue region is obtained using the constraint coming from the upper bound of
$\sum m_{i} < 0.12\,{\rm  eV}$ on the total neutrino mass, whereas, the cyan region is obtained without the total neutrino mass
constraint.
Fig.~\ref{fig:3.1a}, Fig.~\ref{fig:3.1b} and Fig.~\ref{fig:3.1c} represent scatter plot of the 
lightest neutrino mass $m_1$ against the Majorana CP violating phase $\alpha$ for $\delta = 0^{\circ}$, $90^{\circ}$ and $180^{\circ}$, 
respectively. It is observed that correlation between $\alpha$ and $m_1$ looks quite similar for $\delta = 0^{\circ}$ and 
$\delta = 180^{\circ}$. $|M_{e\mu}|$ can vanish for all values of $\alpha$ value with $\delta = 0^{\circ}$ and 
$\delta = 180^{\circ}$. However, with $\delta = 90^{\circ}$, value of $\alpha$ should lie either between $(20^{\circ}, 180^{\circ})$ and
between $(200^{\circ}, 360^{\circ})$ to have vanishing $|M_{e\mu}|$. 
The lightest neutrino mass $m_1 \ge 6.0\times 10^{-3}\,{\rm eV}$ to have vanishing $|M_{e\mu}|$. Similarly, in 
Fig.~\ref{fig:3.1d}, Fig.~\ref{fig:3.1e} and Fig.~\ref{fig:3.1f}, we show the correlation between the 
two Majorana CP violating phases, namely $\alpha$ and $\beta$. It is observed that the correlation between $\alpha$ and $\beta$ does depend
on the value of $\delta$. We have also shown the correlation between $\theta_{23}$ and $\alpha$ with vanishing $|M_{e\mu}|$ in Fig.~\ref{fig:Nmalpth23emuzero_no} of Appendix~\ref{app1}.

\subsubsection{Inverted Ordering}
\begin{figure}[htbp!]
\begin{subfigure}{0.35\textwidth}
\includegraphics[width=\textwidth]{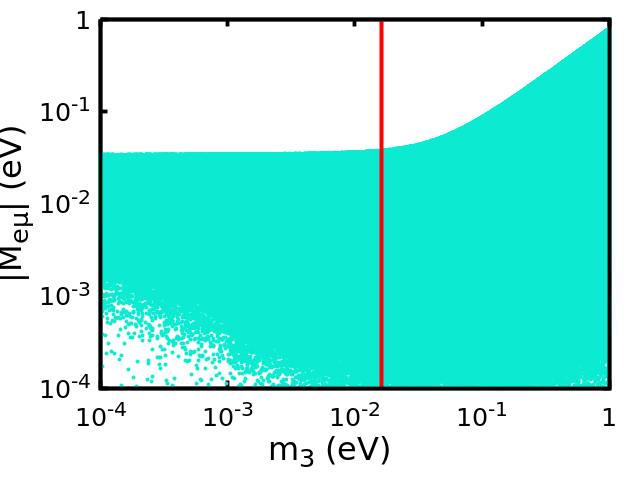}
\end{subfigure}
\captionsetup{justification=raggedright,singlelinecheck=false}
\caption{Variation of $|M_{e\mu}|$ with the lightest neutrino mass $m_3$ and the three CP violating phases $\alpha$, $\beta$ and $\delta$ in 
case of inverted ordering. Total neutrino mass $\sum m_{i} = 0.12\,{\rm  eV}$~\cite{Zhang:2020mox} is shown with the red vertical line.}
\label{fig:4}
\end{figure}

\begin{figure}[htbp!]
\begin{subfigure}{0.30\textwidth}
\includegraphics[width=\textwidth]{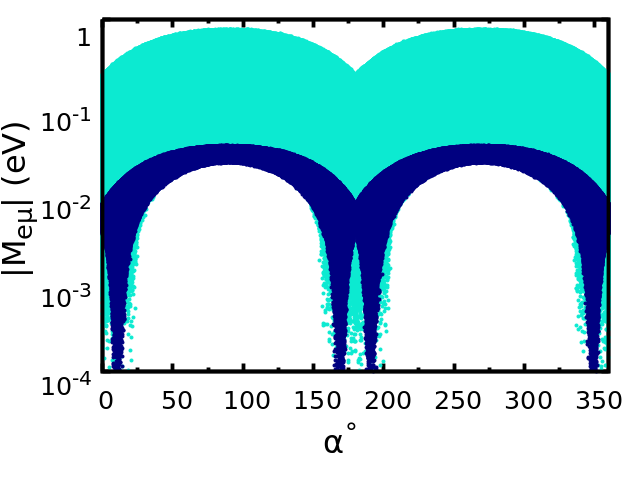}
\caption{}
\label{fig:4b}
\end{subfigure}
\begin{subfigure}{0.30\textwidth}
\includegraphics[width=\textwidth]{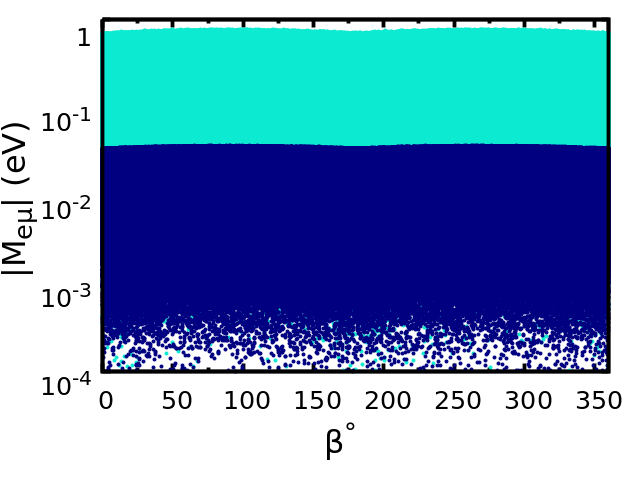}
\caption{}
\label{fig:4c}
\end{subfigure}
\begin{subfigure}{0.30\textwidth}
\includegraphics[width=\textwidth]{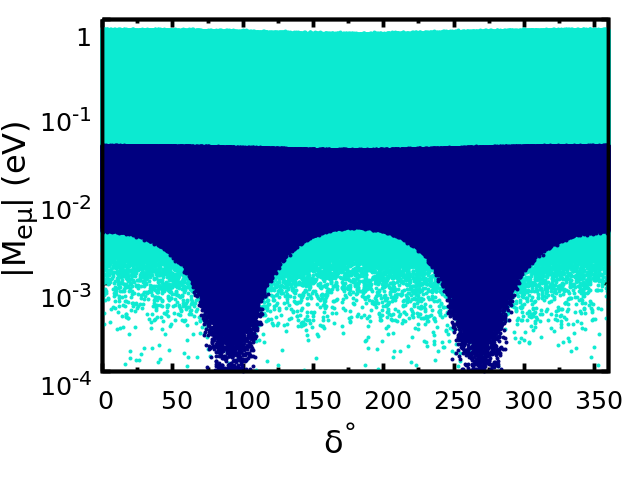}
\caption{}
\label{fig:4d}
\end{subfigure}
\caption{Variation of $|M_{e\mu}|$ with the three CP violating phases $\alpha$, $\beta$ and $\delta$ in 
case of inverted ordering.}
\label{fig:4_cp_phases}
\end{figure}

In Fig.~\ref{fig:4}, we plot $|M_{e\mu}|$ against the lightest neutrino mass $m_3$. The total neutrino mass $\sum m_{i} = 0.12\,{\rm eV}$ is 
shown with a red vertical line. 
In Fig.~\ref{fig:4_cp_phases}, we show $|M_{e\mu}|$ as function of the CP violating phases $\alpha$, 
$\beta$ and $\delta$ with~(blue band) and without~(cyan band) the constraint coming from the total neutrino mass 
$\sum m_{i} < 0.12\,{\rm  eV}$.

Our observations are as follows:
\begin{itemize}
 \item $|M_{e\mu}|$ can vanish for all values of the lightest neutrino mass $m_3$.
 \item We get vanishing $|M_{e\mu}|$ for a very limited parameter range of the Majorana CP violating phase $\alpha$. The allowed range of 
$\alpha$ is found to be $(8^{\circ},\,14^{\circ})$, $(166^{\circ},\,172^{\circ})$, $(188^{\circ},\,194^{\circ})$ and 
$(346^{\circ},\,352^{\circ})$ once we impose $\sum m_i < 0.12\,{\rm eV}$ constraint. Without the total neutrino mass constraint, 
$|M_{e\mu}|$ can vanish if the $\alpha$ value lies within $(0^{\circ}, \,20^{\circ})$, $(160^{\circ}, \,200^{\circ})$ and $(340^{\circ}, \,360^{\circ})$.
 \item There is no constraint on the value of $\beta$, i.e, $|M_{e\mu}|$ can vanish for any value of $\beta$ within $(0,\,360^{\circ})$ 
irrespective of any value of the lightest neutrino mass $m_3$.
\item With $\sum m_i < 0.12\,{\rm eV}$ constraint, $|M_{e\mu}|$ can vanish only if the Dirac CP violating phase $\delta$ lies within 
$(75^{\circ},\,105^{\circ})$ and $(255^{\circ},\,285^{\circ})$. However, without the constraint, it can vanish for any value of $\delta$ 
within $(0,\,360^{\circ})$.
 \end{itemize}

We now proceed to analyse vanishing of $M_{e\mu}$ for different values of the lightest neutrino mass $m_3$. In case of inverted ordering the $M_{e\mu}$ can be written as
\begin{eqnarray}
M_{ e\mu}&=&c_{12}c_{13}(-s_{12}c_{23}-c_{12}s_{23}s_{13}e^{i\delta})\,\sqrt{m_3^2 + \Delta m^2_{13}}+s_{12}c_{13}(c_{12}c_{23}-
s_{12}s_{23}s_{13}e^{i\delta})e^{2i\alpha}\,\sqrt{m_3^2 + \Delta m^2_{13} + \Delta m^2_{21}} \nonumber \\
&&
+ (c_{13}s_{23}s_{13}e^{-i\delta})\,m_3\,e^{2i\beta} 
\end{eqnarray}
\begin{itemize}
\item {\bf Case I:} $m_{3}<<\sqrt{\Delta m^2_{21}}$\\
In this case, we can write $|M_{e\mu}|$ as
\begin{eqnarray}
|M_{ e\mu}| \simeq \sqrt{\Delta m^2_{13}}\Big|c_{12}c_{13}(-s_{12}c_{23}-c_{12}s_{23}s_{13}e^{i\delta}) + s_{12}c_{13}(c_{12}c_{23}-
s_{12}s_{23}s_{13}e^{i\delta})e^{2i\alpha}\Big|
\end{eqnarray}
If we assume $s_{23} \simeq c_{23}$, then $|M_{e\mu}|$ can be simplified as
\begin{eqnarray}
|M_{ e\mu}| \simeq \sqrt{\Delta m^2_{13}}\,c_{13}\,s_{23}\Big|c_{12}s_{12}(e^{2i\alpha} - 1) - e^{i\delta}\,(c_{12}^2 + s_{12}^2\,
e^{2i\alpha})s_{13}\Big|
\end{eqnarray}
 It is clear from 
Fig.~\ref{fig:4b} and Fig.~\ref{fig:4d} that for $\alpha$ value very close to 
$0^{\circ}$ and $180^{\circ}$, and $\delta$ value close to $90^{\circ}$ and $270^{\circ}$, $|M_{e\mu}|$ vanishes. It can be shown from the above equation that at $\sin2\alpha \simeq s_{13}/s_{12}c_{12} 
= 0.31$ or $\alpha \simeq 9^{\circ}$, we can have vanishing $|M_{e\mu}|$.  

\item {\bf Case II:} $m_{3} >> \sqrt{\Delta m^2_{13}}$: \\
In this case
\begin{eqnarray}
\label{memuinv}
&&|M_{ e\mu}| = m_3\Big|c_{12}s_{12}c_{13}c_{23}(e^{2i\alpha} - 1) + c_{13}s_{23}s_{13}\Big[-(c_{12}^2 + s_{12}^2\,e^{2i\alpha}) \nonumber \\
  &&\hspace{1.5cm}+ 
e^{2i(\beta - \delta)}\Big]\,e^{i\delta}\Big|\,.
\end{eqnarray}
As expected, Eq.~\ref{memuinv} and Eq.~\ref{memuno} are exactly same. Hence, $|M_{ e\mu}|$ will vanish if we have $\alpha = 0^{\circ}, 180^{\circ}$ and 
$\beta - \delta = 0^{\circ}, 180^{\circ}$.

\item {\bf Case III:} $m_{3}\approx \sqrt{\Delta m^2_{13}}$: \\
In this case
  \begin{eqnarray}
  \label{eq:16}
   |M_{e\mu}|\simeq \sqrt{\Delta m^2_{13}}\,c_{13}\,s_{23}\,\Big|\sqrt{2}s_{12}c_{12}(e^{2i\alpha}-1)+s_{13}(e^{2i(\beta-\delta)}-\sqrt{2}
c^{2}_{12}-\sqrt{2}s^{2}_{12}e^{2i\alpha})e^{i\delta}\Big|\,, 
  \end{eqnarray}
where we have assumed $s_{23} \simeq c_{23}$. It is evident from 
Fig.~\ref{fig:4d} that the $\delta$ value has to be close to $90^{\circ}$ or $270^{\circ}$ to have vanishing $|M_{e\mu}|$. Now,
from the above equation it is clear that for very small value of $\sin\alpha \simeq 0$ and $\beta - \delta = 0^{\circ},\,180^{\circ}$, the first and the 
second term can be close to zero. It is found numerically that at $\alpha$ value around $2^{\circ}$, $|M_{e\mu}|$ will vanish. It is, indeed, 
very important to note that for inverted ordering, $|M_{ e\mu}|$ will vanish for any value of the lightest neutrino mass.
\end{itemize}

\begin{figure}[htbp!]
\begin{subfigure}{0.30\textwidth}
\includegraphics[width=5cm,height=4cm]{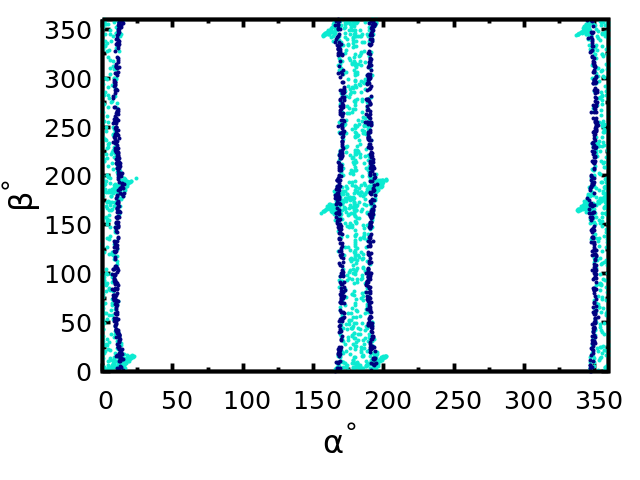}
\caption{}
\label{fig:4.1a}
\end{subfigure}
\begin{subfigure}{0.30\textwidth}
\includegraphics[width=5cm,height=4cm]{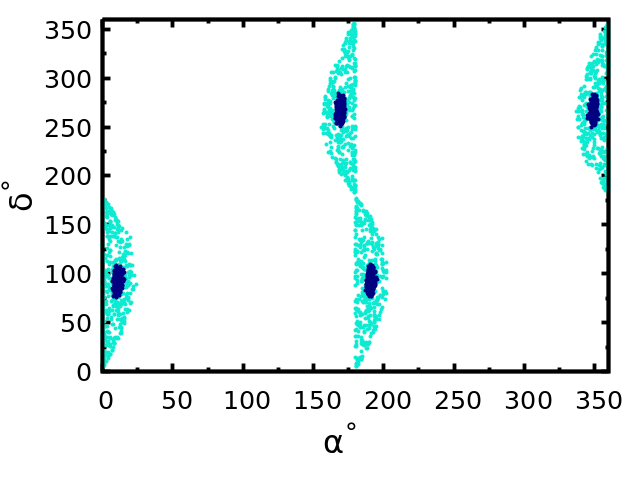}
\caption{}
\label{fig:4.1b}
\end{subfigure}
\begin{subfigure}{0.30\textwidth}
\includegraphics[width=5cm,height=4cm]{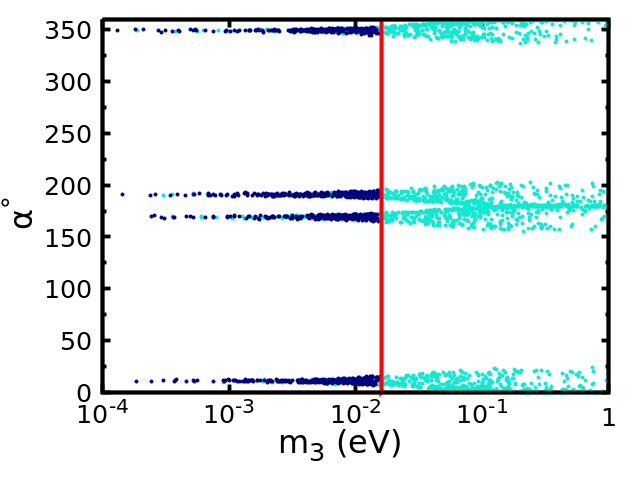}
\caption{}
\label{fig:4.1c}
\end{subfigure}
\captionsetup{justification=raggedright,singlelinecheck=false}
\caption{Correlation between $\alpha$ and $\beta$, $\alpha$ and $\delta$, and between $\alpha$ and $m_{3}$ with vanishing $|M_{e\mu}|$ 
in case of inverted ordering. The red line in Fig.~\ref{fig:4.1c} represents $\sum m_i = 0.12\,{\rm eV}$~\cite{Zhang:2020mox}.}
\label{fig:4.1}
\end{figure}

In Fig.~\ref{fig:4.1}, we show the correlation between $\alpha$ and $\beta$, $\alpha$ and $\delta$, and between $\alpha$ and $m_{3}$ with 
vanishing $|M_{e\mu}|$ in case of inverted ordering. The region shown with blue is obtained by using the constraint coming from 
$\sum m_i < 0.12\,{\rm eV}$, whereas, the cyan region is obtained without imposing the total neutrino mass constraint. It is observed 
that we can have vanishing $|M_{e\mu}|$ for any value of the lightest neutrino mass $m_3$. For
vanishing $|M_{e\mu}|$, $\alpha$ value lies very close to $0^{\circ}$, $180^{\circ}$ and $360^{\circ}$. It is also evident from 
Fig.~\ref{fig:4.1b} that the value of the Dirac CP violating phase $\delta$ will be restricted to a very narrow range around $90^{\circ}$
and $270^{\circ}$ once we impose the $\sum m_i < 0.12\,{\rm eV}$ constraint. However, no such restrictions on $\beta$ appears even with
the total neutrino mass constraint. We have shown the correlation between $\theta_{23}$ and $\alpha$ with vanishing $|M_{e\mu}|$ in case of inverted ordering in Fig.~\ref{fig:Imalpth23emuzero_io} of Appendix~\ref{app1}.

\subsection{$M_{e\tau}$: the $e\tau$ element of the mass matrix}
\begin{figure}[htbp!]
\begin{subfigure}{0.35\textwidth}
\includegraphics[width=\textwidth]{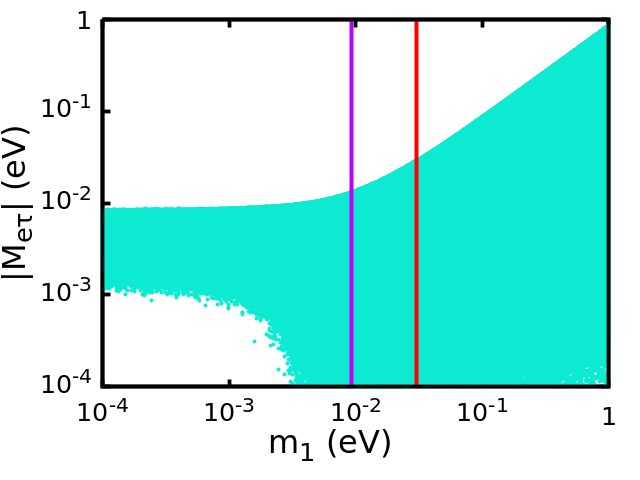}
\label{fig:5a}
\end{subfigure}
\captionsetup{justification=raggedright,singlelinecheck=false}
\caption{Variation of $|M_{e\tau}|$ with the lightest neutrino mass $m_1$ in 
case of normal ordering. Total neutrino mass $\sum m_{i} = 0.12\,{\rm  eV}$~\cite{Zhang:2020mox} and $\sum m_{i} = 0.072\,{\rm  eV}$~\cite{DESI:2024mwx} are 
shown with red and magenta lines, respectively.}
\label{fig:5}
\end{figure}

\begin{figure}[htbp!]
\begin{subfigure}{0.32\textwidth}
\includegraphics[width=\textwidth]{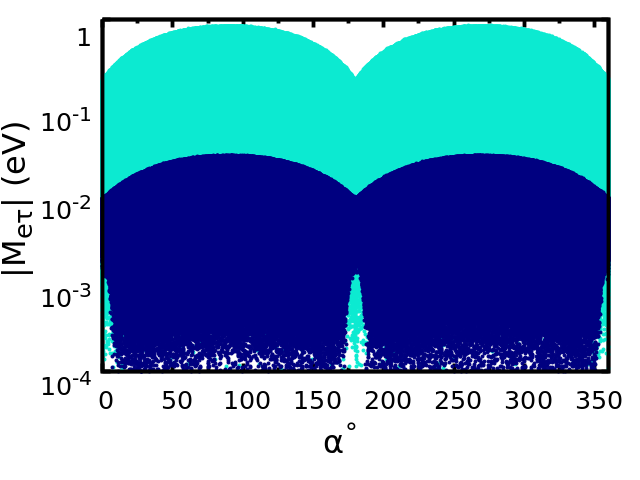}
\caption{}
\label{fig:5b}
\end{subfigure}
\begin{subfigure}{0.32\textwidth}
\includegraphics[width=\textwidth]{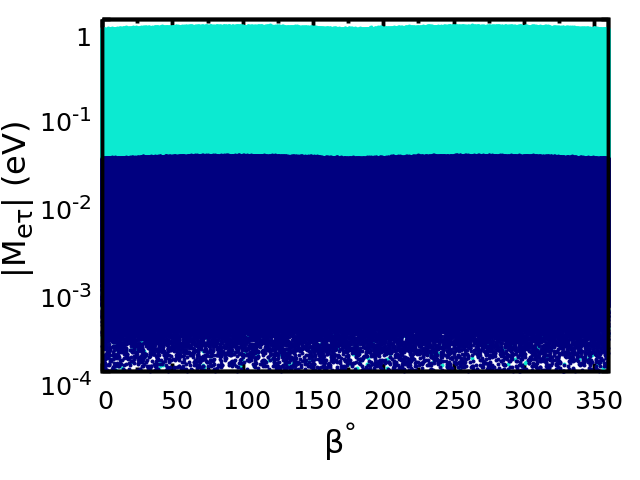}
\caption{}
\label{fig:5c}
\end{subfigure}
\begin{subfigure}{0.32\textwidth}
\includegraphics[width=\textwidth]{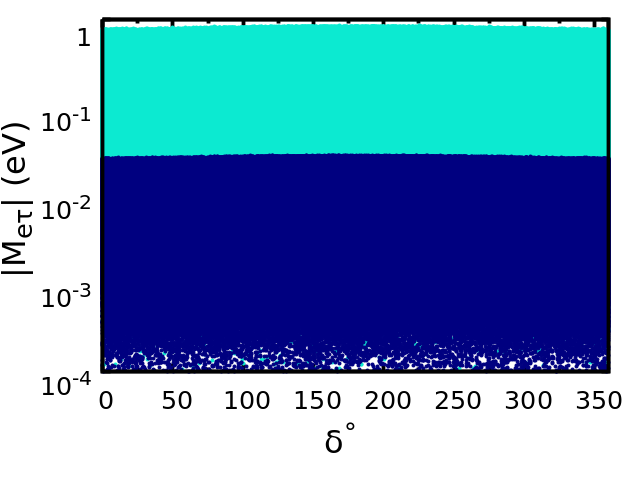}
\caption{}
\label{fig:5d}
\end{subfigure}
\caption{Variation of $|M_{e\tau}|$ with the three CP violating phases $\alpha$, $\beta$ and $\delta$ in 
case of normal ordering.}
\label{fig:5_cp_phases}
\end{figure}

\begin{figure}[htbp!]
\begin{subfigure}{0.35\textwidth}
\includegraphics[width=\textwidth]{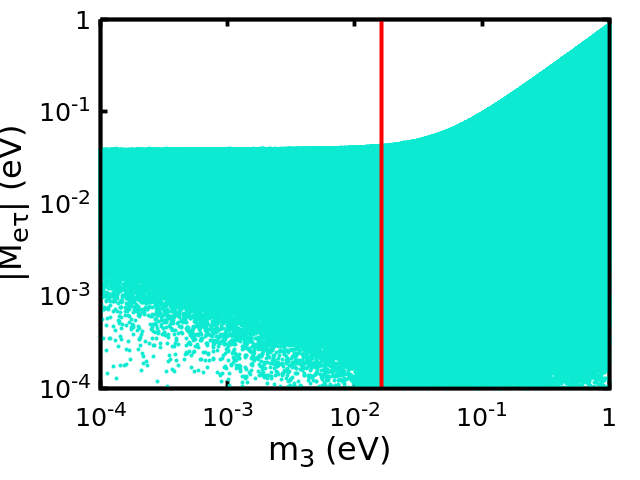}
\caption{}
\label{fig:6a}
\end{subfigure}
\captionsetup{justification=raggedright,singlelinecheck=false}
\caption{Variation of $|M_{e\tau}|$ with the lightest neutrino mass $m_3$ in case of inverted ordering. Total neutrino mass $\sum m_{i} = 0.12\,{\rm  eV}$~\cite{Zhang:2020mox} is shown with the red vertical line.}
\label{fig:6}
\end{figure}

\begin{figure}[htbp!]
\begin{subfigure}{0.325\textwidth}
\includegraphics[width=\textwidth]{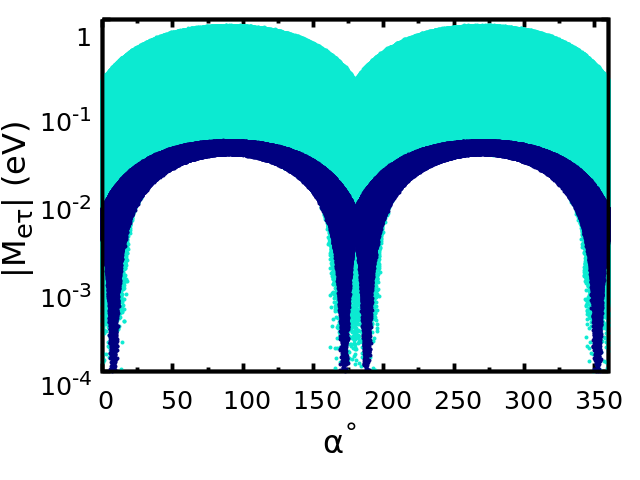}
\caption{}
\label{fig:6b}
\end{subfigure}
\begin{subfigure}{0.325\textwidth}
\includegraphics[width=\textwidth]{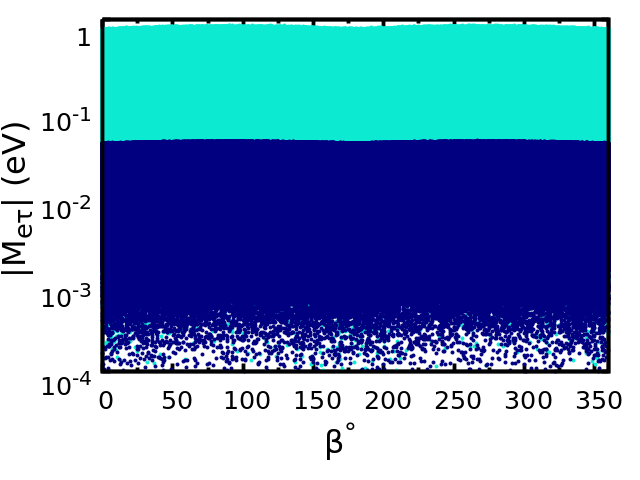}
\caption{}
\label{fig:6c}
\end{subfigure}
\begin{subfigure}{0.325\textwidth}
\includegraphics[width=\textwidth]{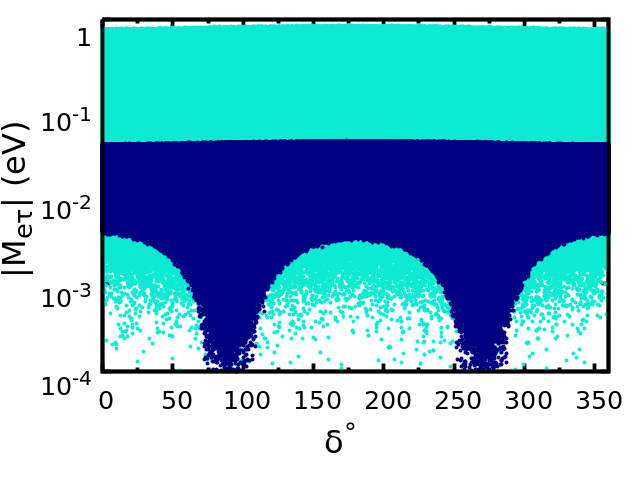}
\caption{}
\label{fig:6d}
\end{subfigure}
\caption{Variation of $|M_{e\tau}|$ with the three CP violating phases $\alpha$, $\beta$ and $\delta$ in 
case of inverted ordering.}
\label{fig:6_cp_phases}
\end{figure}

We plot in Fig.~\ref{fig:5} and Fig.~\ref{fig:6} variation of $|M_{e\tau}|$ with the lightest neutrino mass $m_{1}$~$(m_{3})$ for normal and
inverted orderings, respectively. The total neutrino mass $\sum m_{i} = 0.12\,{\rm  eV}$ and $\sum m_{i} = 0.072\,{\rm  eV}$ are shown with 
red and magenta lines, respectively. We also show the variation of $|M_{e\tau}|$ with CP violating phases $\alpha$, $\beta$ and $\delta$ in 
Fig.~\ref{fig:5_cp_phases} and Fig.~\ref{fig:6_cp_phases}, respectively. The blue band is obtained by imposing the constraint coming from 
$\sum m_{i} < 0.12\,{\rm  eV}$, whereas, the cyan band is obtained without the constraint on the total neutrino mass.
The upper bound on the total neutrino mass imposes constraint on the lightest neutrino mass $m_1$ and on
$|M_{e\tau}|$. Nevertheless, we have vanishing $|M_{e\tau}|$ irrespective of the value of the total neutrino mass. 

For the normal ordering, our observations are as follows:
\begin{itemize}
 
 \item For vanishing $|M_{e\tau}|$, the lightest neutrino mass has to be $m_{1} \geq 3.0\times 10^{-3}\,{\rm eV}$.
 \item For vanishing $|M_{e\tau}|$, the Majorana CP violating phase $\alpha$ has to lie within $(7^{\circ},\,173^{\circ})$ or 
$(187^{\circ},\,353^{\circ})$ if we impose $\sum m_i < 0.12\,{\rm eV}$ constraint. However, $|M_{e\tau}|$ can vanish for any
value of $\alpha$ without the total neutrino mass constraint.
 \item   There is no constraint on the values of $\beta$ and $\delta$ for vanishing $|M_{e\tau}|$, i.e, $|M_{e\tau}|$ can vanish for any
value of  $\beta$ and $\delta$ within $(0,\,360^{\circ})$.
 \end{itemize}

For the inverted ordering, our observations are as follows:
\begin{itemize}
 \item $|M_{e\tau}|$ can vanish for any value of the lightest neutrino mass $m_3$.
 \item For vanishing $|M_{e\tau}|$, $\alpha$ value has to lie within $(6^{\circ},\,10^{\circ})$, $(170^{\circ},\,174^{\circ})$,
$(186^{\circ},\,190^{\circ})$ and $(350^{\circ},\,354^{\circ})$ with $\sum m_i < 0.12\,{\rm eV}$ constraint. However, without the 
constraint, $\alpha$ value has to lie within $(0^{\circ}, \,15^{\circ})$, $(165^{\circ}, \,195^{\circ})$ and $(345^{\circ}, \,360^{\circ})$ to have vanishing $|M_{e\tau}|$. 
 \item  There is no constraint on $\beta$, i.e, $|M_{e\tau}|$ can vanish for any value of $\beta$. 
\item For vanishing $|M_{e\tau}|$, the Dirac CP violating phase $\delta$ has to lie between $(75^{\circ},\,105^{\circ})$ and 
$(255^{\circ},\,285^{\circ})$ if we impose $\sum m_i < 0.12\,{\rm eV}$ constraint. However, $|M_{e\tau}|$ can vanish for any
value of $\delta$ within $(0,\,360^{\circ})$ without the total neutrino mass constraint.
 \end{itemize}

It is important to note that Fig.~\ref{fig:5} and Fig.~\ref{fig:6} are quite similar to the Fig.~\ref{fig:3} and Fig.~\ref{fig:4},
respectively. This feature is a direct consequence of the approximate $\mu - \tau$ symmetry of the neutrino mass matrix in the charged lepton
basis. Let us analyse this feature analytically.
In general, the $M_{e\tau}$ can be written as
 \begin{eqnarray}
 \label{metau}
  M_{ e\tau}=c_{12}c_{13}(s_{12}s_{23}-c_{12}c_{23}s_{13}e^{i\delta})\,m_1+s_{12}c_{13}(-c_{12}s_{23}-s_{12}c_{23}s_{13}e^{i\delta})\,
m_2 e^{2i\alpha}+ (c_{13}c_{23}s_{13}e^{-i\delta})\,m_3 e^{2i\beta}.  
 \end{eqnarray}
 It is evident from Eq.~\ref{memu} and Eq.~\ref{metau} that $M_{e\mu}$ and $M_{e\tau}$ are quite similar to each other. By using the
 transformation $s_{23} \to c_{23}$ and $c_{23} \to -s_{23}$, one can obtain $M_{ e\tau}$ from $M_{e\mu}$. 

\begin{figure}[htbp!]
\begin{subfigure}{0.30\textwidth}
\includegraphics[width=5cm,height=4cm]{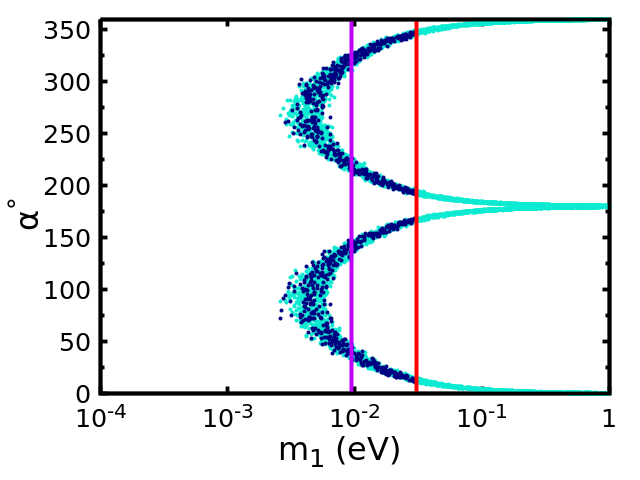}
\caption{}
\label{fig:5.1a}
\end{subfigure}
\begin{subfigure}{0.325\textwidth}
\includegraphics[width=5.2cm,height=4.3cm]{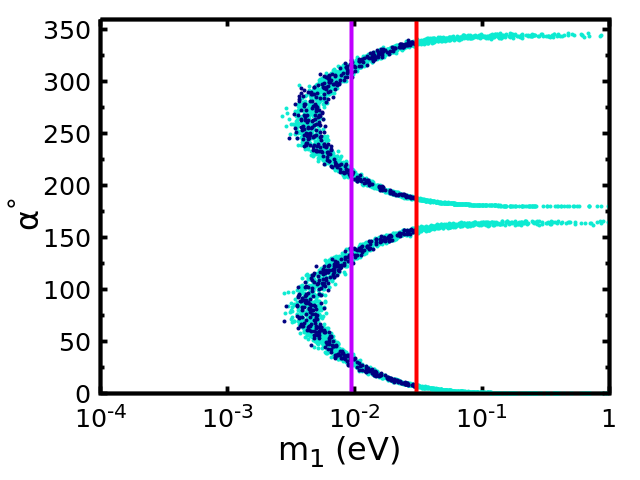}
\caption{}
\label{fig:5.1b}
\end{subfigure}
\begin{subfigure}{0.325\textwidth}
\includegraphics[width=5.2cm,height=4.3cm]{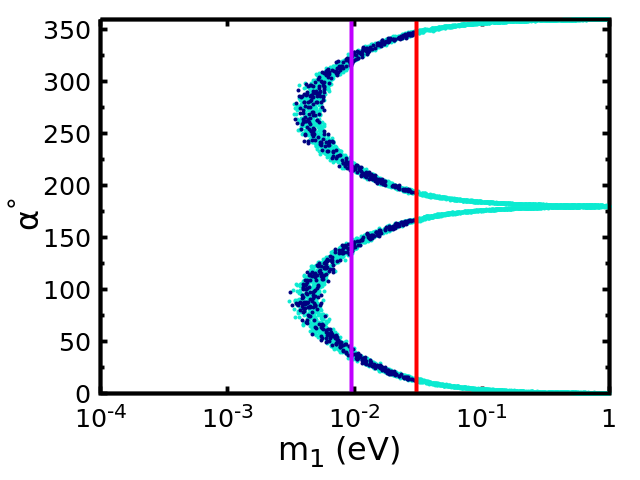}
\caption{}
\label{fig:5.1c}
\end{subfigure}
\begin{subfigure}{0.325\textwidth}
\includegraphics[width=5.2cm,height=4.3cm]{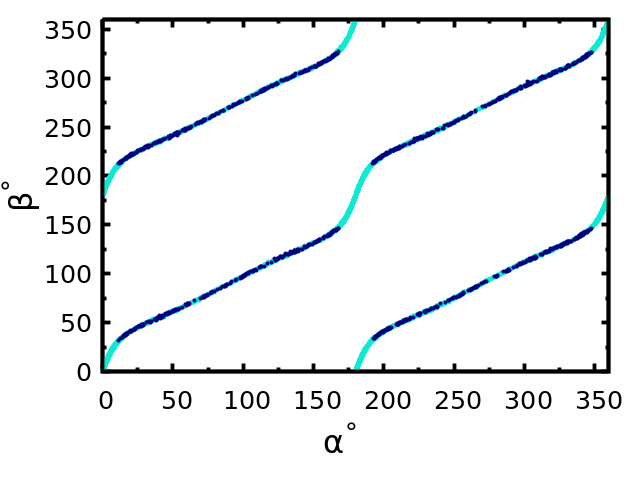}
\caption{}
\label{fig:5.1d}
\end{subfigure}
 \begin{subfigure}{0.325\textwidth}
\includegraphics[width=5.2cm,height=4.3cm]{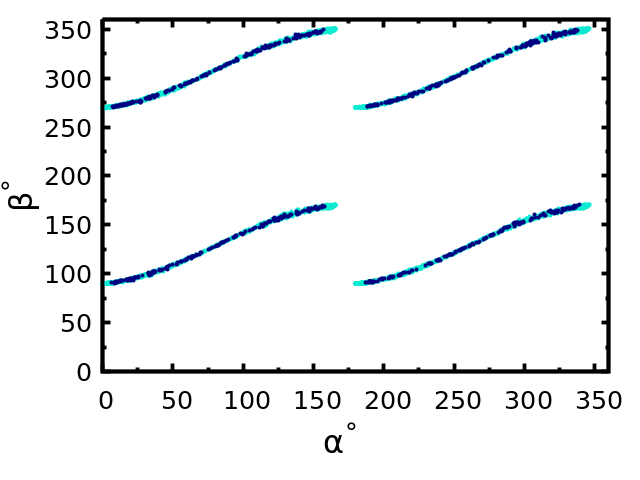}
\caption{}
\label{fig:5.1e}
\end{subfigure}
 \begin{subfigure}{0.325\textwidth}
\includegraphics[width=5.2cm,height=4.3cm]{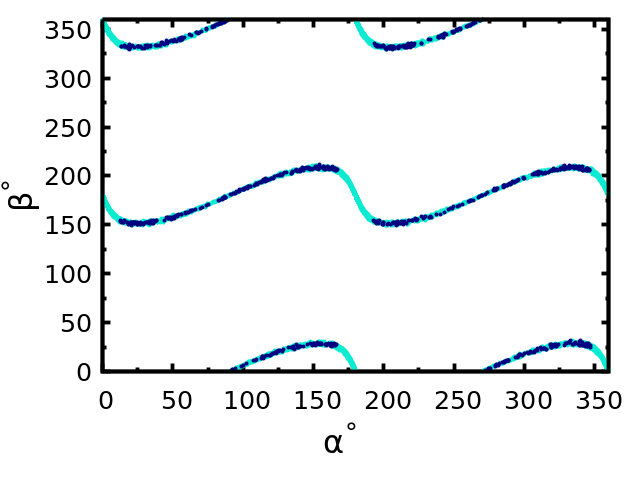}
\caption{}
\label{fig:5.1f}
\end{subfigure}
\captionsetup{justification=raggedright,singlelinecheck=false}
\caption{Correlation between $\beta$ and $\alpha$, and between $\alpha$ and $m_{1}$ for $\delta=0^{\circ}$~(Left panel), 
$\delta=90^{\circ}$~(Middle panel) and $\delta=180^{\circ}$~(Right panel) with vanishing $|M_{e\tau}|$ in case of normal ordering. The red and magenta lines 
represent $\sum m_i = 0.12\,{\rm eV}$~\cite{Zhang:2020mox} and $\sum m_i = 0.072\,{\rm eV}$~\cite{DESI:2024mwx}, respectively.}
\label{fig:5.1}
\end{figure}

\begin{figure}[h!]
\begin{subfigure}{0.30\textwidth}
\includegraphics[width=5cm,height=4cm]{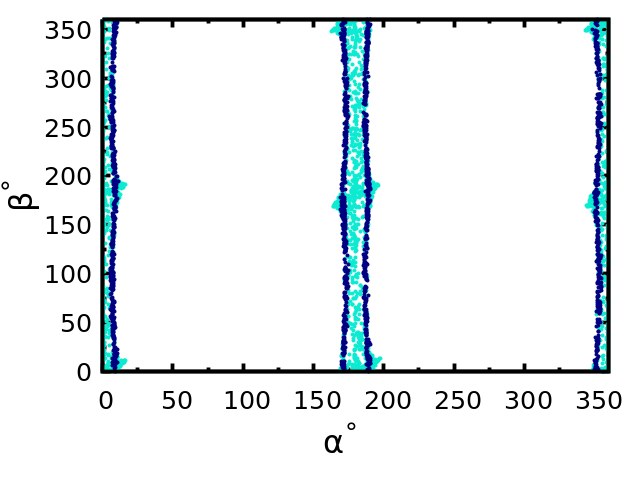}
\caption{}
\label{fig:6.1a}
\end{subfigure}
\begin{subfigure}{0.30\textwidth}
\includegraphics[width=5cm,height=4cm]{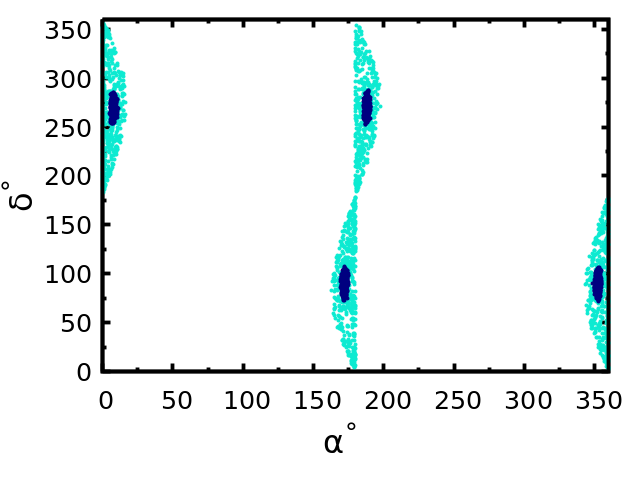}
\caption{} 
\label{fig:6.1b} 
\end{subfigure}
\begin{subfigure}{0.30\textwidth}
\includegraphics[width=5cm,height=4cm]{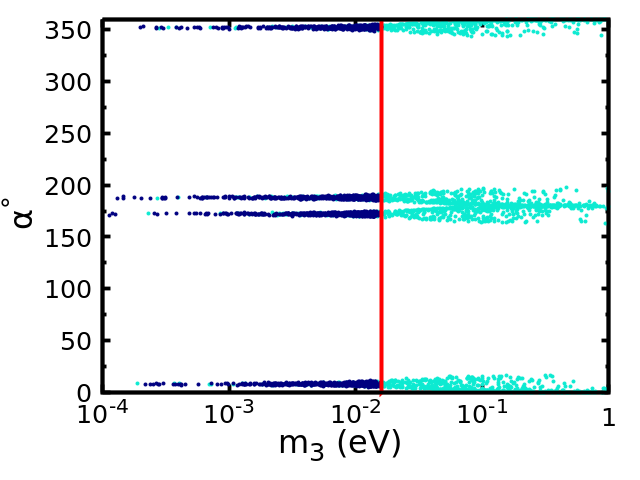}
\caption{}
\label{fig:6.1c}
\end{subfigure}
\captionsetup{justification=raggedright,singlelinecheck=false}
\caption{Correlation between $\alpha$ and $\beta$, $\alpha$ and $\delta$, and between $\alpha$ and $m_{3}$ with vanishing $|M_{e\tau}|$ 
in case of inverted ordering. The red line in Fig.~\ref{fig:6.1c} represents $\sum m_i = 0.12\,{\rm eV}$~\cite{Zhang:2020mox}.}
\label{fig:6.1}
\end{figure}

We plot in Fig.~\ref{fig:5.1} the correlation between $\beta$ and $\alpha$, and between $\alpha$ and
$m_{1}$ for several values of the CP violating phase $\delta$ with vanishing $|M_{e\tau}|$ in case of normal ordering. 
The region shown with blue is obtained by using the constraint coming from $\sum m_i < 0.12\,{\rm eV}$, whereas, the cyan region is 
obtained without imposing the total neutrino mass constraint. It is evident 
from Fig.~\ref{fig:3.1} and Fig.~\ref{fig:5.1} that although the correlation
between $\alpha$ and $m_1$ looks quite similar for vanishing $|M_{e\mu}|$ and for vanishing $|M_{e\tau}|$, the correlation between $\beta$ 
and $\alpha$ in case of vanishing $|M_{e\tau}|$ looks very different from that of vanishing $|M_{e\mu}|$. 
We plot in Fig.~\ref{fig:6.1} the correlation between $\alpha$ and $\beta$, $\alpha$ and $\delta$, and between $\alpha$ and $m_{3}$ with 
vanishing $|M_{e\tau}|$
in case of inverted ordering. The correlation between $\alpha$ and $\beta$ and between $\alpha$ and $m_{3}$ of Fig.~\ref{fig:6.1a} and 
Fig.~\ref{fig:6.1c} looks quite similar to that of
Fig.~\ref{fig:4.1a} and Fig.~\ref{fig:4.1c}, respectively. It is, however, worth mentioning that correlation between $\alpha$ and $\delta$ of 
Fig.~\ref{fig:6.1b} is quite different
from that of Fig.~\ref{fig:4.1b}. It is evident that despite having approximate $\mu - \tau$ symmetry, these matrix elements are 
quite different. The CP violating phases play a very important role here. Furthermore, we show the correlation between $\theta_{23}$ and $\alpha$  for vanishing $|M_{e\tau}|$ in Fig.~\ref{fig:Nmalpth32etauzero_no} and Fig.~\ref{fig:Imalpth23etauzero_io} of Appendix~\ref{app1}, in case of normal and inverted ordering, respectively.  

\subsection{$M_{\mu\mu}$: the $\mu\mu$ element of the mass matrix}
\subsubsection{Normal Ordering:}
\begin{figure}[htbp!]
\begin{subfigure}{0.35\textwidth}
\includegraphics[width=\textwidth]{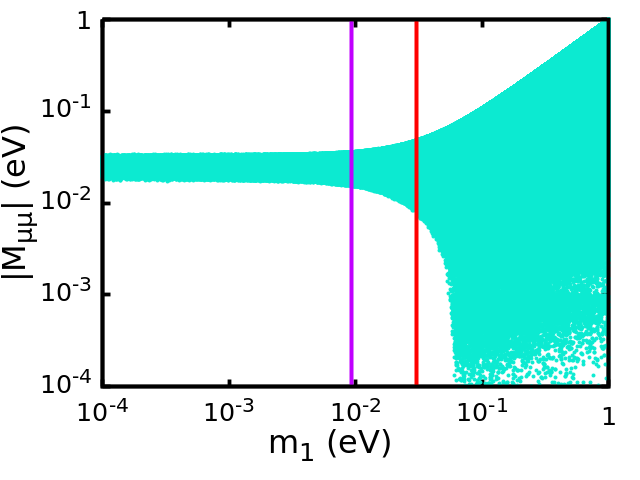}
\label{fig:7a}
\end{subfigure}
\captionsetup{justification=raggedright,singlelinecheck=false}
\caption{Variation of $|M_{\mu\mu}|$ with the lightest neutrino mass $m_1$ in case of normal ordering. Total neutrino mass $\sum m_{i} = 0.12\,{\rm  eV}$~\cite{Zhang:2020mox} and $\sum m_{i} = 0.072\,{\rm  eV}$~\cite{DESI:2024mwx} are shown with red and magenta lines, respectively.}
\label{fig:7}
\end{figure}

We plot in Fig.~\ref{fig:7} the variation of $|M_{\mu\mu}|$ with $m_1$ in case of normal ordering. The total neutrino mass 
$\sum m_{i} = 0.12\,{\rm  eV}$ and $\sum m_{i} = 0.072\,{\rm  eV}$ are shown with red and magenta lines, respectively. 

\begin{figure}[htbp!]
\begin{subfigure}{0.325\textwidth}
\includegraphics[width=\textwidth]{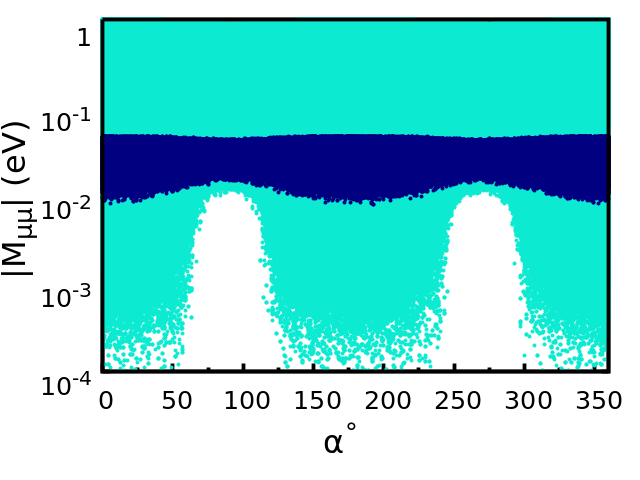}
\caption{}
\label{fig:7b}
\end{subfigure}
\begin{subfigure}{0.325\textwidth}
\includegraphics[width=\textwidth]{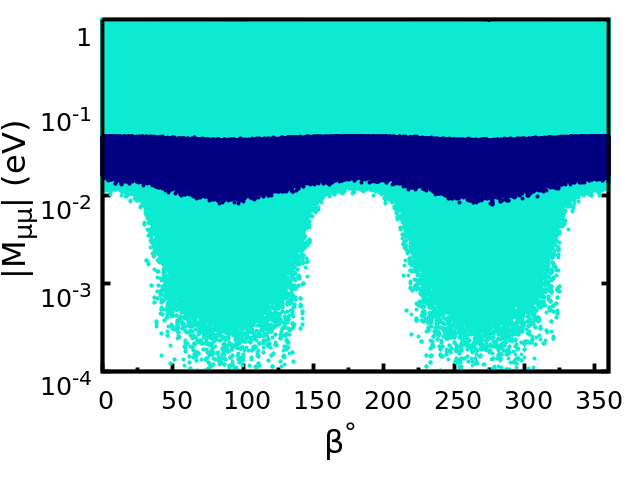}
\caption{}
\label{fig:7c}
\end{subfigure}
\begin{subfigure}{0.325\textwidth}
\includegraphics[width=\textwidth]{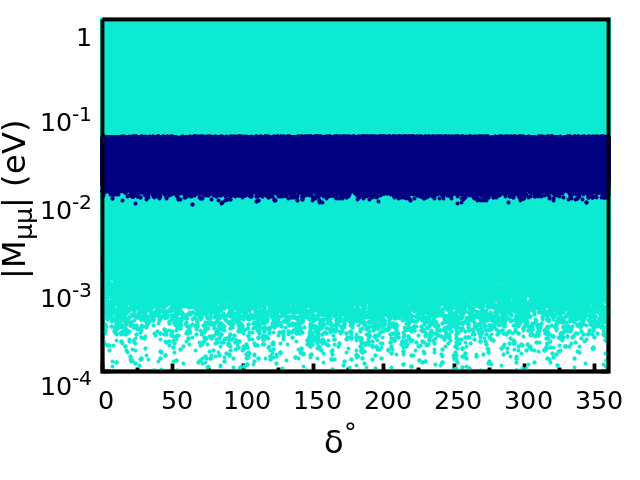}
\caption{}
\label{fig:7d}
\end{subfigure}
\caption{Variation of $|M_{\mu\mu}|$ with the three CP violating phases $\alpha$, $\beta$ and $\delta$ 
in case of normal ordering.}
\label{fig:7_cp_phases}
\end{figure}

We also show the variation of $|M_{\mu\mu}|$ with $\alpha$, $\beta$ and $\delta$ in Fig.~\ref{fig:7_cp_phases}. The blue band is obtained by 
imposing the constraint coming from $\sum m_{i} < 0.12\,{\rm  eV}$, whereas, the cyan band is obtained without the constraint on the total 
neutrino mass. 

For the normal ordering, our observations are as follows:
\begin{itemize}
 
 \item For vanishing $|M_{\mu\mu}|$, the lightest neutrino mass has to be $m_{1} \geq 5.1\times 10^{-2}\,{\rm eV}$. However, if we impose $\sum m_i < 0.12\,{\rm eV}$ constraint, the lowest value of $|M_{\mu\mu}|$ is found to be of $\mathcal{O}(10^{-2})\,{\rm eV}$.
 \item For vanishing $|M_{\mu\mu}|$, the Majorana CP violating phase $\alpha$ has to lie within $(0^{\circ},\,60^{\circ})$, $(120^{\circ},\,240^{\circ})$ and 
$(300^{\circ},\,360^{\circ})$. 
\item $|M_{\mu\mu}|$ can vanish if $\beta$ value  lies within $(45^{\circ},\,130^{\circ})$ and 
$(225^{\circ},\,310^{\circ})$. 

\item   There is no constraint on the values of  $\delta$ for vanishing $|M_{\mu\mu}|$, i.e, $|M_{\mu\mu}|$ can vanish for any
value of $\delta$ within $(0,\,360^{\circ})$.
 \end{itemize}

In case of normal ordering, the $M_{\mu\mu}$ can be expressed as
 \begin{eqnarray}
M_{ \mu\mu}=(-s_{12}c_{23}-c_{12}s_{23}s_{13}e^{i\delta})^2\,m_1+(c_{12}c_{23}-s_{12}s_{23}s_{13}e^{i\delta})^2\,
\sqrt{m_1^2 + \Delta m_{21}^2}\,e^{2i\alpha}+ (c^{2}_{13}s^{2}_{23})\,\sqrt{m_1^2 + \Delta m_{31}^2}\,e^{2i\beta}
\end{eqnarray}
Now, we will discuss few extreme cases analytically.
\begin{itemize}
 \item {\bf Case I:} $m_{1} << \sqrt{\Delta m^2_{21}}$: \\
 In this case, we have
 \begin{equation}
\label{mmumuno}
 |M_{\mu\mu}| \approx \Big|(c_{12}c_{23}-s_{12}s_{23}s_{13}e^{i\delta})^2\sqrt{\Delta m^2_{21}}\,e^{2i\alpha}+c^{2}_{13}s^{2}_{23}
\sqrt{\Delta m^2_{31}}\,e^{2i\beta}\Big|
\end{equation}
The minimum value of $|M_{\mu\mu}|$ can be obtained when there is maximum cancellation between the two terms. This can be achieved by putting
$\alpha = 0^{\circ}$, $\beta = 90^{\circ}$ and $\delta = 180^{\circ}$. Again, we can assume $s_{23} \simeq c_{23}$. Using all these values, we
can write Eq.~\ref{mmumuno} as
\begin{equation}
 |M_{\mu\mu}| \approx c_{23}^2\,\Big|(c_{12} + s_{12}s_{13})^2\sqrt{\Delta m^2_{21}} - c^{2}_{13}\sqrt{\Delta m^2_{31}}\Big| \approx 
c_{23}^2\,c^{2}_{13}\sqrt{\Delta m^2_{31}}
\end{equation}
Hence, in this case, we can not have vanishing $|M_{\mu\mu}|$. 

\item {\bf Case II:} $m_{1} >> \sqrt{\Delta m^2_{31}}$: \\
In this case, we have
\begin{eqnarray}
\label{mumuno}
  &&|M_{ \mu\mu}| = m_1\Big|(-s_{12}c_{23}-c_{12}s_{23}s_{13}e^{i\delta})^2+(c_{12}c_{23}-s_{12}s_{23}s_{13}e^{i\delta})^2\,e^{2i\alpha}+ 
(c^{2}_{13}s^{2}_{23})\,e^{2i\beta}\Big|   
 \end{eqnarray}
 Under the approximation $s_{23}=c_{23}$, we simplify $|M_{ \mu\mu}|$ as
 \begin{eqnarray}
  |M_{ \mu\mu}| &=& m_1\,s_{23}^2\Big|(s_{12}+c_{12}s_{13}e^{i\delta})^2+(c_{12}-s_{12}s_{13}e^{i\delta})^2\,e^{2i\alpha}+ c^{2}_{13}\,e^{2i\beta}\Big|   
 \end{eqnarray}
It is quite clear that if we have $\alpha = 0^{\circ}$, $\delta = 90^{\circ}$ and $\beta = 90^{\circ}$, we have a exact cancellation of all the terms and $|M_{ \mu\mu}|$ will vanish.

\end{itemize}

\begin{figure}[htbp!]
\begin{subfigure}{0.30\textwidth}
\includegraphics[width=5cm,height=4cm]{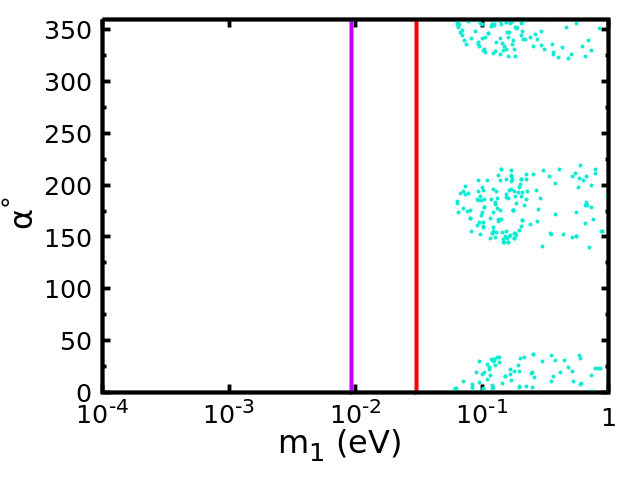}
\caption{}
\label{fig:mumuno_m1.1a}
\end{subfigure}
\begin{subfigure}{0.30\textwidth}
\includegraphics[width=5cm,height=4cm]{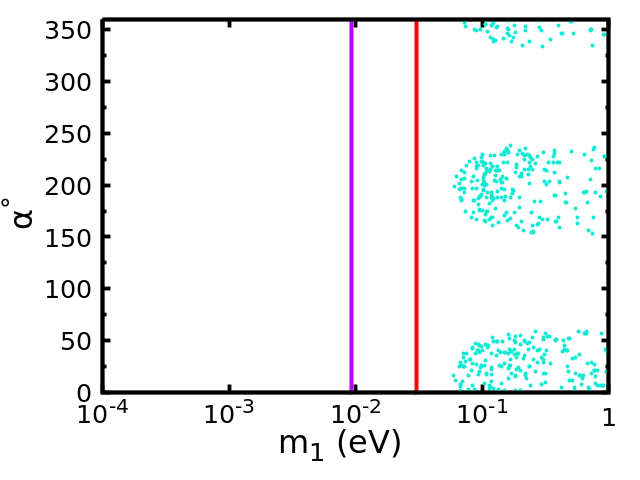}
\caption{}
\label{fig:mumuno_m1.1b}
\end{subfigure}
\begin{subfigure}{0.30\textwidth}
\includegraphics[width=5cm,height=4cm]{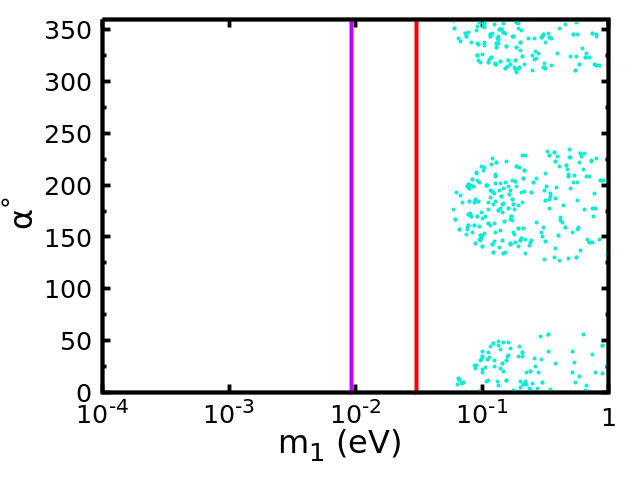}
\caption{}
\label{fig:mumuno_m1.1c}
\end{subfigure}
\begin{subfigure}{0.30\textwidth}
\includegraphics[width=5cm,height=4cm]{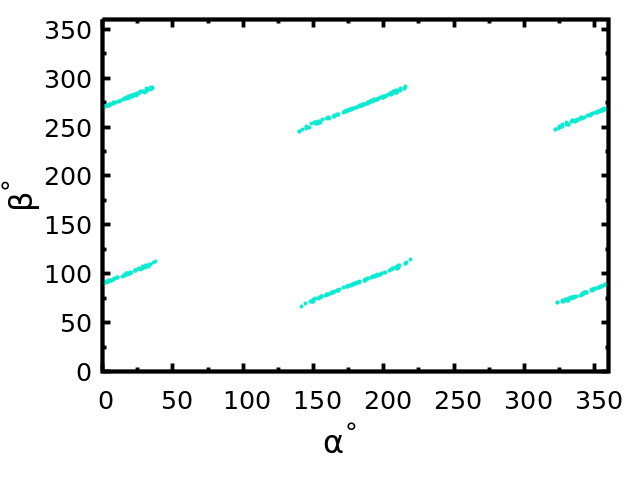}
\caption{}
\label{fig:mumuno_cp.1d}
\end{subfigure} 
\begin{subfigure}{0.30\textwidth}
\includegraphics[width=5cm,height=4cm]{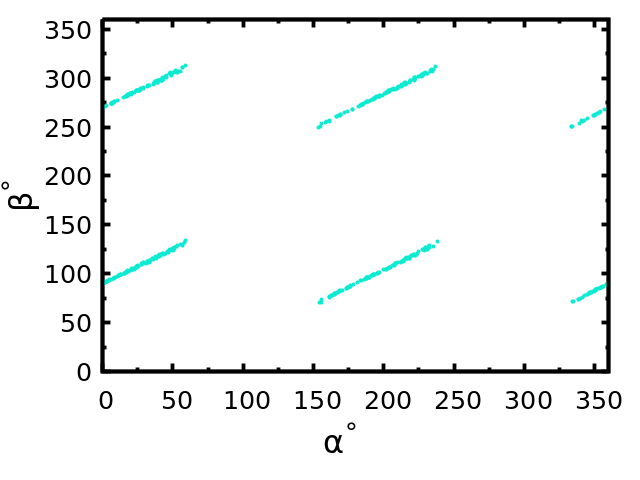}
\caption{}
\label{fig:mumuno_cp.1e}
\end{subfigure} 
\begin{subfigure}{0.30\textwidth}
\includegraphics[width=5cm,height=4cm]{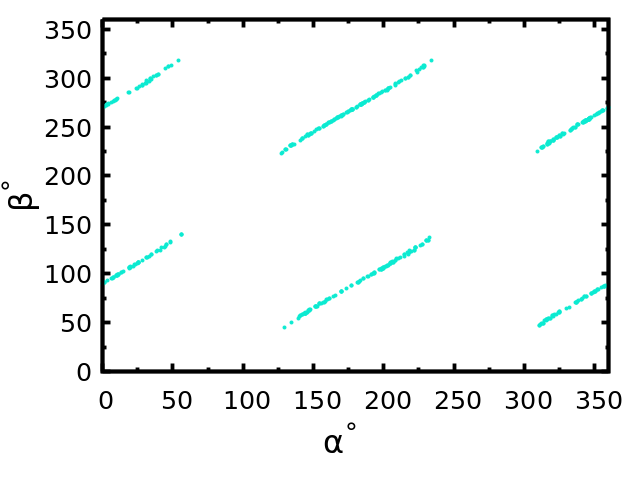}
\caption{}
\label{fig:mumuno_cp.1f}
\end{subfigure} 
\captionsetup{justification=raggedright,singlelinecheck=false}
\caption{Correlation between $\beta$ and $\alpha$, and between $\alpha$ and $m_{1}$ for $\delta=0^{\circ}$~(Left panel), 
$\delta=90^{\circ}$~(Middle panel) and $\delta=180^{\circ}$~(Right panel) with vanishing $|M_{\mu\mu}|$ in case of normal ordering. The red and magenta lines 
represent $\sum m_i = 0.12\,{\rm eV}$~\cite{Zhang:2020mox} and $\sum m_i = 0.072\,{\rm eV}$~\cite{DESI:2024mwx}, respectively.}
\label{fig:mumuno_cp}
\end{figure}

In Fig~\ref{fig:mumuno_cp}, we show correlation between $\beta$ and $\alpha$, and between $\alpha$ and $m_{1}$ with vanishing 
$|M_{\mu\mu}|$ for several values of the Dirac CP violating phase $\delta$. It is evident that for $\delta = 0^{\circ}$, $90^{\circ}$ and 
$180^{\circ}$, $|M_{\mu\mu}|$
can vanish for $\alpha$ value around $0^{\circ}$, $180^{\circ}$ or $360^{\circ}$, and $\beta$ value around $90^{\circ}$ or $270^{\circ}$. 
It is observed that the correlation between $\beta$ and $\alpha$, and between $\alpha$ and $m_{1}$ does not change much with the value of the 
Dirac CP violating phase $\delta$. It is worth mentioning that $|M_{\mu\mu}|$ vanishes only if the atmospheric mixing angle lies in the lower
octant, more precisely value of the atmospheric mixing angle has to be $\theta_{23} < 45.3^{\circ}$ to have vanishing $|M_{\mu\mu}|$. We
have shown it explicitely in Fig.~\ref{fig:Nmalpth23mumzero_no} of Appendix~\ref{app1}.

\subsubsection{Inverted Ordering}
\begin{figure}[htbp!]
\begin{subfigure}{0.35\textwidth}
\includegraphics[width=\textwidth]{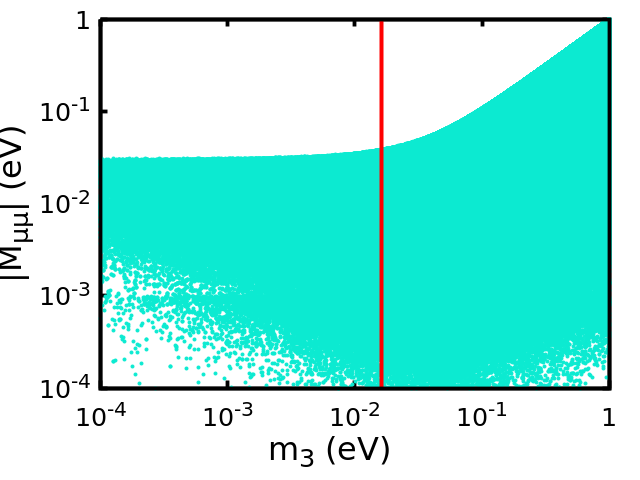}
\end{subfigure}
\captionsetup{justification=raggedright,singlelinecheck=false}
\caption{Variation of $|M_{\mu\mu}|$ with the lightest neutrino mass $m_3$ in case of inverted ordering. Total neutrino mass $\sum m_{i} = 0.12\,{\rm  eV}$~\cite{Zhang:2020mox} is shown with the red vertical line.}
\label{fig:8}
\end{figure}

\begin{figure}[htbp!]
\begin{subfigure}{0.30\textwidth}
\includegraphics[width=\textwidth]{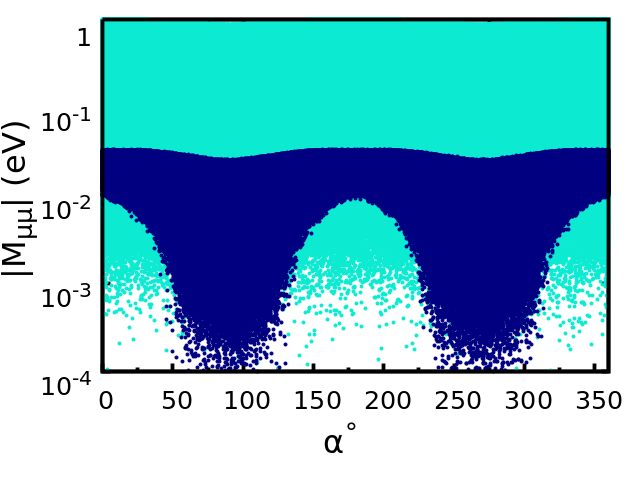}
\caption{}
\label{fig:8b}
\end{subfigure}
\begin{subfigure}{0.30\textwidth}
\includegraphics[width=\textwidth]{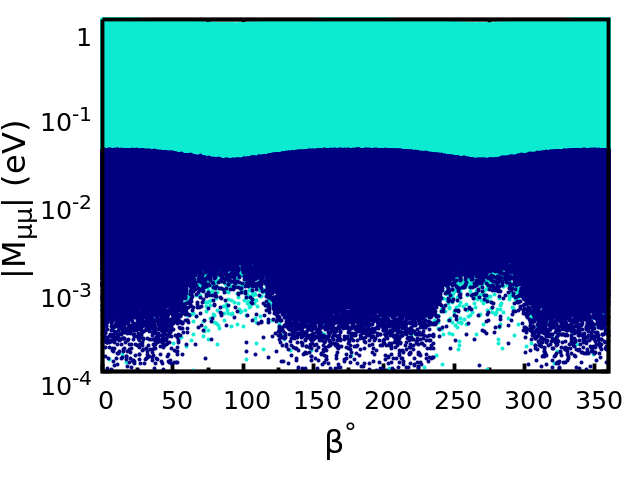}
\caption{}
\label{fig:8c}
\end{subfigure}
\begin{subfigure}{0.30\textwidth}
\includegraphics[width=\textwidth]{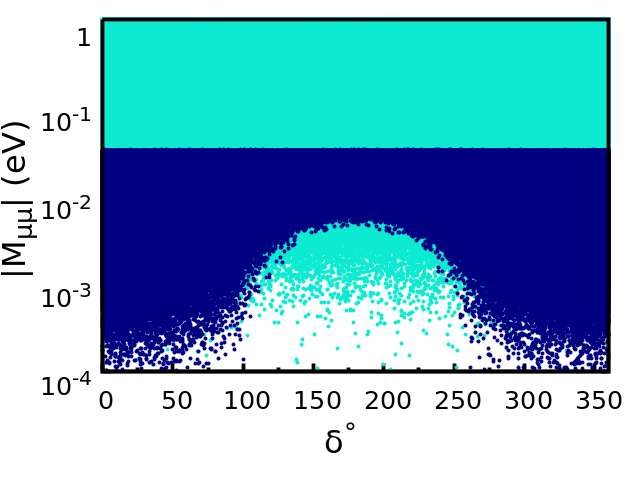}
\caption{}
\label{fig:8d}
\end{subfigure}
\caption{Variation of $|M_{\mu\mu}|$ with the three CP violating phases $\alpha$, $\beta$ and $\delta$ 
in case of inverted ordering.}
\label{fig:8_cp_phases}
\end{figure}

We plot in Fig.~\ref{fig:8} the variation of $|M_{\mu\mu}|$ with $m_3$, for inverted ordering. It is observed that in case of inverted 
ordering we can have vanishing $|M_{\mu\mu}|$. 
The variation of $|M_{\mu\mu}|$ with the CP violating phases $\alpha$, $\beta$ and $\delta$ is shown in Fig.~\ref{fig:8_cp_phases}. The blue
band is obtained once we impose $\sum m_i < 0.12\,{\rm eV}$ constraint coming from the total neutrino mass, whereas, the cyan band is
obtained without the constraint.

In case of inverted ordering, our main observations are as follows:
\begin{itemize}
 \item $|M_{\mu\mu}|$ can vanish for any value of the lightest neutrino mass $m_3$.
 \item  For vanishing $|M_{\mu\mu}|$, the Majorana CP violating phase $\alpha$ has to lie within $(60^{\circ},\,120^{\circ})$ and 
$(240^{\circ},\,300^{\circ})$ if we impose $\sum m_i < 0.12\,{\rm eV}$ constraint. However, without the constraint, we can have
vanishing $|M_{\mu\mu}|$ for all values of $\alpha$ within $(0^{\circ}, 360^{\circ})$.
 \item 
Without the total neutrino mass constraint, it is evident that we can have vanishing $|M_{\mu\mu}|$ for any value of $\beta$ within
$(0^{\circ}, 360^{\circ})$. However, with the constraint, value of $\beta$ has to be within $(0^{\circ},\,55^{\circ})$, 
$(130^{\circ},\,230^{\circ})$ and $(305^{\circ},\,360^{\circ})$ for vanishing $|M_{\mu\mu}|$.
 \item Dirac CP violating phase $\delta$ should lie within $(0^{\circ},\,90^{\circ})$ and $(270^{\circ},\,360^{\circ})$ for vanishing 
$|M_{\mu\mu}|$ with the total neutrino mass constraint. However, there is no such constraint on $\delta$ if we do not impose 
$\sum m_i < 0.12\,{\rm eV}$ constraint.
 \end{itemize}

Let us now analyse vanishing of $|M_{\mu\mu}|$ for several values of the lightest neutrino mass $m_3$. 
In case of inverted ordering, $M_{\mu\mu}$ can be written as
\begin{eqnarray}
M_{ \mu\mu} &=& (-s_{12}c_{23}-c_{12}s_{23}s_{13}e^{i\delta})^2\,\sqrt{m^{2}_{3}+\Delta m^2_{13}}+(c_{12}c_{23}-s_{12}s_{23}s_{13}
e^{i\delta})^2\,e^{2i\alpha}\,\sqrt{m^{2}_{3}+\Delta m^2_{21}+ \Delta m^2_{13}} \nonumber \\
&& + (c^{2}_{13}s^{2}_{23})\,m_3 e^{2i\beta}
\end{eqnarray}

\begin{figure}[htbp!]
\begin{subfigure}{0.35\textwidth}
\includegraphics[width=5cm,height=4cm]{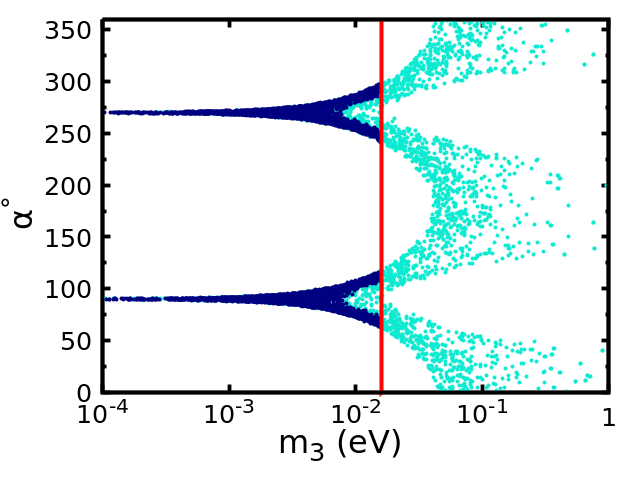}
\caption{}
\label{fig:8.1a}
\end{subfigure}
\begin{subfigure}{0.35\textwidth}
\includegraphics[width=5cm,height=4cm]{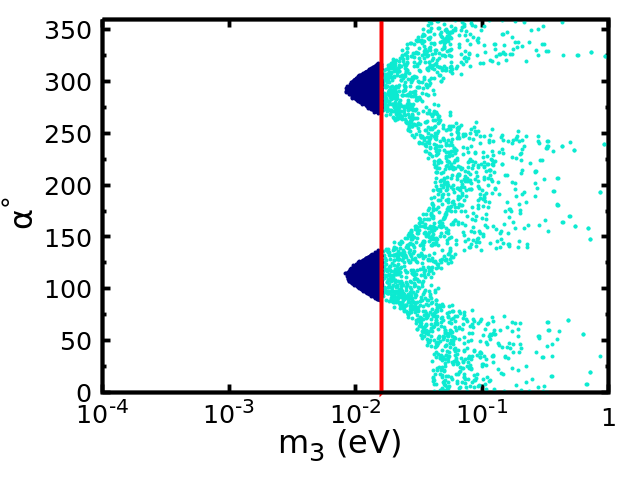}
\caption{}
\label{fig:8.1b}
\end{subfigure}
 \begin{subfigure}{0.35\textwidth}
\includegraphics[width=5cm,height=4cm]{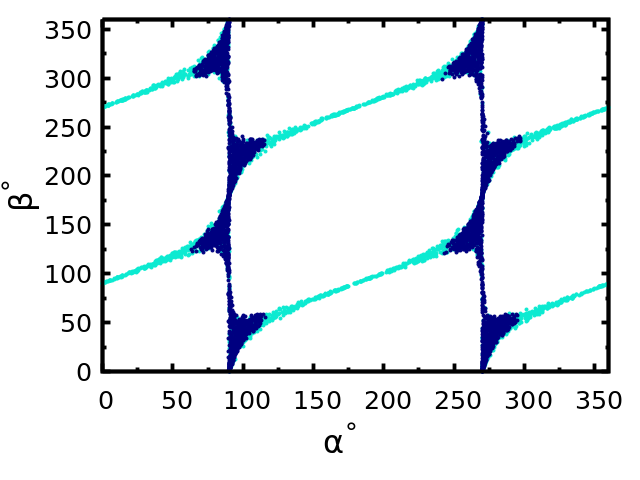}
\caption{}
\label{fig:8.1c}
\end{subfigure} 
 \begin{subfigure}{0.35\textwidth}
\includegraphics[width=5cm,height=4cm]{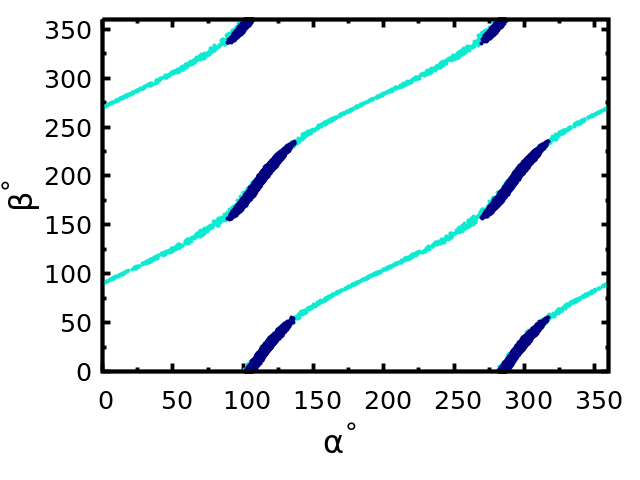}
\caption{}
\label{fig:8.1d}
\end{subfigure} 
\captionsetup{justification=raggedright,singlelinecheck=false}
\caption{Correlation between $\beta$ and $\alpha$, and between $\alpha$ and $m_{3}$ for $\delta=0^{\circ}$~(Left panel) and 
$\delta=90^{\circ}$~(Right panel) with vanishing $|M_{\mu\mu}|$ in case of inverted ordering. The red line  represents $\sum m_i = 0.12\,{\rm eV}$~\cite{Zhang:2020mox}.}
\label{fig:8.1}
\end{figure}

\begin{itemize}
 \item {\bf Case I:} $m_{3} << \sqrt{\Delta m^2_{21}}$: \\
 
 In this case, we have
  \begin{equation}
  \label{eq:26}
   |M_{\mu\mu}|\simeq \sqrt{\Delta m^2_{13}}|(-s_{12}c_{23}-c_{12}s_{23}s_{13}e^{i\delta})^2+(c_{12}c_{23}-s_{12}s_{23}s_{13}
e^{i\delta})^2\,e^{2i\alpha}|. 
  \end{equation}
 If we assume 
$s_{23}\simeq c_{23}$, then $|M_{\mu\mu}|$ can be expressed as
\begin{eqnarray}
|M_{\mu\mu}|\simeq \sqrt{\Delta m^2_{13}}\,s_{23}^2\,\Big|\Big(s_{12} + s_{13}c_{12}\,e^{i\delta}\Big)^2 + \Big(c_{12} - s_{12}s_{13}\,
e^{i\delta}\Big)^2\,e^{2i\alpha}\Big|.
\end{eqnarray}
It is evident that $|M_{\mu\mu}|$ will be minimum if there is maximum cancellation between the two terms. This can be achieved if we use
$\alpha = 90^{\circ},\,270^{\circ}$ and $\delta = 0^{\circ}, 360^{\circ}$. Numerically it is observed that $|M_{\mu\mu}|$ will vanish if 
$\alpha$ lies within $(75^{\circ},\,105^{\circ})$ or $(255^{\circ},\,285^{\circ})$ and $\delta$ lies within $(0^{\circ},\,40^{\circ})$ or 
$(320^{\circ},\,360^{\circ})$. 

 \item {\bf Case II:} $m_{3} >> \sqrt{\Delta m^2_{13}}$:\\ 
 In this case, we have
 \begin{eqnarray}
\label{mumuio}
  &&|M_{ \mu\mu}| = m_3\Big|(-s_{12}c_{23}-c_{12}s_{23}s_{13}e^{i\delta})^2+(c_{12}c_{23}-s_{12}s_{23}s_{13}e^{i\delta})^2\,e^{2i\alpha}+ 
(c^{2}_{13}s^{2}_{23})\,e^{2i\beta}\Big|   
 \end{eqnarray}
As expected, Eq.~\ref{mumuio} looks exactly same as Eq.~\ref{mumuno}. Hence we can have vanishing $|M_{ \mu\mu}|$ in this case.

\item {\bf Case III:} $m_{3}\approx \sqrt{\Delta m^2_{13}}$:\\
In this case, we have
  \begin{eqnarray}
  \label{eq:28}
   &&|M_{\mu\mu}| \simeq c^{2}_{23}\sqrt{\Delta m^2_{13}}\,\Big|\sqrt{2}\,(s^{2}_{12}+c^{2}_{12}e^{2i\alpha})+2\sqrt{2}\,s_{12}s_{13}
c_{12}(1-e^{2i\alpha})e^{i\delta}+\sqrt{2}\,s^{2}_{13}(c^{2}_{12}+s^{2}_{12}e^{2i\alpha})e^{2i\delta}\,\nonumber \\
  &&\hspace{1.5cm} + c^{2}_{13}e^{2i\beta}\Big|\,,
  \end{eqnarray}
where we have assumed $s_{23} \approx c_{23}$. $\delta$ and $\beta$ values 
around $0^{\circ}$ or $360^{\circ}$, $|M_{\mu\mu}|$ will vanish if we have
\begin{eqnarray}
\cos2\alpha\approx -\frac{2\Big[s_{12}s_{13}-c_{12}\Big]^4+\Big[\sqrt{2}\,(s_{13}c_{12}+s_{12})^2+c_{13}^{2}\Big]^{2}}
{2\sqrt{2}\,\Big[s_{12}s_{13}-c_{12}\Big]^{2}\Big[\sqrt{2}\,(s_{13}c_{12}+s_{12})^{2}+c_{13}^{2}\Big]}\approx -1
\end{eqnarray}
Hence, the $|M_{\mu\mu}|$ approaches zero under this condition.
\end{itemize}
Now, we will discuss the correlation of $\alpha$, $\beta$ and $m_3$ for different values of $\delta$ with vanishing $|M_{\mu\mu}|$. We show in 
Fig.~\ref{fig:8.1} the correlation between $\beta$ and $\alpha$, and between $\alpha$ and $m_{3}$ for $\delta = 0^{\circ}$ and $90^{\circ}$, 
respectively. The region shown with blue is obtained by using the constraint coming from $\sum m_i < 0.12\,{\rm eV}$, whereas, the cyan 
region is obtained without imposing the total neutrino mass constraint.   
It is observed that at $\delta = 0^{\circ}$, value of the Majorana CP violating phase $\alpha$ is constrained to lie around $90^{\circ}$ and
$270^{\circ}$ if
the lightest neutrino mass $m_3 \leq 10^{-3}\,{\rm eV}$. However, value of $\alpha$ will deviate from $90^{\circ}$ or $270^{\circ}$ if
the lightest neutrino mass becomes $m_3 \approx 10^{-2}\,{\rm eV}$ or larger. The deviation from $90^{\circ}$ will be more if we keep 
increasing
the value of $m_3$. Value of $\alpha$ will approach $0^{\circ}$, $180^{\circ}$ and $360^{\circ}$ for $m_{3} > 4.94\times 10^{-2}\,{\rm eV}$.
For $\delta = 90^{\circ}$,  with the total neutrino mass $\sum m_i < 0.12\,{\rm eV}$ constraint, the value of $\alpha$ is also restricted to a very limited region in the parameter space for vanishing $|M_{\mu\mu}|$. We have also shown the correlation between $\theta_{23}$ and $\alpha$ for vanishing $|M_{\mu\mu}|$ in Fig.~\ref{fig:Imalpth23mumuzero_io} of Appendix~\ref{app1}.


\subsection{$M_{\mu\tau}$: the $\mu\tau$ element of the mass matrix}
\subsubsection{Normal Ordering:}
\begin{figure}[h!]
\begin{subfigure}{0.35\textwidth}
\includegraphics[width=\textwidth]{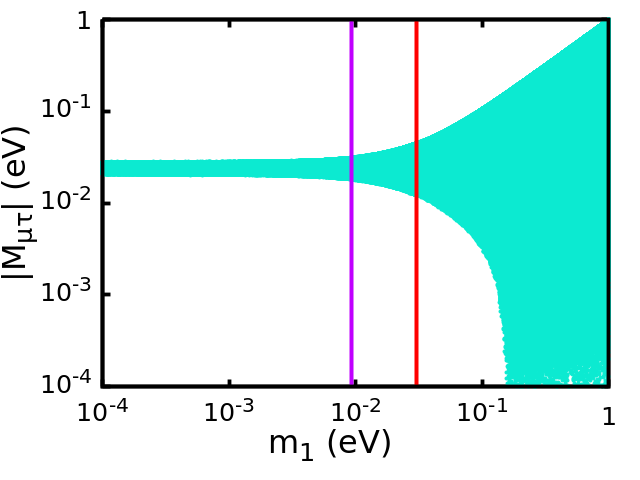}
\end{subfigure}
\captionsetup{justification=raggedright,singlelinecheck=false}
\caption{Variation of $|M_{\mu\tau}|$ with the lightest neutrino mass $m_1$ in case of normal ordering. Total neutrino mass 
$\sum m_{i} = 0.12\,{\rm  eV}$~\cite{Zhang:2020mox} and $\sum m_{i} = 0.072\,{\rm  eV}$~\cite{DESI:2024mwx} are shown with red and magenta 
lines, respectively.}
\label{fig:9}
\end{figure}

Variation of $|M_{\mu\tau}|$ with $m_1$ is shown in Fig.~\ref{fig:9}. It is evident from Fig.~\ref{fig:9} that we can have vanishing 
$|M_{\mu\tau}|$ for large value of $m_1$. 

\begin{figure}[h!]
\begin{subfigure}{0.30\textwidth}
\includegraphics[width=\textwidth]{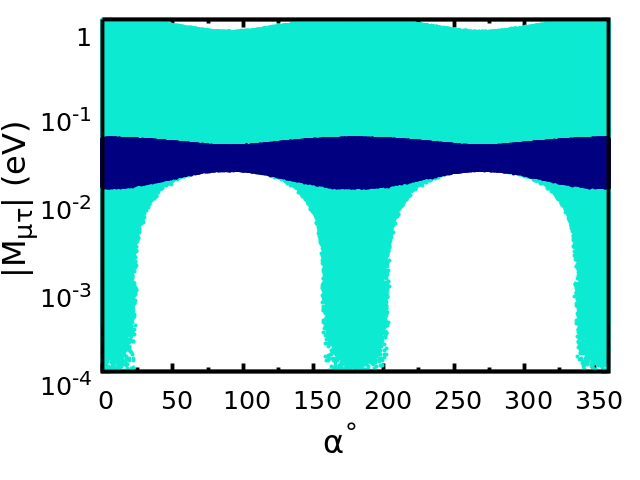}
\caption{}
\label{fig:9b}
\end{subfigure}
\begin{subfigure}{0.30\textwidth}
\includegraphics[width=\textwidth]{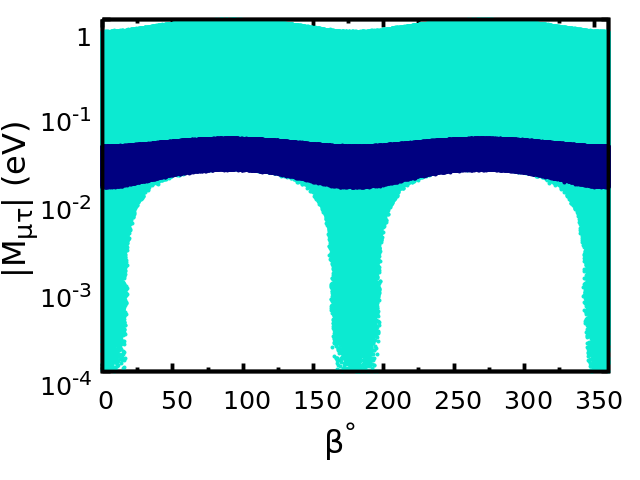}
\caption{}
\label{fig:9c}
\end{subfigure}
\begin{subfigure}{0.30\textwidth}
\includegraphics[width=\textwidth]{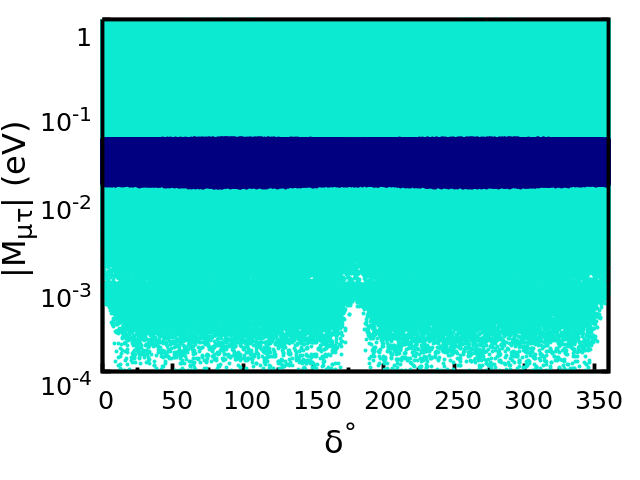}
\caption{}
\label{fig:9d}
\end{subfigure}
\caption{Variation of $|M_{\mu\tau}|$ with the three CP violating phases $\alpha$, $\beta$ and $\delta$ 
in case of normal ordering.}
\label{fig:9_cp_phases}
\end{figure}

We also show the variation of $|M_{\mu\tau}|$ with the CP violating phases $\alpha$, $\beta$ and $\delta$ in Fig.~\ref{fig:9b}, 
Fig.~\ref{fig:9c} and Fig.~\ref{fig:9d}, respectively. The blue region is obtained with $\sum m_i < 0.12\,{\rm eV}$ constraint, whereas,
the cyan region is obtained without any constraint on the total neutrino mass.


Our main observations are as follows.
\begin{itemize}
\item Without the total neutrino mass constraint, $|M_{\mu\tau}|$ can vanish if the lightest neutrino mass $m_1 \ge 0.16\,{\rm eV}$. However,
with the $\sum m_i < 0.12\,{\rm eV}$ constraint, the lowest value of $|M_{\mu\tau}|$ is found to be of
$\mathcal{O}(10^{-2})\,{\rm eV}$.
\item For the vanishing $|M_{\mu\tau}|$ condition, the Majorana CP violating phase $\alpha$ has to lie within $(0^{\circ}, \,25^{\circ})$, $(155^{\circ}, \,205^{\circ})$ and $(335^{\circ},360^{\circ})$, respectively.
\item Value of $\beta$ has to lie within $(0^{\circ}, \,17^{\circ})$, $(163^{\circ}, \,179^{\circ})$ and $(343^{\circ}, \,360^{\circ})$ to have vanishing $|M_{\mu\tau}|$. 
\item Vanishing of $|M_{\mu\tau}|$ excludes $\delta = 0^{\circ}$, $180^{\circ}$ and $360^{\circ}$.
\end{itemize}

In general, for normal ordering, we can write $M_{\mu\tau}$ as
\begin{eqnarray}
M_{ \mu\tau}&=&(-s_{12}c_{23}-c_{12}s_{23}s_{13}e^{i\delta})(s_{12}s_{23}-c_{12}c_{23}s_{13}e^{i\delta})\,m_1+(c_{12}c_{23}-s_{12}s_{23}
s_{13}e^{i\delta})(-c_{12}s_{23}-s_{12}c_{23}s_{13}e^{i\delta})\, \nonumber \\
&& \sqrt{m_1^2 + \Delta m_{21}^2}\,e^{2i\alpha} +(c^{2}_{13}s_{23}c_{23})\,\sqrt{m_1^2 + \Delta m_{31}^2}\,e^{2i\beta},
\end{eqnarray}
We wish to explore few extreme cases analytically.
\begin{itemize}
 \item {\bf Case I:} $m_{1} << \sqrt{\Delta m^2_{21}}$: \\
In this case, we can write $|M_{\mu\tau}|$ as
 \begin{equation}
\label{mmutauno}
 |M_{\mu\tau}| \simeq \Big|(c_{12}c_{23}-s_{12}s_{23}
s_{13}e^{i\delta})(-c_{12}s_{23}-s_{12}c_{23}s_{13}e^{i\delta})\sqrt{\Delta m_{21}^2}\,e^{2i\alpha} +
(c^{2}_{13}s_{23}c_{23})\,\sqrt{\Delta m_{31}^2}\,e^{2i\beta}\Big|
\end{equation}
Assuming $s_{23} \approx c_{23}$, we can write Eq.~\ref{mmutauno} as
 \begin{equation}
\label{mmutauno1}
 |M_{\mu\tau}| \simeq c_{23}^2\Big|-(c^2_{12}-s^2_{12}s^2_{13}e^{2i\delta})\sqrt{\Delta m_{21}^2}\,e^{2i\alpha} +
c^{2}_{13}\,\sqrt{\Delta m_{31}^2}\,e^{2i\beta}\Big|
\end{equation}
Minimum value of $|M_{\mu\tau}|$ corresponds to $\alpha = 0^{\circ}$, $\beta = 0^{\circ}$ and $\delta = 90^{\circ}$ and it is found to be
of $\mathcal{O}(10^{-2})$. This is obvious because $|M_{\mu\tau}| \propto c_{23}^2\,c^{2}_{13}\,\sqrt{\Delta m_{31}^2}$ as the second term in
Eq.~\ref{mmutauno1} will always dominate over the first term. Hence $|M_{\mu\tau}|$ will not vanish in the $m_1 \to 0$ limit. 
  \item {\bf Case II:} $m_{1} >> \sqrt{\Delta m^2_{31}}$: \\
   In this case, $M_{\mu\tau}$ element can be written as 
  \begin{eqnarray}
  \label{mutauno}
   |M_{\mu\tau}| &\simeq& m_1\Big|(-s_{12}c_{23}-c_{12}s_{23}s_{13}e^{i\delta})(s_{12}s_{23}-c_{12}c_{23}s_{13}e^{i\delta})+(c_{12}c_{23}-s_{12}s_{23}s_{13}e^{i\delta}) \nonumber \\
   &&(-c_{12}s_{23}-s_{12}c_{23}s_{13}e^{i\delta})\,e^{2i\alpha} +(c^{2}_{13}s_{23}c_{23})
e^{2i\beta}\Big|. 
  \end{eqnarray}
Assuming $c_{23} \approx s_{23}$, we can simplify Eq.~\ref{mutauno} as
\begin{eqnarray}
   |M_{\mu\tau}| \simeq m_1\,c_{23}^2\Big|\Big(c_{12}^2 + s_{12}^2\,e^{2i\alpha}\Big)\Big(s_{13}^2\,e^{2i\delta} + 1\Big) - \Big(1 + e^{2i\alpha}\Big) + c^{2}_{13}\,e^{2i\beta}\Big|. 
  \end{eqnarray}
  It is clear that if we have $\alpha = \beta = \delta = 0^{\circ}$, all the terms will cancel exactly and we will have vanishing
  $M_{\mu\tau}$ element.
\end{itemize}

We can have vanishing $|M_{\mu\tau}|$ when we have quasi degenerate neutrino mass states, i.e, 
$m_1 \approx m_2 \approx m_3 >> \sqrt{\Delta m^2_{31}}$.

\begin{figure}[htbp]
 \begin{subfigure}{0.40\textwidth}
\includegraphics[width=5cm,height=4cm]{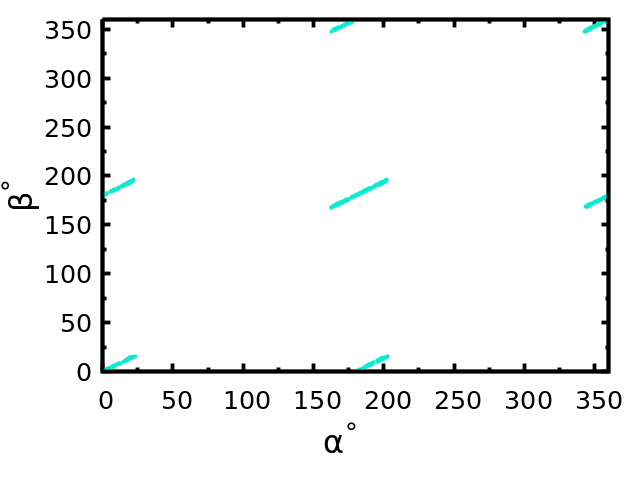}
\caption{}
\label{fig:mutauno_cp}
\end{subfigure} 
\begin{subfigure}{0.40\textwidth}
\includegraphics[width=5cm,height=4cm]{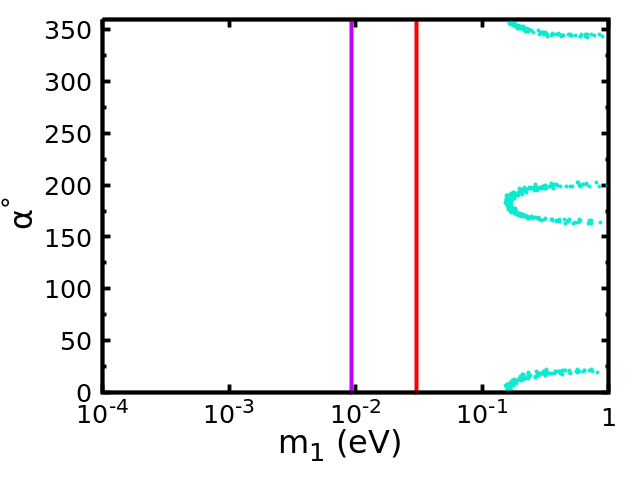}
\caption{}
\label{fig:mutauno_m1}
\end{subfigure}
\captionsetup{justification=raggedright,singlelinecheck=false}
\caption{Correlation between $\beta$ and $\alpha$, and between $\alpha$ and $m_{1}$ for $\delta=90^{\circ}$ with vanishing $|M_{\mu\tau}|$
in case of normal ordering. The red and magenta lines 
represent $\sum m_i = 0.12\,{\rm eV}$~\cite{Zhang:2020mox} and $\sum m_i = 0.072\,{\rm eV}$~\cite{DESI:2024mwx}, respectively.}
\label{fig:mutaud90}
\end{figure}

We show the correlation between $\beta$ and $\alpha$, and between $\alpha$ and $m_{1}$ for $\delta = 90^{\circ}$ with vanishing 
$|M_{\mu\tau}|$ in Fig.~\ref{fig:mutauno_cp} and Fig.~\ref{fig:mutauno_m1}, respectively. It is evident that for vanishing $|M_{\mu\tau}|$,
the value of the lightest neutrino mass should be $m_1 \ge 0.16\,{\rm eV}$. The Majorana CP violating phases $\alpha$ 
and $\beta$ values are also restricted to a very limited region in the parameter space. We have shown the correlation between $\theta_{23}$ and $\alpha$ for vanishing $|M_{\mu\tau}|$ in Fig.~\ref{fig:Nmalpth23mumtautauzero_no} of Appendix~\ref{app1}.

\subsubsection{Inverted Ordering:}
\begin{figure}[htbp!]
\begin{subfigure}{0.35\textwidth}
\includegraphics[width=\textwidth]{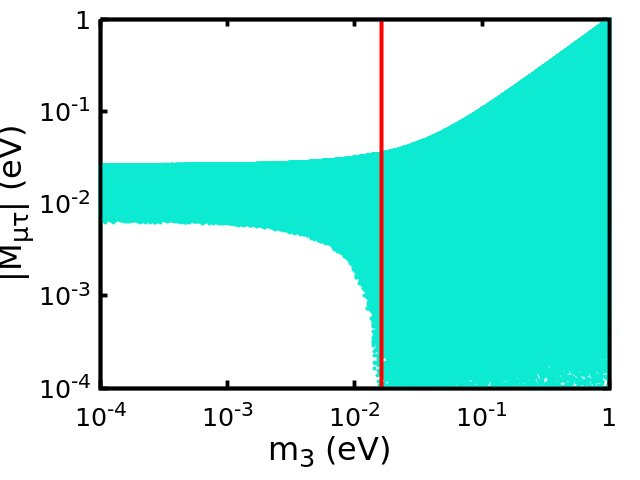}
\end{subfigure}
\captionsetup{justification=raggedright,singlelinecheck=false}
\caption{Variation of $|M_{\mu\tau}|$ with the lightest neutrino mass $m_3$ in case of inverted ordering. Total neutrino mass $\sum m_{i} = 0.12\,{\rm  eV}$~\cite{Zhang:2020mox} is shown with the red vertical line.}
\label{fig:10}
\end{figure}

\begin{figure}[htbp!]
\begin{subfigure}{0.30\textwidth}
\includegraphics[width=\textwidth]{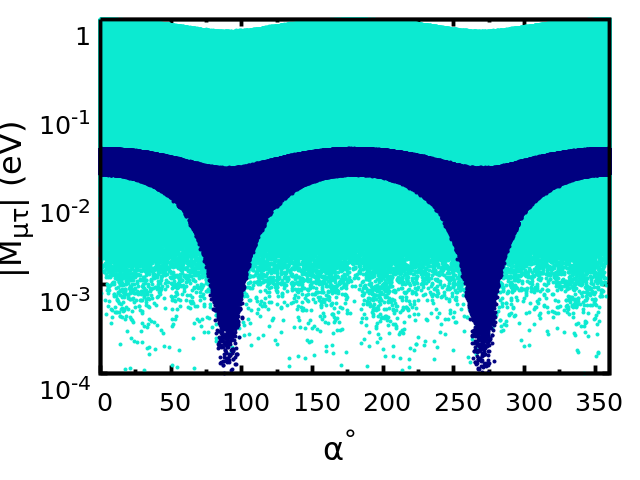}
\caption{}
\label{fig:10b}
\end{subfigure}
\begin{subfigure}{0.30\textwidth}
\includegraphics[width=\textwidth]{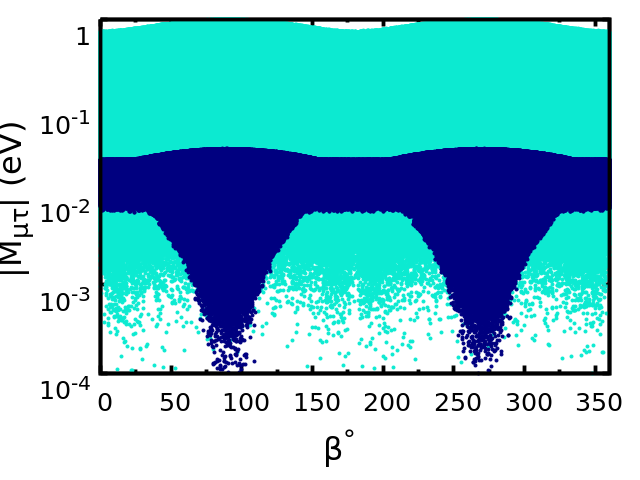}
\caption{}
\label{fig:10c}
\end{subfigure}
\begin{subfigure}{0.30\textwidth}
\includegraphics[width=\textwidth]{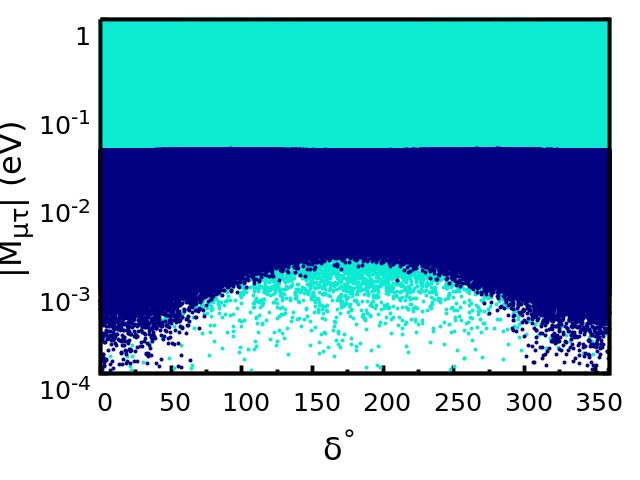}
\caption{}
\label{fig:10d}
\end{subfigure}
\caption{Variation of $|M_{\mu\tau}|$ with the three CP violating phases $\alpha$, $\beta$ and $\delta$ 
in case of inverted ordering.}
\label{fig:10_cp_phases}
\end{figure}

Variation of $|M_{\mu\tau}|$ with $m_3$ is shown in Fig.~\ref{fig:10}.  
We also show the variation of $|M_{\mu\tau}|$ with the CP violating phases $\alpha$, $\beta$ and $\delta$ in Fig.~\ref{fig:10b}, 
Fig.~\ref{fig:10c} and Fig.~\ref{fig:10d}, respectively. The blue region is obtained with the $\sum m_i < 0.12\,{\rm eV}$ constraint, 
whereas, the cyan region is obtained without imposing the total neutrino mass constraint.

For the inverted ordering, our main observations are as follows:
\begin{itemize}
\item We have vanishing $|M_{\mu\tau}|$ if the lightest neutrino mass $m_3  > 1.2\times10^{-2}\,{\rm eV}$.
\item For vanishing $|M_{\mu\tau}|$,  the Majorana CP violating phases $\alpha$ and $\beta$ values are restricted to lie around 
$90^{\circ}$ or $270^{\circ}$ once we impose $\sum m_i < 0.12\,{\rm eV}$ constraint. However, without the total neutrino mass constraint,
we get vanishing $|M_{\mu\tau}|$ for any values of $\alpha$ and $\beta$ within $(0^{\circ}, 360^{\circ})$.
 \item Similarly, for vanishing $|M_{\mu\tau}|$, the Dirac CP violating phase $\delta$ should be around $0^{\circ}$ or $360^{\circ}$ if we
impose $\sum m_i < 0.12\,{\rm eV}$ constraint. However, there is no such restrictions on the value of $\delta$ without the constraint.
We can have vanishing $|M_{\mu\tau}|$ for any value of $\delta$ within $(0^{\circ}, 360^{\circ})$.
 \end{itemize}

Let us now discuss vanishing of $M_{\mu\tau}$ for different values of $m_3$. In case of inverted ordering, we can write 
$M_{\mu\tau}$ as
\begin{eqnarray}
M_{ \mu\tau}&=&(-s_{12}c_{23}-c_{12}s_{23}s_{13}e^{i\delta})(s_{12}s_{23}-c_{12}c_{23}s_{13}e^{i\delta})\,\sqrt{m_3^2 + \Delta m_{13}^2}+
(c_{12}c_{23}-s_{12}s_{23}s_{13}e^{i\delta})(-c_{12}s_{23}-s_{12}c_{23}s_{13}e^{i\delta})\nonumber \\
&&\sqrt{m_3^2 + \Delta m_{13}^2 + \Delta m_{21}^2}\,e^{2i\alpha} + (c^{2}_{13}s_{23}c_{23})\,m_3 e^{2i\beta},
\end{eqnarray}

\begin{itemize}
\item {\bf Case I:} $m_{3} \le \sqrt{\Delta m^2_{21}}$: \\
  \begin{eqnarray}
  \label{eq:30}
   |M_{\mu\tau}| &\simeq& \sqrt{\Delta m^2_{13}}\Big|(-s_{12}c_{23}-c_{12}s_{23}s_{13}e^{i\delta})(s_{12}s_{23}-c_{12}c_{23}s_{13}
e^{i\delta}) \nonumber \\
  && +(c_{12}c_{23}-s_{12}s_{23}s_{13}e^{i\delta})(-c_{12}s_{23}-s_{12}c_{23}s_{13}e^{i\delta})\,e^{2i\alpha}\Big|. 
  \end{eqnarray}
For this case, the allowed range of $|M_{\mu\tau}|$ is found to be ($6.0\times 10^{-3},2.5\times 10^{-2}$\,{\rm eV}). Assuming 
$s_{23} \approx c_{23}$, we can simplify Eq.~\ref{eq:30} as
\begin{eqnarray}
   |M_{\mu\tau}| \simeq \sqrt{\Delta m^2_{13}}\,c_{23}^2\Big|\Big(c_{12}^2 + s_{12}^2\,e^{2i\alpha}\Big)\Big(s_{13}^2\,e^{2i\delta} + 1\Big)
 - \Big(1 + e^{2i\alpha}\Big)\Big|. 
  \end{eqnarray}
  It is evident that, irrespective of the value of $\delta$, $s_{13}^2\,e^{2i\delta} + 1 \approx 1$, since $s_{13}^2 \approx 0.02$. Again, if 
we assume $\alpha = 90^{\circ}$, then we get vanishing $|M_{\mu\tau}|$ only if $c_{12} \approx s_{12}$, i.e, if we have $\theta_{12} = 
45^{\circ}$.
 Hence, under this condition, $|M_{\mu\tau}|$ can not vanish. 

  \item {\bf Case II:} $m_{3} >> \sqrt{\Delta m^2_{13}}$: \\
  In this case, $|M_{\mu\tau}|$ can be written as
  \begin{eqnarray}
  \label{mutauio}
   |M_{\mu\tau}| &\simeq& m_3\Big|(-s_{12}c_{23}-c_{12}s_{23}s_{13}e^{i\delta})(s_{12}s_{23}-c_{12}c_{23}s_{13}e^{i\delta})+(c_{12}c_{23}-s_{12}s_{23}s_{13}e^{i\delta}) \nonumber \\
   &&(-c_{12}s_{23}-s_{12}c_{23}s_{13}e^{i\delta})\,e^{2i\alpha} +(c^{2}_{13}s_{23}c_{23})
e^{2i\beta}\Big|. 
  \end{eqnarray}
It is clear that Eq.~\ref{mutauio} is exactly same as Eq.~\ref{mutauno}. Hence $|M_{\mu\tau}|$ will vanish in this case.

\item {\bf Case III:} $m_{3}\approx \sqrt{\Delta m^2_{13}}$:
  \begin{eqnarray}
\label{mutau}
   &&|M_{\mu\tau}|\simeq \sqrt{\Delta m^2_{13}}\,c_{23}^2\Big|-\sqrt{2}\,(s^{2}_{12}+c^{2}_{12}e^{2i\alpha})
+\sqrt{2}\,s^{2}_{13}(c^{2}_{12}+s^{2}_{12}\,e^{2i\alpha})e^{2i\delta}+c^{2}_{13}\,e^{2i\beta}\Big|. 
  \end{eqnarray}
For this case, it can be shown that by using $\beta = 90^{\circ}$ and $\delta = 0^{\circ}$ in Eq.~\ref{mutau}, we get vanishing 
$|M_{\mu\tau}|$ if
\begin{equation}
 \cos2\alpha\approx-\frac{2(s^{2}_{13}s^{2}_{12}-c^{2}_{12})^2+(\sqrt{2}s^{2}_{13}c^{2}_{12}-\sqrt{2}s^{2}_{13}-c^2_{13})^2}
{2\sqrt{2}(s^{2}_{13}s^{2}_{12}-c^{2}_{12})(\sqrt{2}s^{2}_{13}c^{2}_{12}-\sqrt{2}s^{2}_{13}-c^2_{13})} \approx -1
\end{equation}
  
\end{itemize}

\begin{figure}[htbp!]
\begin{subfigure}{0.33\textwidth}
\includegraphics[width=5cm,height=4cm]{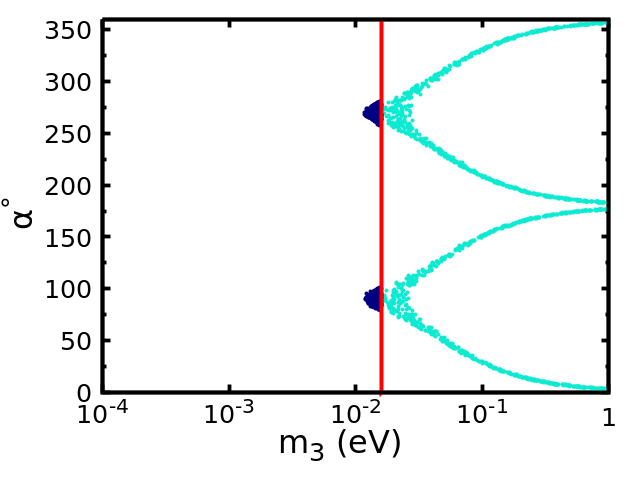}
\caption{}
\label{fig:10.1a}
\end{subfigure}
\begin{subfigure}{0.33\textwidth}
\includegraphics[width=5cm,height=4cm]{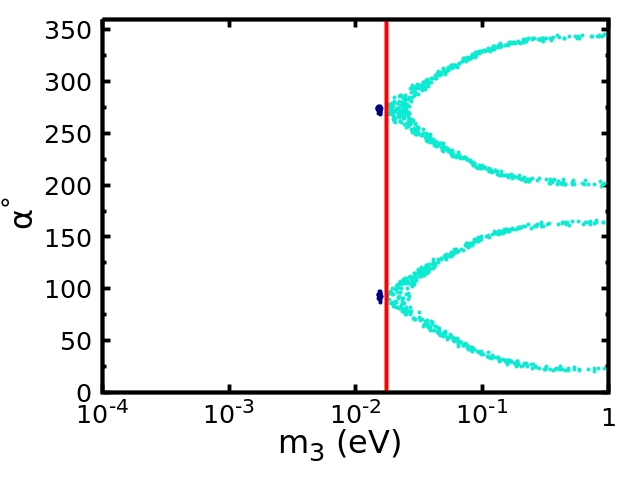}
\caption{}
\label{fig:10.1b}
\end{subfigure}
\begin{subfigure}{0.33\textwidth}
\includegraphics[width=5cm,height=4cm]{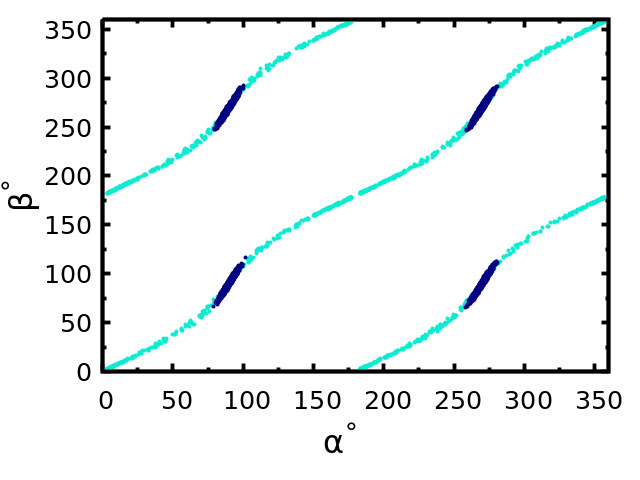}
\caption{}
\label{fig:10.1c}
\end{subfigure} 
\begin{subfigure}{0.33\textwidth}
\includegraphics[width=5cm,height=4cm]{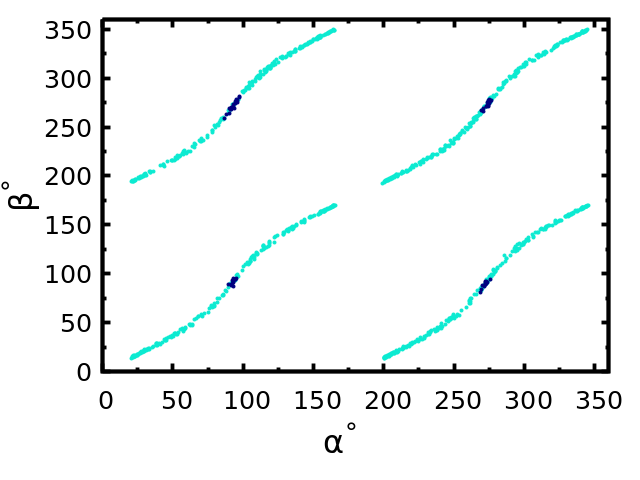}
\caption{}
\label{fig:10.1d}
\end{subfigure} 
\captionsetup{justification=raggedright,singlelinecheck=false}
\caption{Correlation between $\beta$ and $\alpha$, and between $\alpha$ and $m_{3}$ for $\delta=0^{\circ}$~(Left panel) and 
$\delta=90^{\circ}$~(Right panel) with vanishing $|M_{\mu\tau}|$ in case of inverted ordering. The red line  represents $\sum m_i = 0.12\,{\rm eV}$~\cite{Zhang:2020mox}.}
\label{fig:10.1}
\end{figure}

Now, we will discuss correlation between various parameters for vanishing $|M_{\mu\tau}|$. We plot in Fig.~\ref{fig:10.1} the correlation 
between $\beta$ and $\alpha$, and between $\alpha$ and $m_{3}$ for $\delta = 0^{\circ}$ and $90^{\circ}$, 
respectively. The blue region is obtained once we impose $\sum m_i < 0.12\,{\rm eV}$ constraint, whereas, the cyan region is obtained
without the constraint coming from the total neutrino mass.
It is observed that $|M_{\mu\tau}|$ will vanish only if the lightest neutrino mass $m_3 \ge 10^{-2}\,{\rm eV}$ for the 
Dirac CP violating phase $\delta= 0^{\circ}$ and $90^{\circ}$. We find that $\alpha$, $\beta$ and $m_3$ are restricted to a very limited
region in the parameter space if we impose the $\sum m_i < 0.12\,{\rm eV}$ constraint. However, without the constraint, all values of
$\alpha$ and $\beta$ within $(0^{\circ}, 360^{\circ})$ are allowed for vanishing $|M_{\mu\tau}|$. We show the correlation between $\theta_{23}$ and $\alpha$ for vanishing $|M_{\mu\tau}|$ in Fig.~\ref{fig:Imalpth23mutautauzero_io} of Appendix~\ref{app1}.
 
\subsection{$M_{\tau\tau}$: the $\tau\tau$ element of the mass matrix}
\subsubsection{Normal Ordering:}
\begin{figure}[ht!]
\begin{subfigure}{0.35\textwidth}
\includegraphics[width=\textwidth]{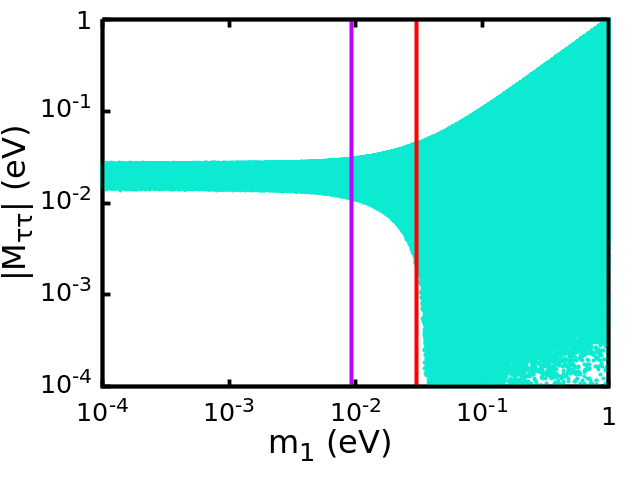}
\end{subfigure}
\captionsetup{justification=raggedright,singlelinecheck=false}
\caption{Variation of $|M_{\tau\tau}|$ with the lightest neutrino mass $m_1$ in case of normal ordering. Total neutrino mass $\sum m_{i} = 0.12\,{\rm  eV}$~\cite{Zhang:2020mox} and $\sum m_{i} = 0.072\,{\rm  eV}$~\cite{DESI:2024mwx} are shown with red and magenta lines, respectively.}
\label{fig:11}
\end{figure}

We show in Fig.~\ref{fig:11} the variation of $|M_{\tau\tau}|$ against the lightest neutrino mass~$m_1$ for normal ordering case. We also show 
the total mass $\sum m_{i} = 0.12\,{\rm eV}$ and $\sum m_{i} = 0.072\,{\rm eV}$ with a red and a magenta lines, respectively. 

\begin{figure}[ht!]
\begin{subfigure}{0.30\textwidth}
\includegraphics[width=\textwidth]{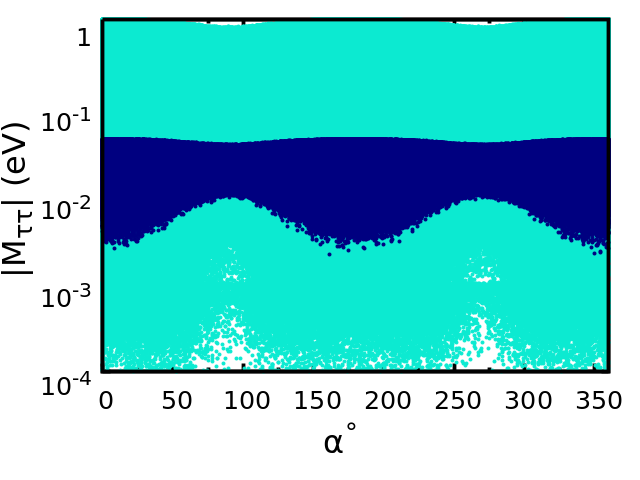}
\caption{}
\label{fig:11b}
\end{subfigure}
\begin{subfigure}{0.30\textwidth}
\includegraphics[width=\textwidth]{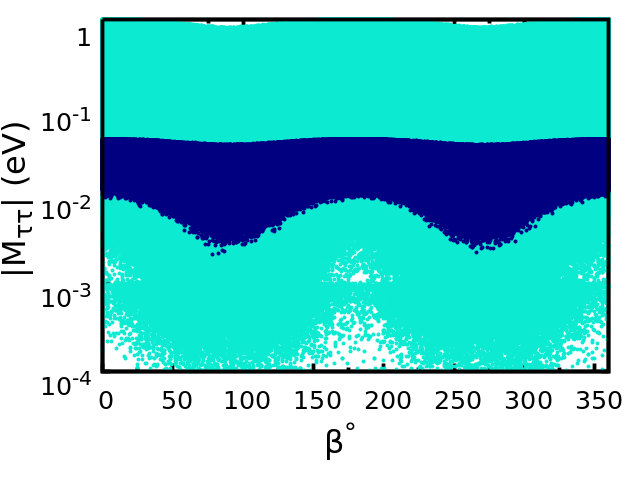}
\caption{}
\label{fig:11c}
\end{subfigure}
\begin{subfigure}{0.30\textwidth}
\includegraphics[width=\textwidth]{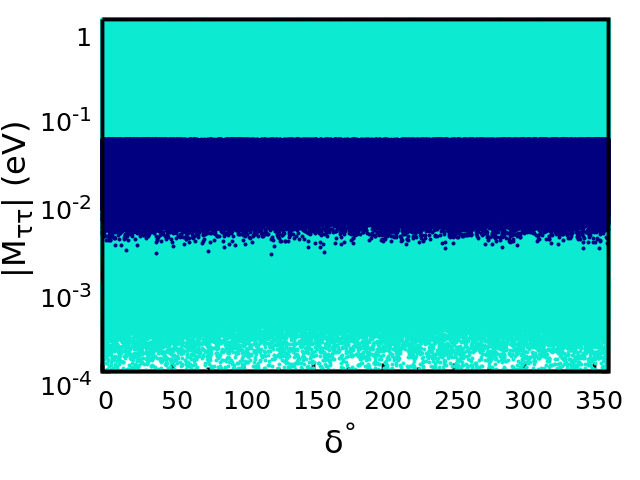}
\caption{}
\label{fig:11d}
\end{subfigure}
\caption{Variation of $|M_{\tau\tau}|$ with the three CP violating phases $\alpha$, $\beta$ and $\delta$ 
in case of normal ordering.}
\label{fig:11_cp_phases}
\end{figure}

We also show the variation of $|M_{\tau\tau}|$ against the CP violating phases $\alpha$, $\beta$ and $\delta$ in Fig.~\ref{fig:11_cp_phases}.
The blue region is obtained by using the constraint on the total neutrino mass $\sum m_{i} < 0.12\,{\rm  eV}$, whereas, the cyan
region is obtained without imposing the total neutrino mass constraint.

Our main observations are as follows:
\begin{itemize}
\item It is clear from Fig.~\ref{fig:11}that, we can have vanishing $|M_{\tau\tau}|$ for $m_{1}>3.4\times 10^{-2}$ eV, in case of normal ordering. However, we can not have 
vanishing $|M_{\tau\tau}|$ if we impose the total neutrino mass constraint $\sum m_i < 0.12\,{\rm eV}$.
\item Without the total neutrino mass constraint, it is evident that we can have vanishing $|M_{\tau\tau}|$ for any value of
the CP violating phases $\alpha$, $\beta$ and $\delta$ within $(0^{\circ}, 360^{\circ})$. 
\end{itemize}

Now, we will discuss vanishing of $|M_{\tau\tau}|$ for few extreme values of $m_{1}$. In case of normal ordering, $M_{\tau\tau}$ can be
written as
\begin{eqnarray}
M_{ \tau\tau}=(s_{12}s_{23}-c_{12}c_{23}s_{13}e^{i\delta})^2\,m_1+(-c_{12}s_{23}-s_{12}c_{23}s_{13}e^{i\delta})^2\,
\sqrt{m_1^2 + \Delta m_{21}^2}\,e^{2i\alpha}+ (c^{2}_{13}c^{2}_{23})\,\sqrt{m_1^2 + \Delta m_{31}^2}\, e^{2i\beta}.
\end{eqnarray}

\begin{itemize}
 \item  {\bf Case I:} $m_{1}<<\sqrt{\Delta m^2_{21}}$: \\
In this case, we have
\begin{equation}
\label{mtautauno}
 |M_{\tau\tau}| \approx \Big|(c_{12}s_{23}+s_{12}s_{13}c_{23}e^{i\delta})^2\sqrt{\Delta m^2_{21}}\,e^{2i\alpha}+c^{2}_{13}c^{2}_{23}
\sqrt{\Delta m^2_{31}}\,e^{2i\beta}\Big|
\end{equation}
Minimum value of $|M_{\tau\tau}|$ can be obtained if there is maximum cancellation between the two terms in Eq.~\ref{mtautauno}. The lowest
value of $|M_{\tau\tau}|$ corresponds to $\alpha = 0^{\circ}$, $\beta = 90^{\circ}$ and $\delta = 0^{\circ}$ and it is obtained to be of
$\mathcal{O}(10^{-2})$. Assuming $s_{23} \approx c_{23}$, we get
\begin{equation}
\label{mtautauno1}
 |M_{\tau\tau}| \approx c_{23}^2\,\Big|(c_{12}+s_{12}s_{13})^2\sqrt{\Delta m^2_{21}} - c^{2}_{13}\sqrt{\Delta m^2_{31}}\Big| \approx 
c_{23}^2\,c^{2}_{13}\sqrt{\Delta m^2_{31}}\,.
\end{equation}
Hence, we can not have vanishing $|M_{\tau\tau}|$.

\item  {\bf Case II:} $m_{1} >> \sqrt{\Delta m^2_{31}}$: \\
Assuming $s_{23} \approx c_{23}$, we can write $|M_{\tau\tau} |$ as
  \begin{eqnarray}
\label{mtautauno2}
    && |M_{\tau\tau} |\simeq m_1\,c_{23}^2\Big|\Big(1 + e^{2i\alpha}\Big) - \Big(c_{12}^2 + s_{12}^2\,e^{2i\alpha}\Big)\Big(1 - s_{13}^2\,
e^{2i\delta}\Big) - 2\,s_{12}\,c_{12}\,s_{13}\,e^{i\delta}\Big(1 - e^{2i\alpha}\Big) + c_{13}^2\,e^{2i\beta}\Big|
  \end{eqnarray}
  It is evident that $|M_{\tau\tau}|$ can vanish for $\alpha = 0^{\circ}$ and $\beta = \delta = 90^{\circ}$.

\item  {\bf Case III:} $m_{1} \approx \sqrt{\Delta m^2_{31}}$:\\
In this case, assuming $s_{23} \approx c_{23}$, we can write $|M_{\tau\tau} |$ 
  \begin{eqnarray}
\label{mtautauno3}
|M_{\tau\tau} |\simeq \sqrt{\Delta m^2_{31}}\,c_{23}^2\Big|(s_{12}-c_{12}s_{13}e^{i\delta})^2+(c_{12}+s_{12}s_{13}e^{i\delta})^2 \,e^{2i\alpha}+ \sqrt{2}\,c^{2}_{13}\,e^{2i\beta}\Big|   
  \end{eqnarray}

It can be shown that with $\alpha =0^{\circ}$, $\beta = 90^{\circ}$ and $\delta = 0^{\circ}$, we can have vanishing $|M_{\tau\tau}|$ in this
region.
\end{itemize}

\begin{figure}[htbp!]
\begin{subfigure}{0.30\textwidth}
\includegraphics[width=5cm,height=4cm]{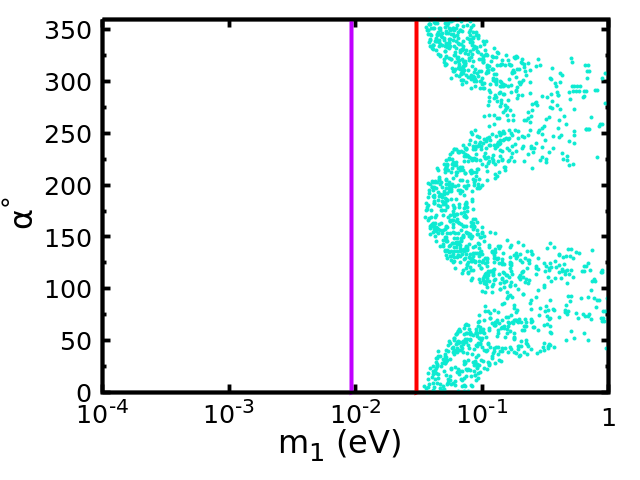}
\caption{}
\label{fig:tautauno_m1.1a}
\end{subfigure}
\begin{subfigure}{0.30\textwidth}
\includegraphics[width=5cm,height=4cm]{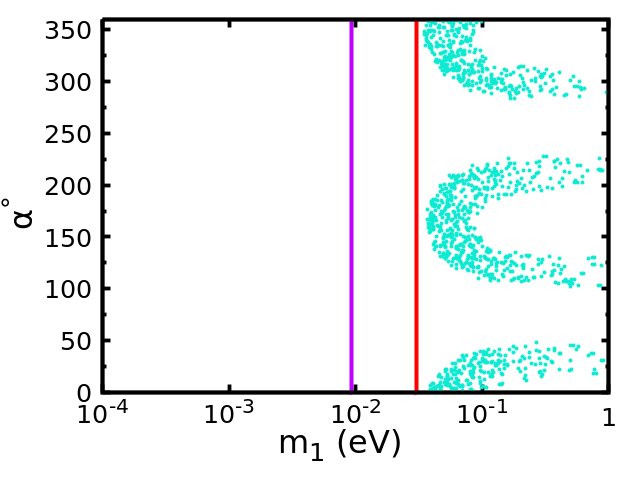}
\caption{}
\label{fig:tautauno_m1.1b}
\end{subfigure}
\begin{subfigure}{0.30\textwidth}
\includegraphics[width=5cm,height=4cm]{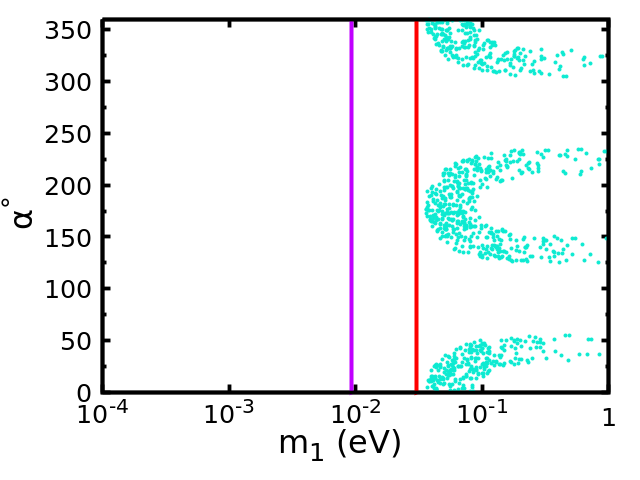}
\caption{}
\label{fig:tautauno_m1.1c}
\end{subfigure}
\begin{subfigure}{0.30\textwidth}
\includegraphics[width=5cm,height=4cm]{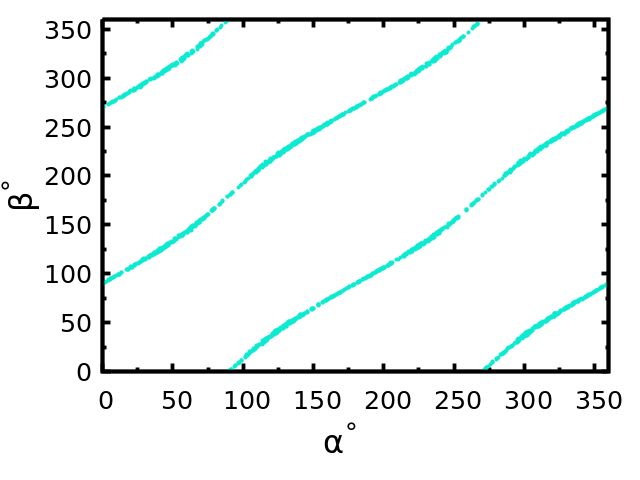}
\caption{}
\label{fig:tautauno_cp.1d}
\end{subfigure} 
\begin{subfigure}{0.30\textwidth}
\includegraphics[width=5cm,height=4cm]{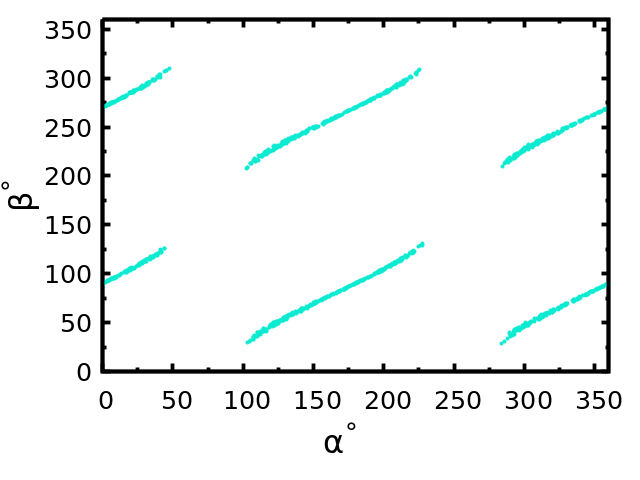}
\caption{}
\label{fig:tautauno_cp.1e}
\end{subfigure} 
\begin{subfigure}{0.30\textwidth}
\includegraphics[width=5cm,height=4cm]{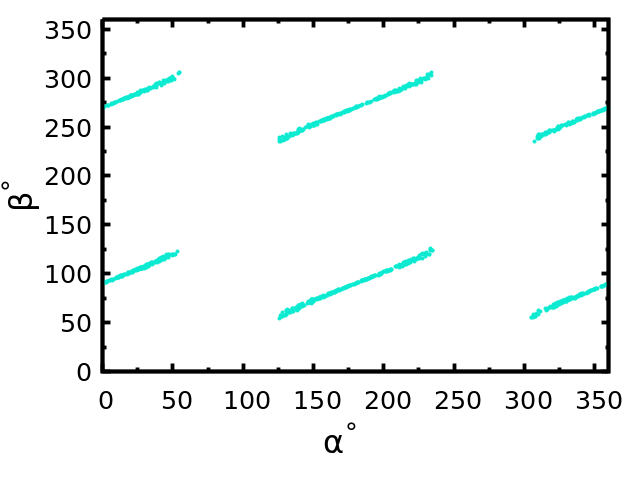}
\caption{}
\label{fig:tautauno_cp.1f}
\end{subfigure} 
\captionsetup{justification=raggedright,singlelinecheck=false}
\caption{Correlation between $\beta$ and $\alpha$, and between $\alpha$ and $m_{1}$ for $\delta=0^{\circ}$~(Left panel), 
$\delta=90^{\circ}$~(Middle panel) and $\delta=180^{\circ}$~(Right panel) with vanishing $|M_{\tau\tau}|$ in case of normal ordering. The red and magenta lines 
represent $\sum m_i = 0.12\,{\rm eV}$~\cite{Zhang:2020mox} and $\sum m_i = 0.072\,{\rm eV}$~\cite{DESI:2024mwx}, respectively.}
\label{fig:tautauno_cp}
\end{figure}

In Fig~\ref{fig:tautauno_cp}, we show correlation between $\beta$ and $\alpha$, and between $\alpha$ and $m_{1}$ with vanishing 
$|M_{\tau\tau}|$ for several values of the Dirac CP violating phase $\delta$. It is evident that for $\delta = 0^{\circ}$, $|M_{\tau\tau}|$
can vanish for any value of $\alpha$ and $\beta$ within $(0^{\circ}, 360^{\circ})$. However, for $\delta = 90^{\circ}$ and 
$180^{\circ}$, the value of $\alpha$ can not be $90^{\circ}$ or $270^{\circ}$ to have vanishing $|M_{\tau\tau}|$. Similarly, $\beta$ value
can not be $0^{\circ}$, $180^{\circ}$ or $360^{\circ}$ to have vanishing $|M_{\tau\tau}|$ for this case. It is worth mentioning that $|M_{\tau\tau}|$ vanishes only if $\theta_{23}$ lies in the upper octant, more precisely value of the $\theta_{23}$ has to be $\theta_{23} > 44.3^{\circ}$ to have vanishing $|M_{\tau\tau}|$. We
have shown it in Fig.~\ref{fig:Nmalpth23tautauzero_no} of Appendix~\ref{app1}.

\subsubsection{Inverted Ordering:}
\begin{figure}[htbp]
\begin{subfigure}{0.35\textwidth}
\includegraphics[width=\textwidth]{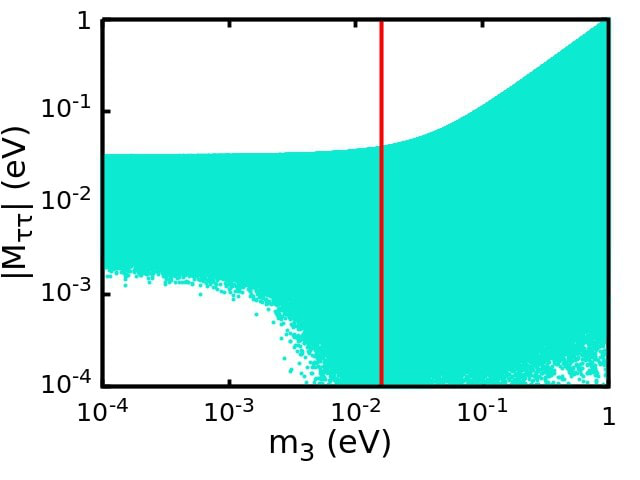}
\end{subfigure}
\captionsetup{justification=raggedright,singlelinecheck=false}
\caption{Variation of $|M_{\tau\tau}|$ with the lightest neutrino mass $m_3$ in case of inverted ordering. Total neutrino mass $\sum m_{i} = 0.12\,{\rm  eV}$~\cite{Zhang:2020mox} is shown with the red vertical line.}
\label{fig:12}
\end{figure}

\begin{figure}[htbp]
\begin{subfigure}{0.32\textwidth}
\includegraphics[width=5.2cm,height=4.4cm]{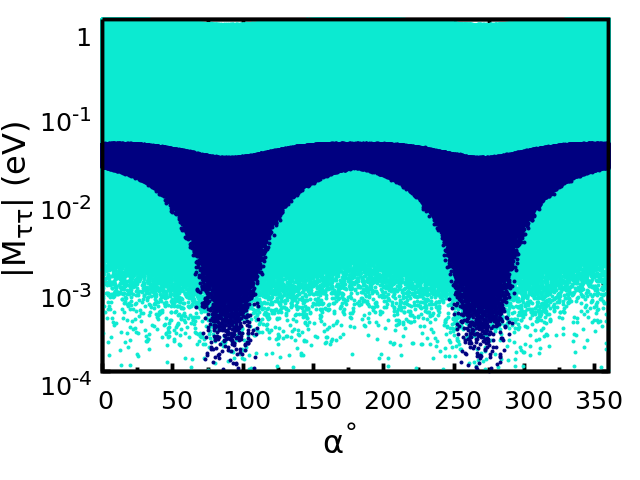}
\caption{}
\label{fig:12b}
\end{subfigure}
\begin{subfigure}{0.32\textwidth}
\includegraphics[width=5.2cm,height=4.4cm]{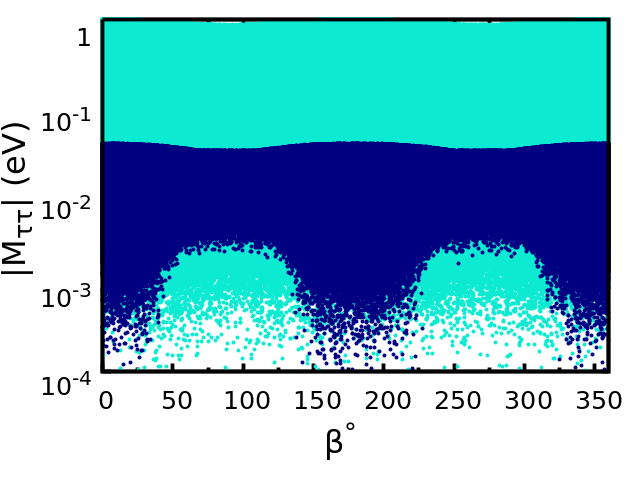}
\caption{}
\label{fig:12c}
\end{subfigure}
\begin{subfigure}{0.32\textwidth}
\includegraphics[width=5.2cm,height=4.4cm]{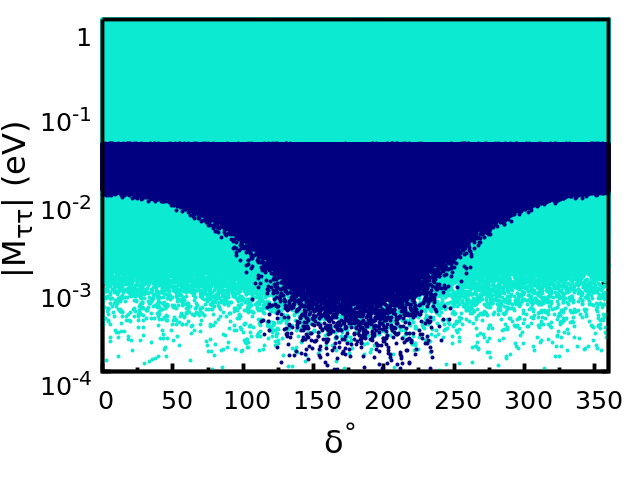}
\caption{}
\label{fig:12d}
\end{subfigure}
\caption{Variation of $|M_{\tau\tau}|$ with the three CP violating phases $\alpha$, $\beta$ and $\delta$ in case of inverted ordering.}
\label{fig:12_cp_phases}
\end{figure}

We show in Fig.~\ref{fig:12} the variation of $|M_{\tau\tau}|$ against the lightest neutrino mass~$m_3$ for the inverted ordering. Upper limit 
on the total neutrino mass $\sum m_i = 0.12\,{\rm eV}$ is shown with a red vertical line.
We also show the variation of $|M_{\tau\tau}|$ against the CP violating phases $\alpha$, $\beta$ and $\delta$ in Fig.~\ref{fig:12_cp_phases}.
The blue band is obtained by using the constraint on the total neutrino mass $\sum m_{i} < 0.12\,{\rm  eV}$, whereas, the cyan band is
obtained without using the total neutrino mass constraint.

In case of inverted ordering, our observations are as follows:
\begin{itemize}
 \item For vanishing $|M_{\tau\tau}|$, the lightest neutrino mass should be $m_{3}\geq 4.0\times 10^{-3}\,\,{\rm eV}$.
 \item  For vanishing $|M_{\tau\tau}|$, $\alpha$ should lie within $(80^{\circ},\,110^{\circ})$ and $(260^{\circ},\,290^{\circ})$ if we 
impose $\sum m_{i} < 0.12\,{\rm  eV}$ constraint. However, without the constraint, we can have vanishing $|M_{\tau\tau}|$ for any value of $\alpha$ within 
$(0^{\circ},\,360^{\circ})$.
 \item  $|M_{\tau\tau}|$ will vanish if $\beta$ value lies within $(0^{\circ},\,25^{\circ})$, $(155^{\circ},\,205^{\circ})$ and
$(335^{\circ},\,360^{\circ})$ with the total neutrino mass constraint. Without the constraint, $|M_{\tau\tau}|$ can vanish for any value of $\beta$
within $(0^{\circ},\,360^{\circ})$.
 \item  For vanishing $|M_{\tau\tau}|$, the Dirac CP violating phase $\delta$ is restricted to lie within $(130^{\circ},\,240^{\circ})$ if we 
impose the total neutrino mass constraint. However, without the constraint, all values of $\delta$ within $(0^{\circ}, 360^{\circ})$ are
allowed.
 \end{itemize}

 Now, we will discuss vanishing of $|M_{\tau\tau}|$ for different values of $m_3$. In case of inverted ordering, the general expression for 
$M_{\tau\tau}$ is
 \begin{eqnarray}
     M_{ \tau\tau} &=&(s_{12}s_{23}-c_{12}c_{23}s_{13}e^{i\delta})^2\,\sqrt{m^{2}_{3}+\Delta m^2_{13}}+(-c_{12}s_{23}-s_{12}c_{23}s_{13}
e^{i\delta})^2\,\sqrt{m^{2}_{3}+\Delta m^2_{13} + \Delta m^2_{21}}\,e^{2i\alpha} \nonumber \\
     &&+(c^{2}_{13}c^{2}_{23})\,m_3 e^{2i\beta}
 \end{eqnarray}
 It is clear that we can obtain $M_{\tau\tau}$ from $M_{\mu\mu}$ by interchanging $s_{23}$ with $c_{23}$ and $c_{23}$ with $-s_{23}$. 

\begin{itemize}
    
\item {\bf Case I:} $m_{3}<<\sqrt{\Delta m^2_{21}}$: \\
In this case, we have
  \begin{eqnarray}
  \label{tautauio}
   |M_{\tau\tau}|\simeq \sqrt{\Delta m^2_{13}}\Big|(s_{12}s_{23}-c_{12}c_{23}s_{13}e^{i\delta})^2+(-c_{12}s_{23}-s_{12}c_{23}s_{13}
e^{i\delta})^2\,e^{2i\alpha}\Big|. 
  \end{eqnarray}
  For the case, the allowed range of $|M_{\tau\tau}|$ is found to be $(1.0\times 10^{-3},3.0\times 10^{-2})\,{\rm eV}$.
  If we assume $s_{23} \approx c_{23}$, then Eq.~\ref{tautauio} becomes
  \begin{eqnarray}
   |M_{\tau\tau} &|\simeq& \sqrt{\Delta m^2_{13}}\,c_{23}^2\Big|\Big(1 + e^{2i\alpha}\Big) - \Big(c_{12}^2 + s_{12}^2\,e^{2i\alpha}\Big)
\Big(1 - s_{13}^2\,e^{2i\delta}\Big) - 2\,s_{12}\,c_{12}\,s_{13}\,e^{i\delta}\Big(1 - e^{2i\alpha}\Big)\Big|
  \end{eqnarray}
It is clear that for $\alpha = 0^{\circ}$ and $180^{\circ}$, the first term $1 + e^{2i\alpha} = 2$, second term $\approx 1$ and the third term
will be $0$ irrespective of the $\delta$ value. So, there will not be exact cancellation of all the terms in this case. Minimum value of 
$|M_{\tau\tau}|$ can be obtained if we assume $\alpha = 90^{\circ}$ and $\delta = 180^{\circ}$. With these choice of parameter values, we 
obtain the minimum of 
$|M_{\tau\tau}|$ to be of $\mathcal O(10^{-3})\,{\rm eV}$. Hence, in the limit $m_3 \to 0$, we can not have vanishing $|M_{\tau\tau}|$. This 
is in contrast to $|M_{\mu\mu}|$ although there is an approximate $\mu - \tau$ symmetry.
  
\item  {\bf Case II:} $m_{3} >> \sqrt{\Delta m^2_{13}}$: \\
In this case
  \begin{eqnarray}
  \label{mtautauio}
   &&|M_{\tau\tau} |\simeq m_3\,c_{23}^2\Big|\Big(1 + e^{2i\alpha}\Big) - \Big(c_{12}^2 + s_{12}^2\,e^{2i\alpha}\Big)\Big(1 - s_{13}^2\,
e^{2i\delta}\Big) - 2\,s_{12}\,c_{12}\,s_{13}\,e^{i\delta}\Big(1 - e^{2i\alpha}\Big)\nonumber \\
  &&\hspace{1.5cm} + c_{13}^2\,e^{2i\beta}\Big|\,,
  \end{eqnarray}
where we have assumed $s_{23} \approx c_{23}$. It is clear that Eq.~\ref{mtautauio} and Eq.~\ref{mtautauno2} are exactly same. Hence, we can 
have vanishing $|M_{\tau\tau}|$ if we have $\alpha = 0^{\circ}$, $\beta = \delta = 90^{\circ}$.

  \item {\bf Case III:} $m_{3}\approx \sqrt{\Delta m^2_{13}}$:\\
  In this case, we have
  \begin{eqnarray}
  \label{eq:36}
   |M_{\tau\tau}| \simeq c_{23}^2\,\sqrt{\Delta m^2_{13}}\,\Big|\sqrt{2}\,(s^{2}_{12}+c^{2}_{12}e^{2i\alpha})-2\sqrt{2}s_{12}s_{13}c_{12}
(1-e^{2i\alpha})e^{i\delta}+\sqrt{2}s^{2}_{13}(c^{2}_{12}+s^{2}_{12}e^{2i\alpha})e^{2i\delta}+c^{2}_{13}e^{2i\beta}\Big|\,,
  \end{eqnarray}
where we have assumed $s_{23} \approx c_{23}$. For the $\delta$ value around 
$180^{\circ}$ and $\beta$ value around $0^{\circ}$, $|M_{\tau\tau}|$ vanishes if we have 
\begin{eqnarray}
    \cos2\alpha\approx -\frac{2\Big[s_{12}+s_{13}c_{12}\Big]^4+\Big[\sqrt{2}\,(c_{12}-s_{12}s_{13})^2+c_{12}^{2}\Big]^{2}}
{2\sqrt{2}\,\Big[s_{12}+s_{13}c_{12}\Big]^{2}\Big[\sqrt{2}\,(c_{12}-s_{12}s_{13})^2+c_{12}^{2}\Big]}\approx -1
\end{eqnarray}
  
\end{itemize}

\begin{figure}[htbp]
 \begin{subfigure}{0.40\textwidth}
\includegraphics[width=5.4cm,height=4.5cm]{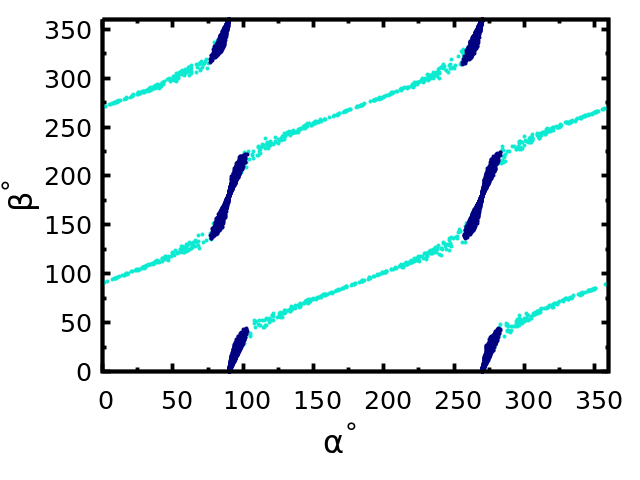}
\caption{}
\label{fig:12.1a}
\end{subfigure} 
\begin{subfigure}{0.40\textwidth}
\includegraphics[width=5.4cm,height=4.5cm]{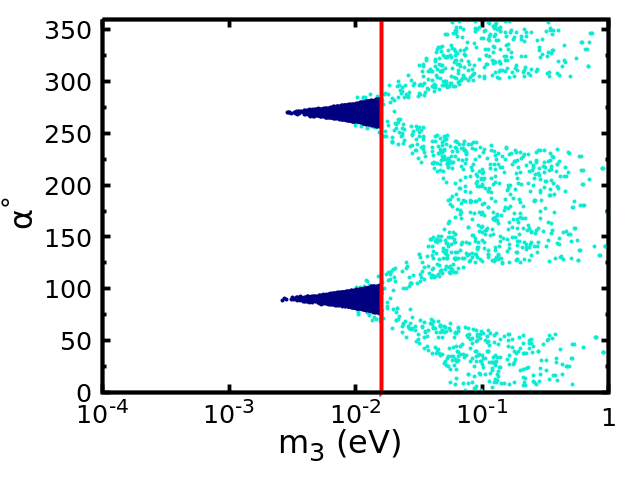}
\caption{}
\label{fig:12.1b}
\end{subfigure}
\captionsetup{justification=raggedright,singlelinecheck=false}
\caption{Correlation between $\beta$ and $\alpha$, and between $\alpha$ and $m_{3}$ for $\delta=180^{\circ}$ with vanishing $|M_{\tau\tau}|$
in case of inverted ordering. The red line  represents $\sum m_i = 0.12\,{\rm eV}$~\cite{Zhang:2020mox}.}
\label{fig:12.1}
\end{figure}

We show the correlation between $\beta$ and $\alpha$, and between $\alpha$ and $m_{3}$ for $\delta = 180^{\circ}$ with vanishing 
$|M_{\tau\tau}|$ in Fig.~\ref{fig:12.1a} and Fig.~\ref{fig:12.1b}, respectively. The blue band represents the correlation once the 
$\sum m_i < 0.12\,{\rm eV}$ constraint coming from the total neutrino mass is imposed, whereas, the cyan band represents the correlation
between the parameters without the total neutrino mass constraint. 

It is evident that the value of the lightest neutrino mass $m_3 \ge \,4.0\times 10^{-3}\,\,{\rm eV}$ to have vanishing $|M_{\tau\tau}|$. 
It is clear that for vanishing $|M_{\tau\tau}|$, the Majorana CP violating phase $\alpha$ has to lie very close to $90^{\circ}$ or 
$270^{\circ}$ for $\delta = 180^{\circ}$ if $m_3 < 10^{-2}\,{\rm eV}$. However, for $m_3 > 10^{-2}\,{\rm eV}$, value of $\alpha$
will gradually deviate from $90^{\circ}$ or $270^{\circ}$. The deviation from $90^{\circ}$ or $270^{\circ}$ will be more if we keep 
increasing the value of $m_3$. We also show the correlation between $\theta_{23}$ and $\alpha$ for vanishing $|M_{\tau\tau}|$ in Fig.~\ref{fig:Imalpth23tautauzero_io} of Appendix~\ref{app1}.

\bigskip
\section{Two texture zeros}
In this section we will explore the phenomenological implications of two texture zeros in the neutrino mass matrix using a bottom-up approach
using the latest oscillation data for normal and inverted mass ordering.
  
\subsection{Normal Ordering}
\subsubsection{Vanishing $|M_{\tau\tau}|$:}
\begin{figure}[htbp!]
   \begin{subfigure}{0.32\textwidth}
\includegraphics[width=5.2cm,height=4.4cm]{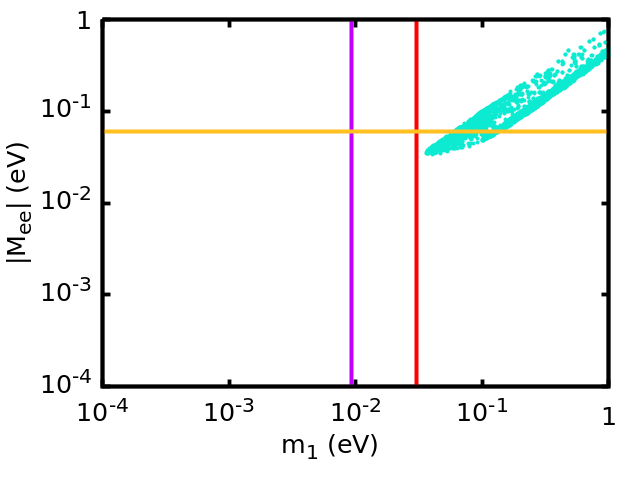}
\caption{}
\label{fig:21.21a}
\end{subfigure}
\begin{subfigure}{0.32\textwidth}
\includegraphics[width=5.2cm,height=4.4cm]{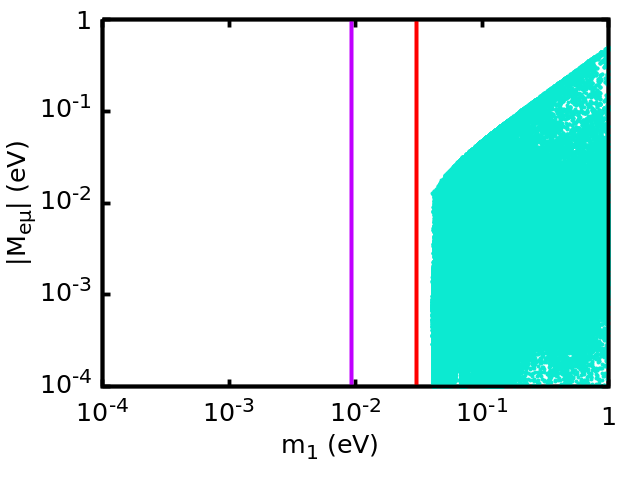}
\caption{}
\label{fig:21.21b}
\end{subfigure}
\begin{subfigure}{0.32\textwidth}
\includegraphics[width=5.2cm,height=4.4cm]{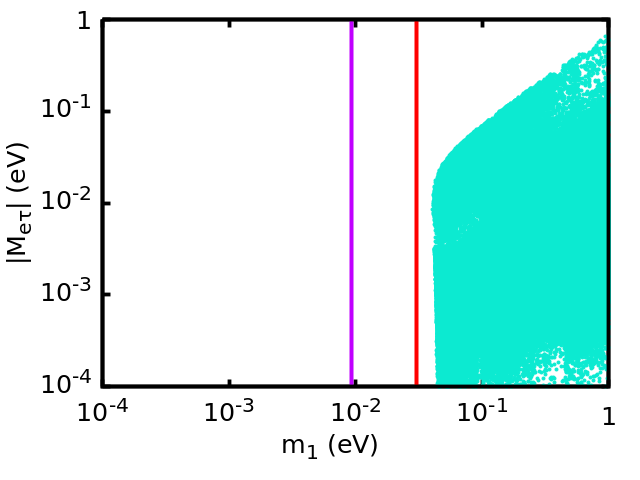}
\caption{}
\label{fig:21.21c}
\end{subfigure}
\begin{subfigure}{0.32\textwidth}
\includegraphics[width=5.2cm,height=4.4cm]{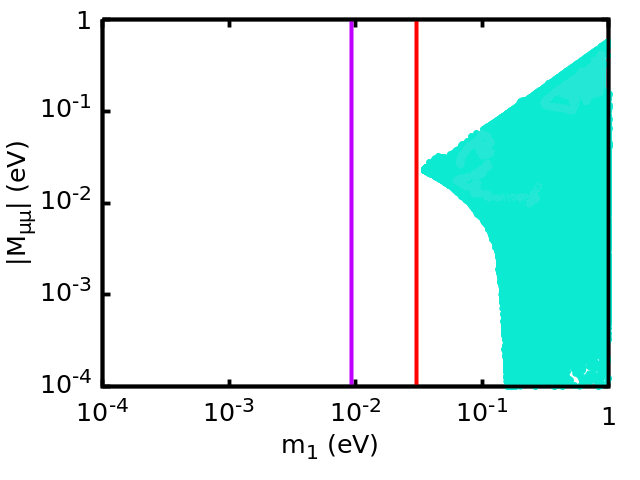}
\caption{}
\label{fig:21.21d}
\end{subfigure}
\begin{subfigure}{0.32\textwidth}
\includegraphics[width=5.2cm,height=4.4cm]{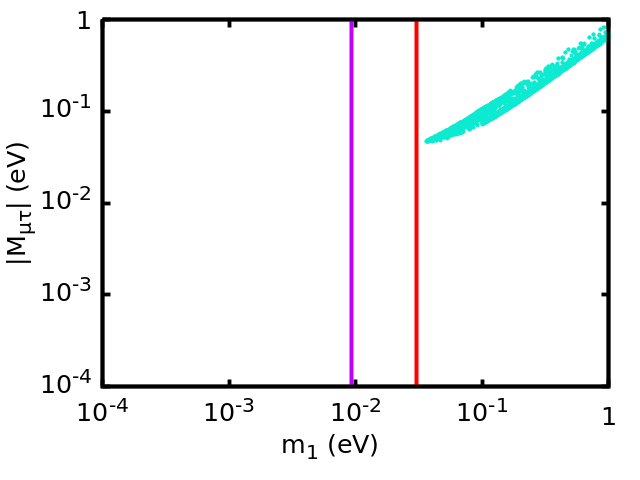}
\caption{}
\label{fig:21.21e}
\end{subfigure}
\captionsetup{justification=raggedright,singlelinecheck=false}
 \caption{ Correlation between elements of the neutrino mass matrix and the lightest neutrino mass $m_{1}$ with vanishing $|M_{\tau\tau}|$ in 
case of normal ordering. The red and magenta lines represent $\sum m_i = 0.12\,{\rm eV}$~\cite{Zhang:2020mox} and 
$\sum m_i = 0.072\,{\rm eV}$~\cite{DESI:2024mwx}, respectively. The yellow line represents 
$|M_{ee}|= 0.06\,{\rm eV}$~\cite{KamLAND-Zen:2016pfg}.}
\label{fig21:tautauzero_no}
\end{figure}

We show the variation of $|M_{ee}|$, $|M_{e\mu}|$, $|M_{e\tau}|$, $|M_{\mu\mu}|$ and 
$|M_{\mu\tau}|$ with the lightest neutrino mass $m_{1}$ for vanishing $|M_{\tau\tau}|$ in Fig.~\ref{fig:21.21a}, Fig.~\ref{fig:21.21b}, 
Fig.~\ref{fig:21.21c}, Fig.~\ref{fig:21.21d}, and Fig.~\ref{fig:21.21e}, respectively. The bound on the total neutrino mass 
$\sum m_i < 0.12\,{\rm eV}$ and $\sum m_i < 0.072\,{\rm eV}$ are represented by red and magenta lines, respectively. Moreover,
the upper bound of $M_{ee} = 0.06\,{\rm eV}$ is represented by yellow line in Fig.~\ref{fig:21.21a}.

\begin{figure}[htbp]
 \begin{subfigure}{0.32\textwidth}
\includegraphics[width=5.2cm,height=4.4cm]{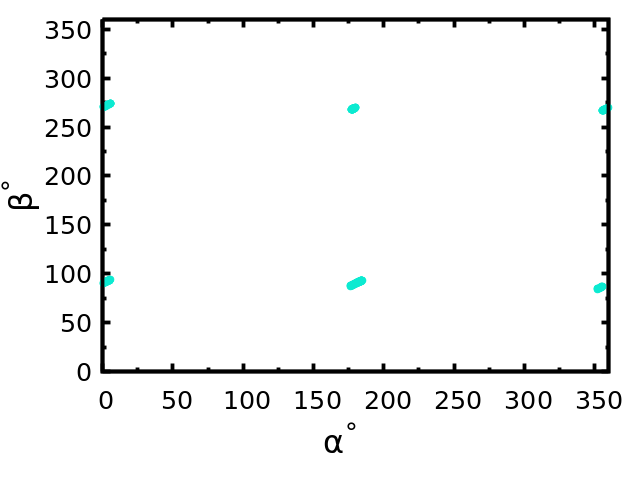}
\caption{}
\label{fig:emutautauzeroaplbeta}
\end{subfigure}
\begin{subfigure}{0.32\textwidth}
\includegraphics[width=5.2cm,height=4.4cm]{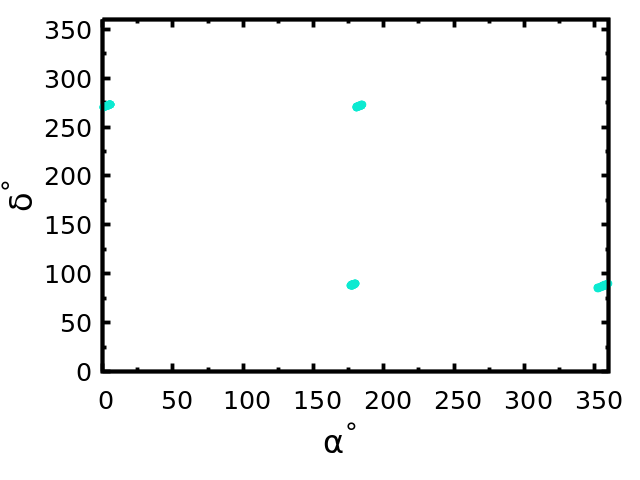}
\caption{}
\label{fig:emutautauzeroapldelta}
\end{subfigure}
\begin{subfigure}{0.32\textwidth}
\includegraphics[width=5.2cm,height=4.4cm]{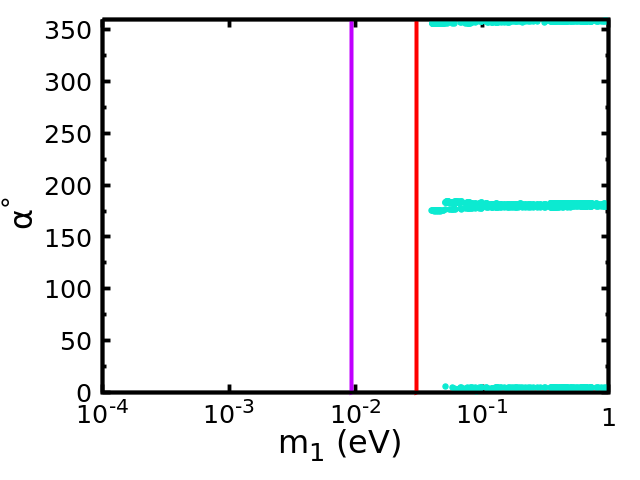}
\caption{}
\label{fig:emutautauzeroaplham1}
\end{subfigure}
\captionsetup{justification=raggedright,singlelinecheck=false}
\caption{Correlation between $\alpha$ and $\beta$, $\alpha$ and $\delta$, and between $\alpha$ and $m_{1}$ with vanishing $|M_{e\mu}|$ and 
$|M_{\tau\tau}|$ in case of normal ordering.}
\label{fig:emutautauzero_no}
\end{figure}

\begin{figure}[htbp]
 \begin{subfigure}{0.32\textwidth}
\includegraphics[width=5.2cm,height=4.4cm]{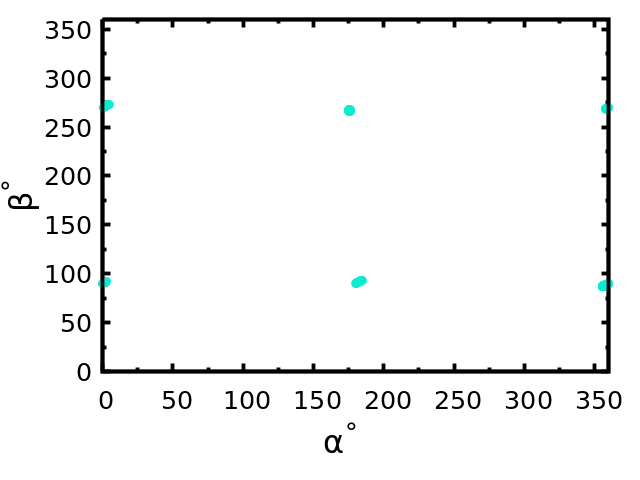}
\caption{}
\label{fig:etautautauzeroaplbeta_no}
\end{subfigure}
\begin{subfigure}{0.32\textwidth}
\includegraphics[width=5.2cm,height=4.4cm]{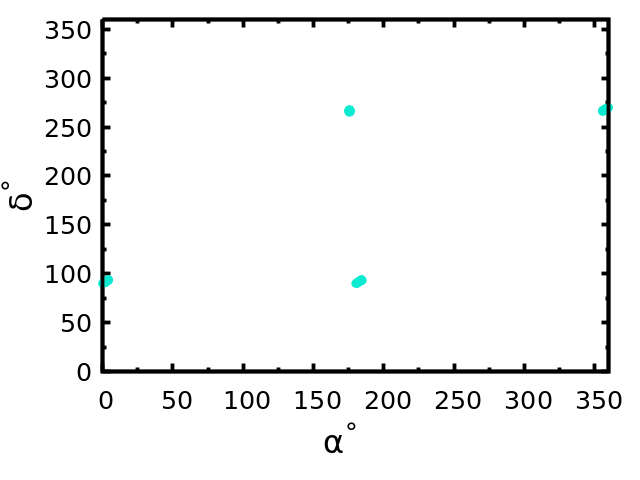}
\caption{}
\label{fig:etautautauzeroapldelta_no}
\end{subfigure}
\begin{subfigure}{0.32\textwidth}
\includegraphics[width=5.2cm,height=4.4cm]{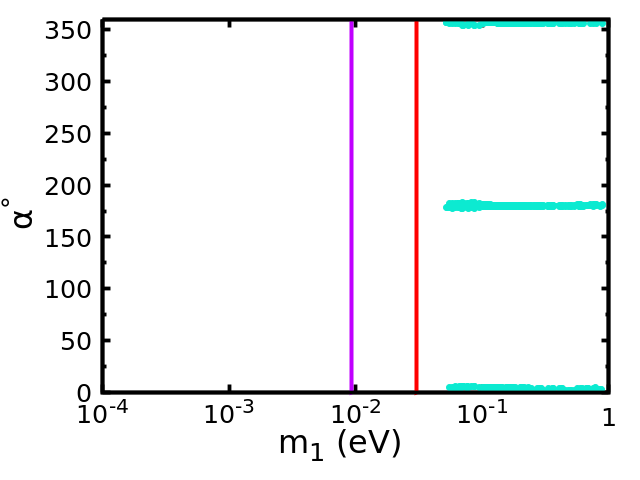}
\caption{}
\label{fig:etautautauzeroaplham1_no}
\end{subfigure}
\captionsetup{justification=raggedright,singlelinecheck=false}
\caption{Correlation between $\alpha$ and $\beta$, $\alpha$ and $\delta$, and between $\alpha$ and $m_{1}$ with vanishing $|M_{e\tau}|$ and 
$|M_{\tau\tau}|$ in case of normal ordering.}
\label{fig:etautautauzero_no}
\end{figure}

From Fig.~\ref{fig21:tautauzero_no}, it is evident that $|M_{e\mu}|$,  
$|M_{e\tau}|$ and $|M_{\mu\mu}|$ can vanish simultaneously with $|M_{\tau\tau}|$, however, $|M_{ee}|$,  and $|M_{\mu\tau}|$ can not. This is in agreement with the result reported in Ref.~\cite{Singh:2019baq}. We also show the correlation between $\alpha$ and $\beta$, $\alpha$ and $\delta$, and between $m_{1}$ and $\alpha$ with vanishing $|M_{e\mu}|$ and $|M_{\tau\tau}|$ in Figs.~\ref{fig:emutautauzeroaplbeta}, 
\ref{fig:emutautauzeroapldelta}, \ref{fig:emutautauzeroaplham1}, respectively. It is observed that $(|M_{e\mu}| = 0, |M_{\tau\tau}| =0)$ 
puts severe constraint on the CP violating phases $\alpha$, $\beta$ and $\delta$. Value of $\alpha$ is restricted to a very narrow region 
around $0^{\circ}$, $180^{\circ}$ and $360^{\circ}$. Similarly, for vanishing $|M_{e\mu}|$ and $|M_{\tau\tau}|$, value of $\beta$ and $\delta$
should lie around $90^{\circ}$ or $270^{\circ}$, respectively. It should also be noted that we can have simultaneous vanishing of 
$|M_{e\mu}|$ and $|M_{\tau\tau}|$ if the lightest neutrino mass $m_1 > 3.9 \times 10^{-2}\, {\rm eV}$ which is slightly above the cosmological
upper bound of $\sum m_i < 0.12\,{\rm eV}$. We also show the correlation between $\alpha$ and $\beta$, $\alpha$ and $\delta$, and between 
$m_{1}$ and $\alpha$ with vanishing $|M_{e\tau}|$ and $|M_{\tau\tau}|$ in Figs.~\ref{fig:etautautauzeroaplbeta_no}, 
\ref{fig:etautautauzeroapldelta_no}, \ref{fig:etautautauzeroaplham1_no}, respectively. 
The correlation between $\alpha$ and $\beta$ and between $\alpha$ and $m_1$ of Fig.~\ref{fig:emutautauzeroaplbeta} and
Fig.~\ref{fig:emutautauzeroaplham1} looks quite similar to that of Fig.~\ref{fig:etautautauzeroaplbeta_no} and
Fig.~\ref{fig:etautautauzeroaplham1_no}, respectively. It is, however, worth mentioning that correlation between $\alpha$ and $\delta$ of
Fig.~\ref{fig:emutautauzeroapldelta} is quite different from that of Fig.~\ref{fig:etautautauzeroapldelta_no}. It is important to note that $|M_{e\mu}|$ and $|M_{\tau\tau}|$ can vanish simultaneously only if $\theta_{23}$ lies in the upper octant, i.e, the value of $\theta_{23}$ has to be $\theta_{23} > 45^{\circ}$ to have vanishing $|M_{e\mu}|$ and $|M_{\tau\tau}|$. We
have shown it in Fig.~\ref{fig:Nmalpth23emutautauzerofinal_no} of Appendix~\ref{app2}. Similarly, $|M_{e\tau}|$ and $|M_{\tau\tau}|$ can vanish simultaneously only if $\theta_{23}$ lies in the upper octant, i.e, the value of $\theta_{23}$ has to be $\theta_{23} > 45^{\circ}$ to have vanishing $|M_{e\tau}|$ and $|M_{\tau\tau}|$. We
have shown it in Fig.~\ref{fig:Nmalpth23etautautauzerofinal_no} of Appendix~\ref{app2}.

\begin{figure}[h!]
 \begin{subfigure}{0.32\textwidth}
\includegraphics[width=5.2cm,height=4.4cm]{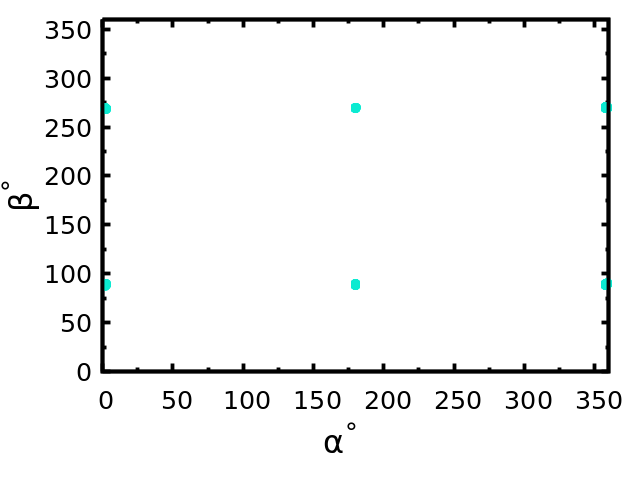}
\caption{}
\label{fig:mumutautauzeroaplbeta_no}
\end{subfigure}
\begin{subfigure}{0.32\textwidth}
\includegraphics[width=5.2cm,height=4.4cm]{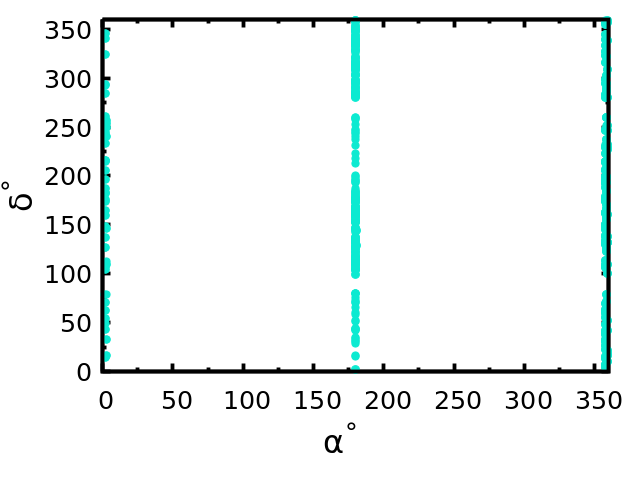}
\caption{}
\label{fig:mumutautauzeroapldelta_no}
\end{subfigure}
\begin{subfigure}{0.32\textwidth}
\includegraphics[width=5.2cm,height=4.4cm]{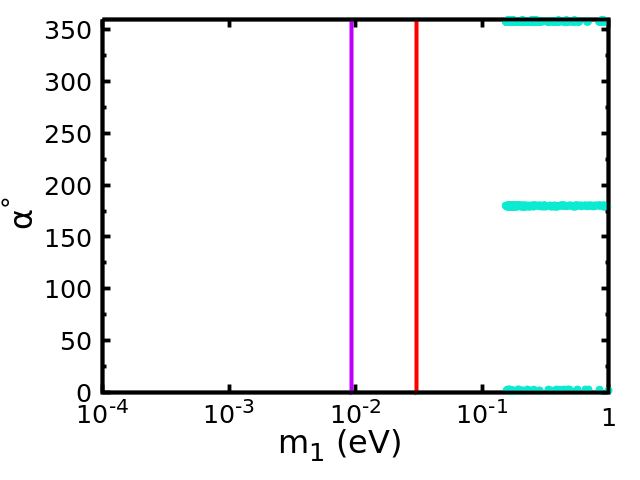}
\caption{}
\label{fig:mumutautauzeroaplham1_no}
\end{subfigure}
\captionsetup{justification=raggedright,singlelinecheck=false}
\caption{Correlation between $\alpha$ and $\beta$, $\alpha$ and $\delta$, and between $\alpha$ and $m_{1}$ with vanishing $|M_{\mu\mu}|$ and 
$|M_{\tau\tau}|$ in case of normal ordering.}
\label{fig:mumutautauzero_no}
\end{figure}

We also show the correlation between $\alpha$ and $\beta$, $\alpha$ and $\delta$, and between 
$m_{1}$ and $\alpha$ with vanishing $|M_{\mu\mu}|$ and $|M_{\tau\tau}|$ in Fig.~\ref{fig:mumutautauzeroaplbeta_no}, 
Fig.~\ref{fig:mumutautauzeroapldelta_no} and Fig.~\ref{fig:mumutautauzeroaplham1_no}, respectively. From Fig.~\ref{fig:mumutautauzero_no}, it 
is evident that $\alpha$ is restricted to a very limitted region around $0^{\circ}$, $180^{\circ}$ and $360^{\circ}$ for
vanishing $|M_{\mu\mu}|$ and $|M_{\tau\tau}|$. Similarly, $\beta$ value should be around $90^{\circ}$ and $270^{\circ}$ to have
vanishing $|M_{\mu\mu}|$ and $|M_{\tau\tau}|$. However, there is no such restriction on the Dirac CP violating phase $\delta$. Moreover, 
we can have simultaneous vanishing of $|M_{\mu\mu}|$ and $|M_{\tau\tau}|$ if the lightest neutrino mass $m_1 > 0.15\, {\rm eV}$. It is important to note that $|M_{\mu\mu}|$ and $|M_{\tau\tau}|$ can vanish simultaneously if the value of $\theta_{23}$ is around $45^{\circ}$. We have shown it in Fig.~\ref{fig:Nmalpth23mumutautauzerofinal_no} of Appendix~\ref{app2}. 

The lowest value of effective Majorana mass term is found to be $M_{ee} = 3.4\times 10^{-2}$ eV for vanishing $|M_{\tau\tau}|$, which is 
consistent with the upper limit of $M_{ee}$ obtained from the EXO-200, CUORE, and KamLAND-Zen experiments. Notably, this value corresponds to 
$m_{1} = 4.0\times 10^{-2}\,{\rm eV}$, which exceeds the upper bound of $\sum m_i < 0.12\,{\rm eV}$ as reported by the DESI and Planck 
data along with BAO observations. For $M_{ee}< 0.06$ eV, the value of $\sum m_i$ lies in the range $(0.13, 0.39)\,{\rm eV}$.

\subsubsection{Vanishing $|M_{\mu\tau}|$:}
\begin{figure}[htbp!]
   \begin{subfigure}{0.24\textwidth}
\includegraphics[width=\textwidth]{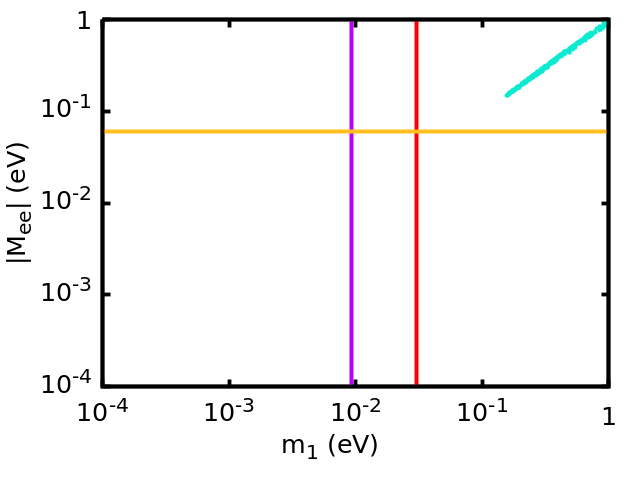}
\caption{}
\label{fig:22.22a}
\end{subfigure}
\begin{subfigure}{0.24\textwidth}
\includegraphics[width=\textwidth]{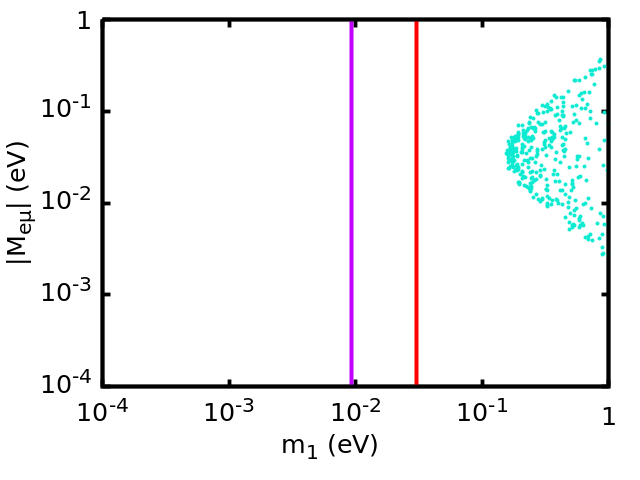}
\caption{}
\label{fig:22.22b}
\end{subfigure}
\begin{subfigure}{0.24\textwidth}
\includegraphics[width=\textwidth]{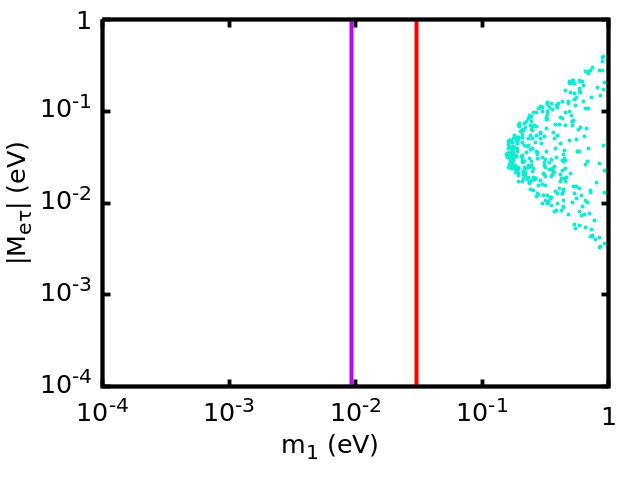}
\caption{}
\label{fig:22.22c}
\end{subfigure}
\begin{subfigure}{0.24\textwidth}
\includegraphics[width=\textwidth]{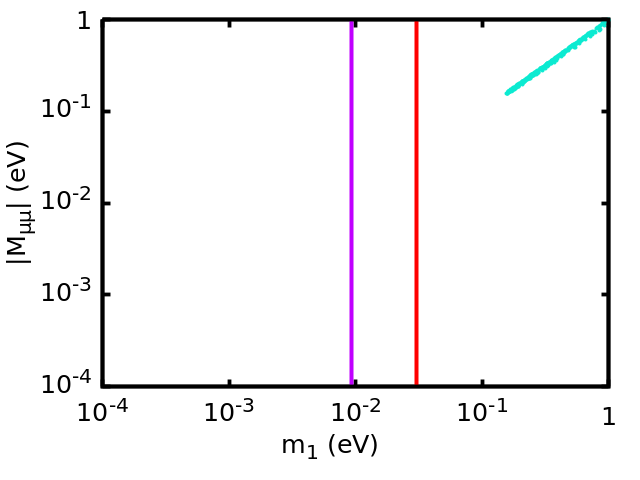}
\caption{}
\label{fig:22.22d}
\end{subfigure}
 \captionsetup{justification=raggedright,singlelinecheck=false}
 \caption{ Correlation of $|M_{ee}|$, $|M_{e\mu}|$, $|M_{e\tau}|$ and $|M_{\mu\mu}|$ with the lightest neutrino mass $m_{1}$ for 
vanishing $|M_{\mu\tau}|$ in case of normal ordering. The red and magenta lines represent $\sum m_i = 0.12\,{\rm eV}$~\cite{Zhang:2020mox} 
and $\sum m_i = 0.072\,{\rm eV}$~\cite{DESI:2024mwx}, respectively. In Fig.~\ref{fig:22.22a} yellow line represents $|M_{ee}|= 0.06\,{\rm eV}$
obtained from KamLAND-Zen experiment~\cite{KamLAND-Zen:2016pfg}.}
\label{fig22:mutauzero_no}
\end{figure}

Variation of $|M_{ee}|$, $|M_{e\mu}|$, $|M_{e\tau}|$, and  $|M_{\mu\mu}|$ with the lightest neutrino mass $m_{1}$ for vanishing 
$|M_{\mu\tau}|$ is shown in Fig.~\ref{fig:22.22a}, Fig.~\ref{fig:22.22b}, Fig.~\ref{fig:22.22c}, and Fig.~\ref{fig:22.22d}, respectively. 
We also show the bound on the total neutrino mass $\sum m_i < 0.12\,{\rm eV}$ and $\sum m_i < 0.072\,{\rm eV}$ with red and magenta lines, 
respectively. In Fig.~\ref{fig:22.22a}, the upper bound of $M_{ee} = 0.06\,{\rm eV}$ is represented by yellow line.

From Fig.~\ref{fig22:mutauzero_no}, it is observed that $|M_{ee}|$, $|M_{e\mu}|$, $|M_{e\tau}|$, and $|M_{\mu\mu}|$ do not vanish for
vanishing $|M_{\mu\tau}|$. The lowest value of effective Majorana mass term $|M_{ee}|$ is obtained to be $0.14\,{\rm eV}$ for vanishing 
$|M_{\mu\tau}|$. This result is consistent with the upper limit for $M_{ee} < (0.06, 0.31)\,{\rm eV}$ reported by the EXO-200, CUORE, and 
KamLAND-Zen experiments. Considering the upper bound of $M_{ee}< 0.31\,{\rm eV}$ obtained from the CUORE experiment, the value of 
$\sum m_i$ is found to be in the range $(0.47, 0.97)\,{\rm eV}$.

\subsubsection{Vanishing $|M_{\mu\mu}|$:} 
\begin{figure}[htbp]
 \begin{subfigure}{0.32\textwidth}
\includegraphics[width=5.2cm,height=4.4cm]{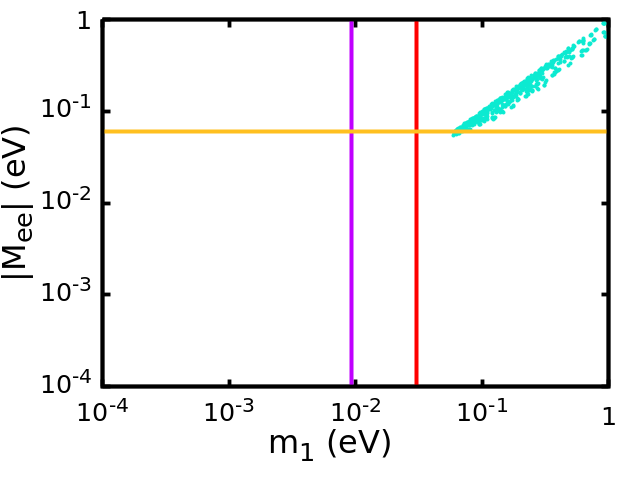}
\caption{}
\label{fig:mumuzeromee_no}
\end{subfigure}
\begin{subfigure}{0.32\textwidth}
\includegraphics[width=5.2cm,height=4.4cm]{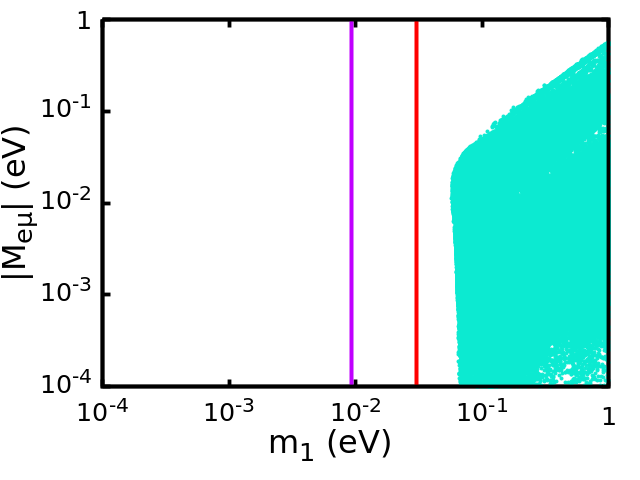}
\caption{}
\label{fig:m1mumuzeromemu_no}
\end{subfigure}
\begin{subfigure}{0.32\textwidth}
\includegraphics[width=5.2cm,height=4.4cm]{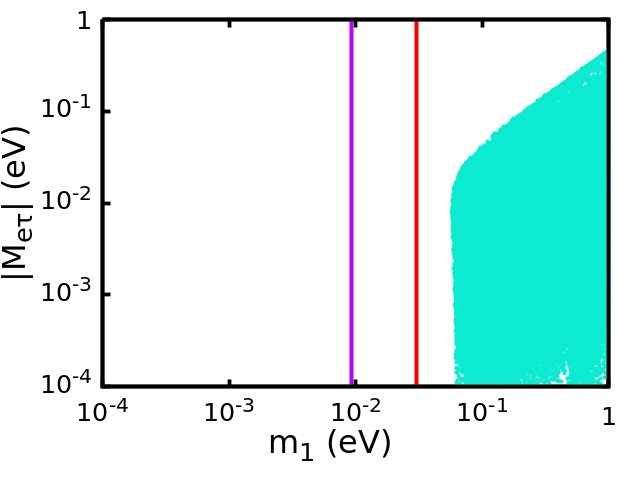}
\caption{}
\label{fig:mumuzerometau_no}
\end{subfigure}
\captionsetup{justification=raggedright,singlelinecheck=false}
 \caption{ Correlation between $|M_{ee}|$ and $m_{1}$, $|M_{e\mu}|$ and $m_{1}$, and between $|M_{e\tau}|$ and $m_{1}$ with vanishing 
$|M_{\mu\mu}|$ in case of normal ordering. The red and magenta lines represent $\sum m_i = 0.12\,{\rm eV}$~\cite{Zhang:2020mox} 
and $\sum m_i = 0.072\,{\rm eV}$~\cite{DESI:2024mwx}, respectively. In Fig.~\ref{fig:mumuzeromee_no} yellow line represents $|M_{ee}|= 0.06\,{\rm eV}$
obtained from KamLAND-Zen experiment~\cite{KamLAND-Zen:2016pfg}.}
\label{fig:mumuzero_no}
\end{figure}

We show the variation of $|M_{ee}|$, $|M_{e\mu}|$, and $|M_{e\tau}|$ with $m_{1}$ for vanishing $|M_{\mu\mu}|$ in Fig.~\ref{fig:mumuzeromee_no}, 
Fig.~\ref{fig:m1mumuzeromemu_no}, and Fig.~\ref{fig:mumuzerometau_no}, respectively. The bound on the total neutrino mass $\sum m_i < 0.12\,{\rm eV}$ and $\sum m_i < 0.072\,{\rm eV}$  are
represented with red and magenta lines, 
respectively. The upper bound of $M_{ee} = 0.06\,{\rm eV}$ is represented by a yellow line.

\begin{figure}[htbp]
 \begin{subfigure}{0.32\textwidth}
\includegraphics[width=5cm,height=4cm]{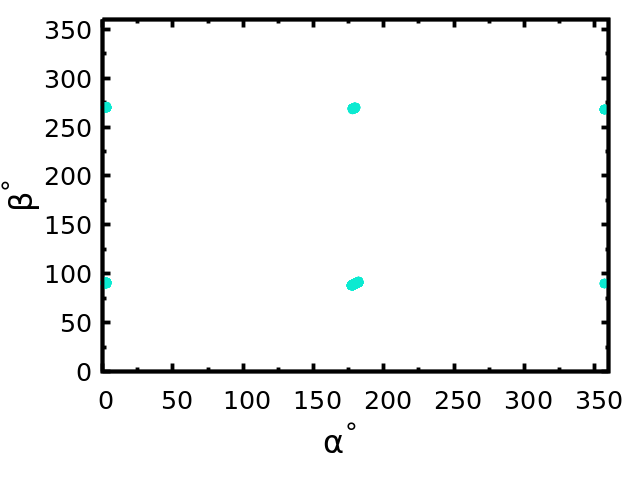}
\caption{}
\label{fig:mumuemuzeroaplbeta_no}
\end{subfigure}
\begin{subfigure}{0.32\textwidth}
\includegraphics[width=5cm,height=4cm]{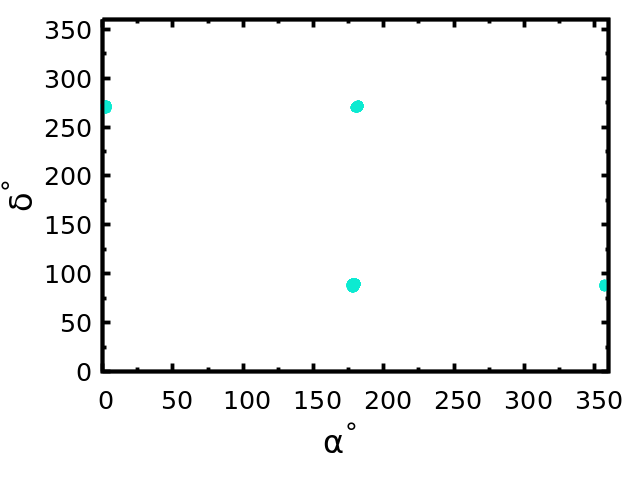}
\caption{}
\label{fig:mumuemuzeroapldelta_no}
\end{subfigure}
\begin{subfigure}{0.32\textwidth}
\includegraphics[width=5cm,height=4cm]{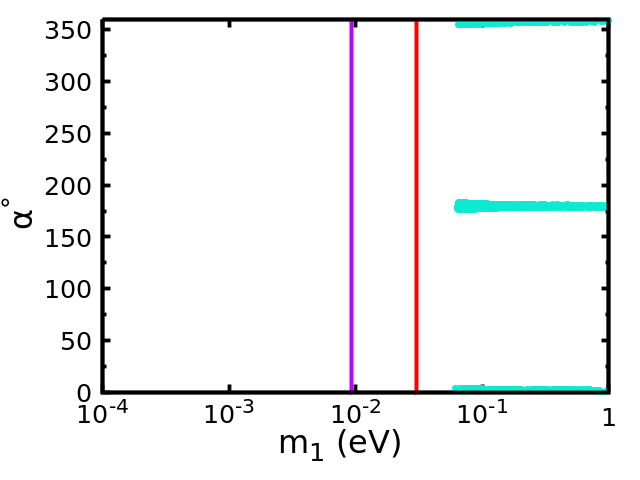}
\caption{}
\label{fig:mumuemuzeroaplham1_no}
\end{subfigure}
\captionsetup{justification=raggedright,singlelinecheck=false}
\caption{Correlation between $\alpha$ and $\beta$, $\alpha$ and $\delta$, and between $\alpha$ and $m_{1}$ with vanishing $|M_{e\mu}|$ and 
$|M_{\mu\mu}|$ in case of normal ordering.}
\label{fig:mumuemuzero_no}
\end{figure}

\begin{figure}[h!]
 \begin{subfigure}{0.32\textwidth}
\includegraphics[width=5cm,height=4cm]{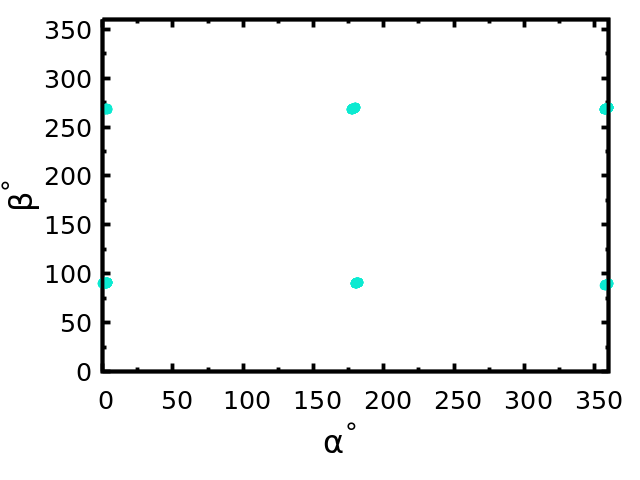}
\caption{}
\label{fig:mumuetauzeroaplbeta_no}
\end{subfigure}
\begin{subfigure}{0.32\textwidth}
\includegraphics[width=5cm,height=4cm]{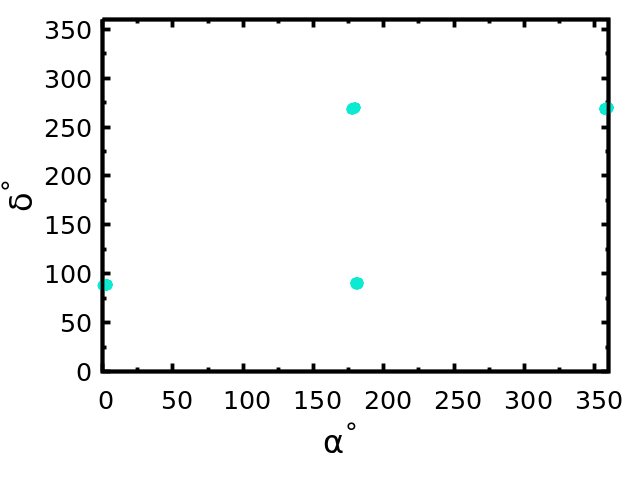}
\caption{}
\label{fig:mumuetauzeroapldelta_no}
\end{subfigure}
\begin{subfigure}{0.32\textwidth}
\includegraphics[width=5cm,height=4cm]{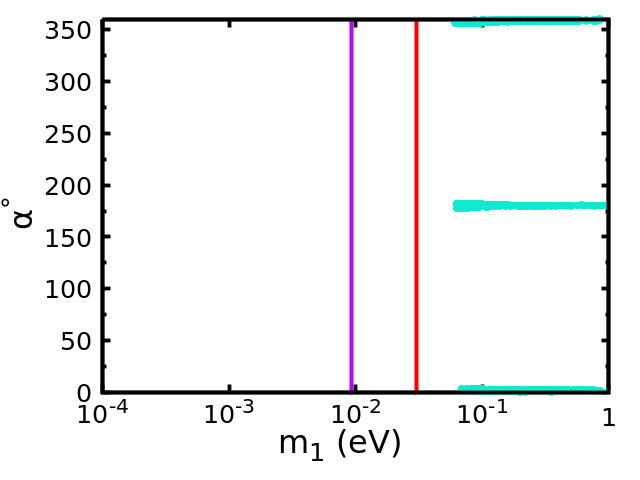}
\caption{}
\label{fig:mumuetauzeroaplham1_no}
\end{subfigure}
\captionsetup{justification=raggedright,singlelinecheck=false}
\caption{Correlation between $\alpha$ and $\beta$, $\alpha$ and $\delta$, and between $\alpha$ and $m_{1}$ with vanishing $|M_{e\tau}|$ and 
$|M_{\mu\mu}|$ in case of normal ordering.}
\label{fig:mumuetauzero_no}
\end{figure}

From Fig.~\ref{fig:mumuzero_no}, it is observed that $|M_{e\mu}|$ and $|M_{e\tau}|$ can vanish simultaneously with $|M_{\mu\mu}|$, however, $|M_{ee}|$ can not. We show the correlation between $\alpha$ and $\beta$, $\alpha$ and $\delta$, and between $m_{1}$ and $\alpha$ with vanishing $|M_{e\mu}|$ and 
$|M_{\mu\mu}|$ in Figs.~\ref{fig:mumuemuzeroaplbeta_no}, \ref{fig:mumuemuzeroapldelta_no}, \ref{fig:mumuemuzeroaplham1_no}, respectively. We can have vanishing $|M_{e\mu}|$ and $|M_{\mu\mu}|$ for $m_{1}>6.0\times 10^{-2}\,{\rm eV}$. Value of
$\alpha$ is restricted to a very narrow region around $0^{\circ}$, $180^{\circ}$ and $360^{\circ}$. Similarly, value of $\beta$ and $\delta$ 
has to lie around $90^{\circ}$ or $270^{\circ}$ to have vanishing $|M_{e\mu}|$ and $|M_{\mu\mu}|$.
Similar correlation plots between $\alpha$ and $\beta$, $\alpha$ and $\delta$, and between $m_{1}$ and $\alpha$ with vanishing $|M_{e\tau}|$ 
and $|M_{\mu\mu}|$ are shown in Figs.~\ref{fig:mumuetauzeroaplbeta_no}, \ref{fig:mumuetauzeroapldelta_no}, \ref{fig:mumuetauzeroaplham1_no}, 
respectively. The correlation between $\alpha$ and $\beta$ and between $\alpha$ and $m_1$ of Fig.~\ref{fig:mumuemuzeroaplbeta_no} and 
Fig.~\ref{fig:mumuemuzeroaplham1_no} looks quite similar to that of Fig.~\ref{fig:mumuetauzeroaplbeta_no} and 
Fig.~\ref{fig:mumuetauzeroaplham1_no}, respectively. However, the correlation between $\alpha$ and $\delta$ of 
Fig.~\ref{fig:mumuemuzeroapldelta_no} is quite different from that of Fig.~\ref{fig:mumuetauzeroapldelta_no}. It is worth mentioning that $|M_{e\mu}|$ and $|M_{\mu\mu}|$ can vanish simultaneously only if $\theta_{23}$ lies in the lower octant, i.e, the value of $\theta_{23}$ has to be $\theta_{23} < 45^{\circ}$ to have vanishing $|M_{e\mu}|$ and $|M_{\mu\mu}|$. We
have shown it in Fig.~\ref{fig:Nmalpth23mumuemuzerofinal_no} of Appendix~\ref{app2}. Similarly, $|M_{e\tau}|$ and $|M_{\mu\mu}|$ can vanish simultaneously only if $\theta_{23}$ lies in the lower octant, i.e, the value of $\theta_{23}$ has to be $\theta_{23} < 45^{\circ}$ to have vanishing $|M_{e\tau}|$ and $|M_{\mu\mu}|$. We
have shown it in Fig.~\ref{fig:Nmalpth23mumetauzerofinal_no} of Appendix~\ref{app2}.

\subsubsection{Vanishing $|M_{e\tau}|$:}
\begin{figure}[htbp!]
   \begin{subfigure}{0.40\textwidth}
\includegraphics[width=5.5cm,height=4.2cm]{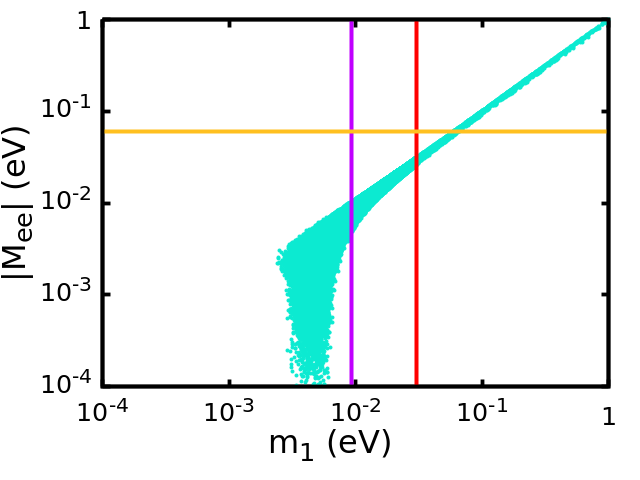}
\caption{}
\label{fig:3.2}
\end{subfigure}
\begin{subfigure}{0.40\textwidth}
\includegraphics[width=5.5cm,height=4.2cm]{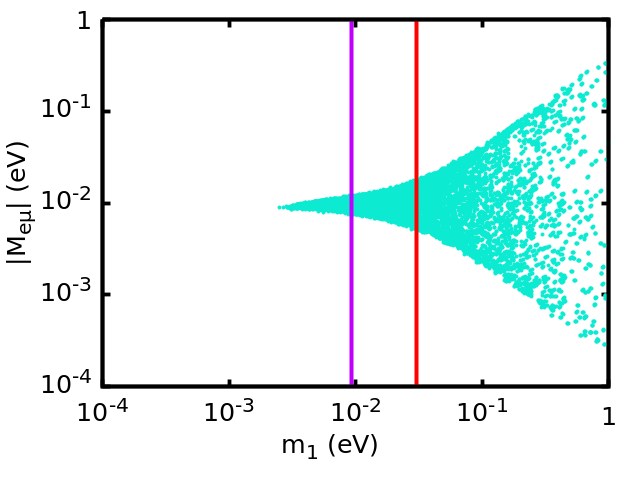}
\caption{}
\label{fig:23.23b}
\end{subfigure}
 \captionsetup{justification=raggedright,singlelinecheck=false}
 \caption{ Correlation between $|M_{ee}|$ and $m_{1}$, and between $|M_{e\mu}|$ and $m_1$ with vanishing $|M_{e\tau}|$ in case of normal 
ordering. The red and magenta lines 
represent $\sum m_i = 0.12\,{\rm eV}$~\cite{Zhang:2020mox} and $\sum m_i = 0.072\,{\rm eV}$~\cite{DESI:2024mwx}, respectively.  The yellow 
line in Fig.~\ref{fig:3.2} represents $|M_{ee}|= 0.06\,{\rm eV}$~\cite{KamLAND-Zen:2016pfg}.}
\label{fig23:etauzero_no}
\end{figure}

Variation of $|M_{ee}|$ and $|M_{e\mu}|$ with $m_{1}$ for vanishing $|M_{e\tau}|$ is shown in Fig.~\ref{fig:3.2} and Fig.~\ref{fig:23.23b}, 
respectively. We show the bound on the total neutrino mass $\sum m_i < 0.12\,{\rm eV}$ and $\sum m_i < 0.072\,{\rm eV}$ by red and magenta 
lines, respectively. In Fig.~\ref{fig:3.2}, the yellow line represents $M_{ee} = 0.06\,{\rm eV}$.

From Fig.~\ref{fig23:etauzero_no}, it is observed that $|M_{ee}|$ and $|M_{e\tau}|$ can vanish simultaneously, however, $|M_{e\mu}|$ and 
$|M_{e\tau}|$ can not. For vanishing $|M_{ee}|$ and $|M_{e\tau}|$, the lightest neutrino mass $m_1$ must lie within 
$(4.0,\, 6.0)\times 10^{-3}\,{\rm eV}$. It should be noted that we can have $(M_{ee} = 0,\,M_{e\tau} = 0)$ with 
$\sum m_i < 0.12\,{\rm eV}$ as well as with $\sum m_i < 0.072\,{\rm eV}$. For $M_{ee} < 0.06$ eV, the value of $\sum m_i$ lies within the 
range $(0.06, 0.21)\,{\rm eV}$.

\begin{figure}[htbp!]
 \begin{subfigure}{0.32\textwidth}
\includegraphics[width=5cm,height=4cm]{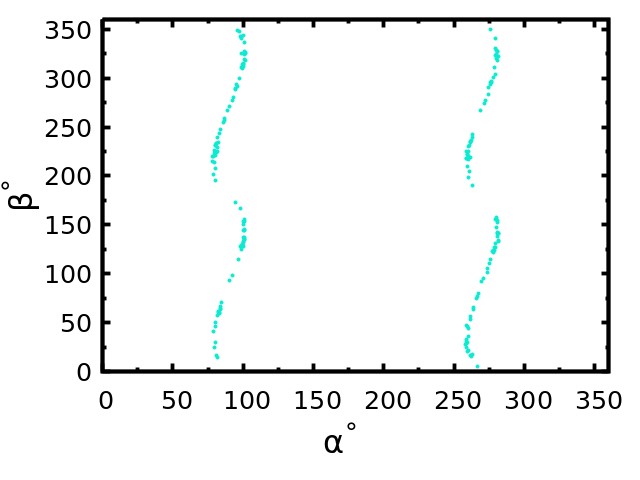}
\caption{}
\label{fig:6.3a}
\end{subfigure}
\begin{subfigure}{0.32\textwidth}
\includegraphics[width=5cm,height=4cm]{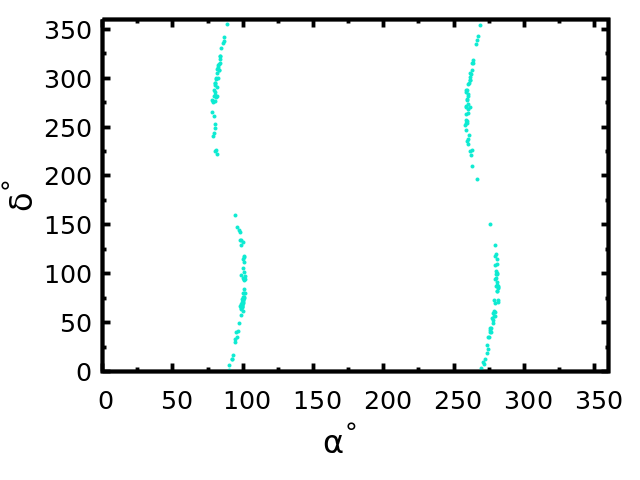}
\caption{}
\label{fig:6.3b}
\end{subfigure}
\begin{subfigure}{0.32\textwidth}
\includegraphics[width=5cm,height=4cm]{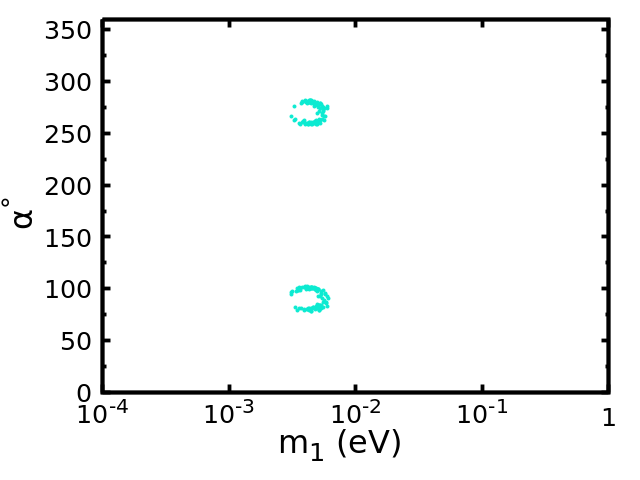}
\caption{}
\label{fig:6.3c}
\end{subfigure}
\captionsetup{justification=raggedright,singlelinecheck=false}
\caption{Correlation between $\alpha$ and $\beta$, $\alpha$ and $\delta$, and between $\alpha$ and $m_{1}$ with vanishing $|M_{ee}|$ and 
$|M_{e\tau}|$ in case of normal ordering.}
\label{fig:6.3}
\end{figure}

We show the correlation of several neutrino observables in Fig.~\ref{fig:6.3} with simultaneous vanishing of $|M_{ee}|$ and $|M_{e\tau}|$.
The value of the lightest neutrino mass $m_1$ should lie within $(3.0,\,6.0)\times 10^{-3}\,{\rm eV}$ and the Majorana CP violating phase 
$\alpha$ should lie within $(80^{\circ},\,100^{\circ})$ and $(260^{\circ},\,280^{\circ})$ to have vanishing $|M_{ee}|$ and $|M_{e\tau}|$.
It is observed that two texture zero condition puts severe constraint on $m_1$ and $\alpha$. However, there is no such constraint on $\delta$ 
and $\beta$. $|M_{ee}|$ and $|M_{e\tau}|$ can vanish for any value of $\delta$ and $\beta$ within $(0,\,360^{\circ})$. We also show the correlation between $\theta_{23}$ and $\alpha$ with vanishing $|M_{ee}|$ and $|M_{e\tau}|$ in  Fig.~\ref{fig:Nmalpth23meeetauzerofinal_no} of Appendix~\ref{app2}.

\subsubsection{Vanishing $|M_{e\mu}|$:} 
\begin{figure}[h!]
\includegraphics[width=5.5cm,height=4.2cm]{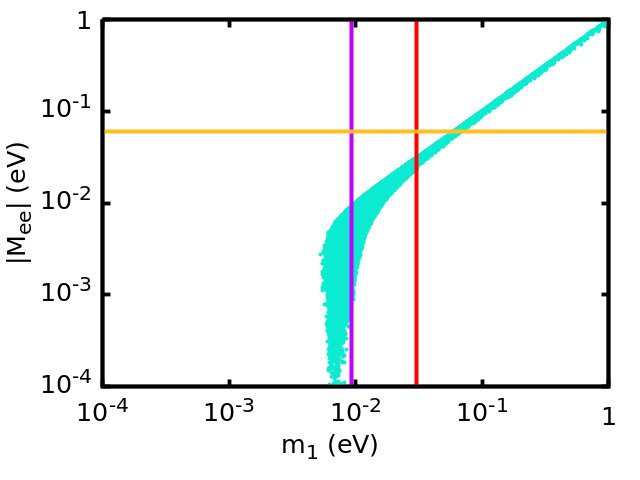}
\captionsetup{justification=raggedright,singlelinecheck=false}
\caption{Correlation between $|M_{ee}|$ and $m_{1}$ with vanishing $|M_{e\mu}|$ in case of normal ordering. The red and magenta lines 
represent $\sum m_i = 0.12\,{\rm eV}$~\cite{Zhang:2020mox} and $\sum m_i = 0.072\,{\rm eV}$~\cite{DESI:2024mwx}, respectively. The yellow 
line represents $|M_{ee}|= 0.06\,{\rm eV}$~\cite{KamLAND-Zen:2016pfg}.}
\label{fig23:emuzero_no}
\end{figure}

In Fig.~\ref{fig23:emuzero_no}, we plot $|M_{ee}|$ against the lightest neutrino mass $m_1$ for vanishing $|M_{e\mu}|$. The red and magenta 
lines correspond to $\sum m_i = 0.12\,{\rm eV}$ and $\sum m_i = 0.072\,{\rm eV}$, respectively.  In Fig.~\ref{fig23:emuzero_no}, we show
the upper bound of $M_{ee} = 0.06\,{\rm eV}$ by a yellow line. For $M_{ee}< 0.06$ eV, the value of $\sum m_i$ lies in the range 
$(0.07, 0.20)\,{\rm eV}$.  

\begin{figure}[htbp!]
 \begin{subfigure}{0.32\textwidth}
\includegraphics[width=5cm,height=4cm]{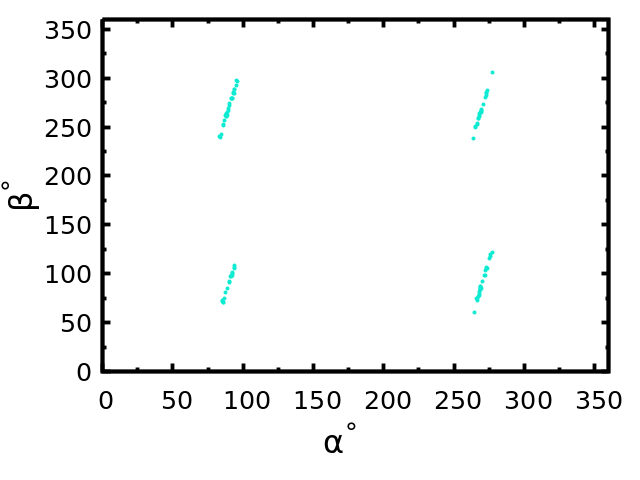}
\caption{}
\label{fig:3.3a}
\end{subfigure}
\begin{subfigure}{0.32\textwidth}
\includegraphics[width=5cm,height=4cm]{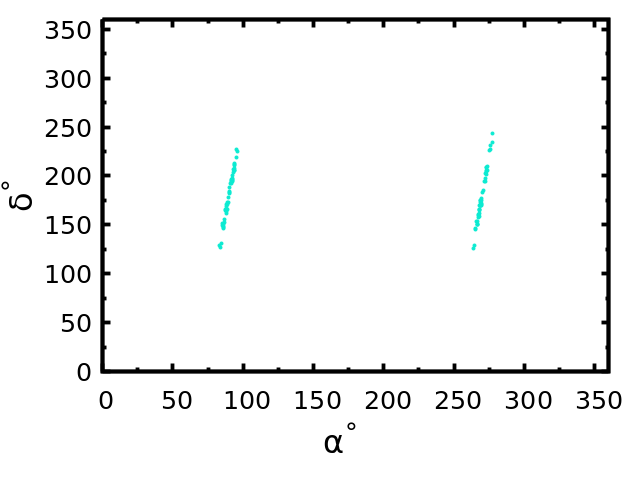}
\caption{}
\label{fig:3.3b}
\end{subfigure}
\begin{subfigure}{0.32\textwidth}
\includegraphics[width=5cm,height=4cm]{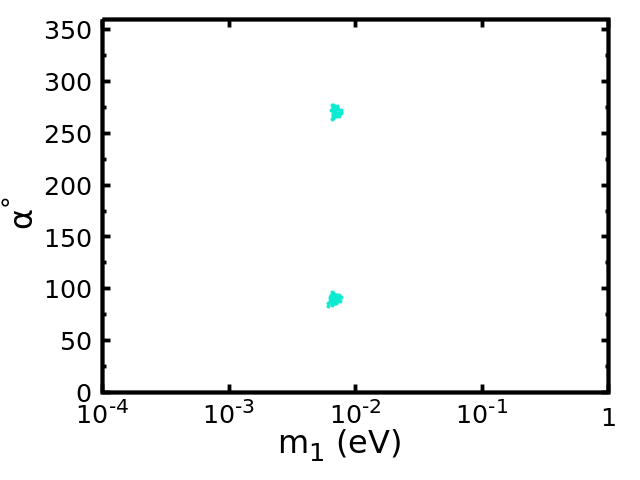}
\caption{}
\label{fig:3.3c}
\end{subfigure}
\captionsetup{justification=raggedright,singlelinecheck=false}
\caption{Correlation between $\alpha$ and $\beta$, $\alpha$ and $\delta$, and between $\alpha$ and $m_{1}$ with vanishing $|M_{ee}|$ and 
$|M_{e\mu}|$ in case of normal ordering.}
\label{fig:3.3}
\end{figure}

From Fig.~\ref{fig23:emuzero_no}, it is evident that $|M_{ee}|$ and $|M_{e\mu}|$ can vanish simultaneously.
It should be noted that we can have $(M_{ee} = 0,\,M_{e\mu} = 0)$ with
$\sum m_i < 0.12\,{\rm eV}$ as well as with $\sum m_i < 0.072\,{\rm eV}$.
We also show the correlation between $\alpha$ and $\beta$, $\alpha$ and $\delta$, and 
$\alpha$ and 
$m_{1}$ with vanishing $|M_{ee}|$ and $|M_{e\mu}|$ in Fig.~\ref{fig:3.3a}, Fig.~\ref{fig:3.3b} and Fig.~\ref{fig:3.3c}, respectively. 
We get vanishing $|M_{ee}|$ and $|M_{e\mu}|$ if the lightest neutrino mass $m_1 \in (6.50,\,8.20)\times 10^{-3}\,{\rm eV}$ and the Majorana CP 
violating phase $\alpha \in (85^{\circ},\, 95^{\circ})$ or $(265^{\circ},\,275^{\circ})$. Moreover, value of $\beta$ should lie within
$(60^{\circ},\,120^{\circ})$ or $(240^{\circ},\,300^{\circ})$ and the Dirac CP violating phase $\delta$ should lie within 
$(125^{\circ},\,245^{\circ})$ to have vanishing $|M_{ee}|$ and $|M_{e\mu}|$. It is evident that two texture zero condition puts a severe 
constraint on the lightest neutrino mass $m_1$ and the CP violating phases $\alpha$, $\beta$ and $\delta$. We show the correlation between $\theta_{23}$ and $\alpha$ with vanishing $|M_{ee}|$ and $|M_{e\mu}|$ in  Fig.~\ref{fig:Nmalpth23meeemufinal_no} of Appendix~\ref{app2}.

\subsection{Inverted Ordering}
\subsubsection{Vanishing $|M_{\tau\tau}|$:} 
\begin{figure}[htbp!]
   \begin{subfigure}{0.32\textwidth}
\includegraphics[width=5.2cm,height=4.4cm]{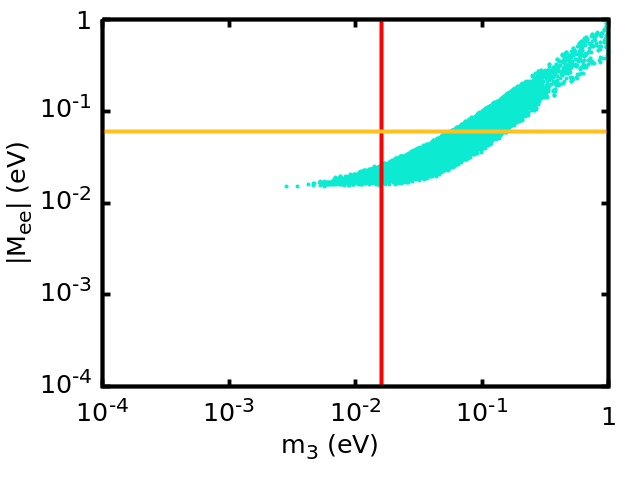}
\caption{}
\label{fig:24.24a}
\end{subfigure}
\begin{subfigure}{0.32\textwidth}
\includegraphics[width=5.2cm,height=4.4cm]{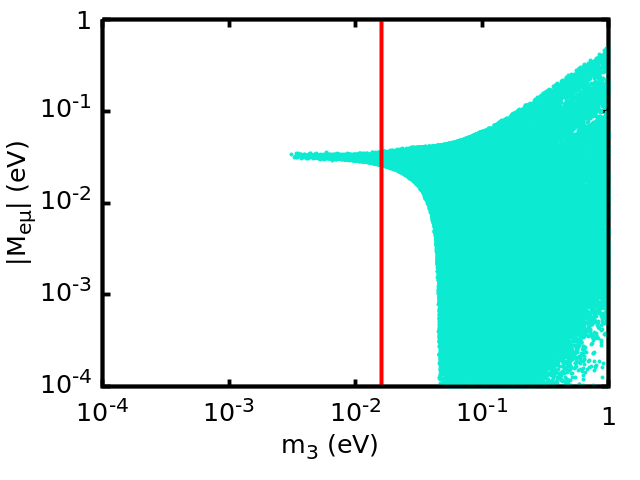}
\caption{}
\label{fig:24.24b}
\end{subfigure}
\begin{subfigure}{0.32\textwidth}
\includegraphics[width=5.2cm,height=4.4cm]{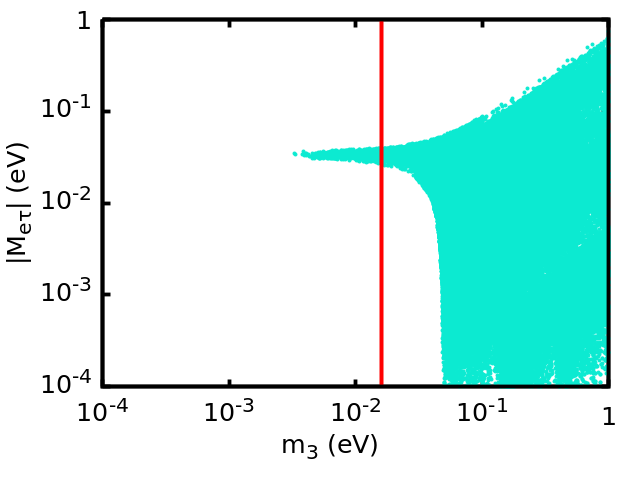}
\caption{}
\label{fig:24.24c}
\end{subfigure}
\begin{subfigure}{0.32\textwidth}
\includegraphics[width=5.2cm,height=4.4cm]{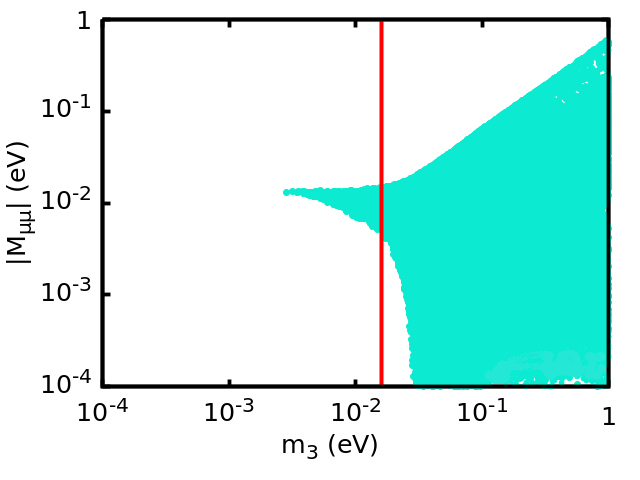}
\caption{}
\label{fig:24.24d}
\end{subfigure}
\begin{subfigure}{0.32\textwidth}
\includegraphics[width=5.2cm,height=4.4cm]{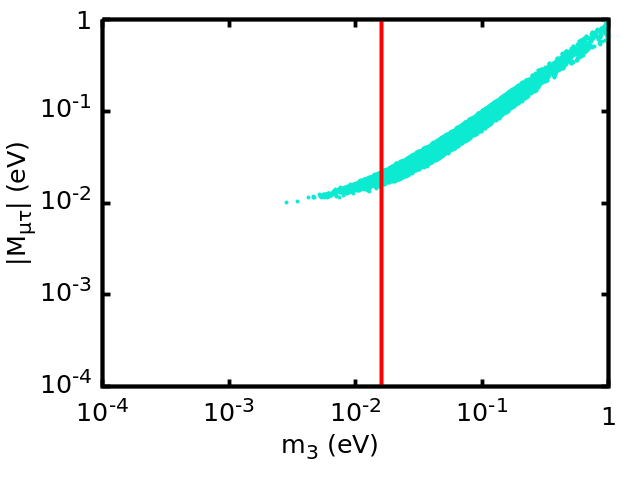}
\caption{}
\label{fig:24.24e}
\end{subfigure}
 \captionsetup{justification=raggedright,singlelinecheck=false}
 \caption{ Correlation of $|M_{ee}|$, $|M_{e\mu}|$, $|M_{e\tau}|$, $|M_{\mu\mu}|$ and $|M_{\mu\tau}|$ with lightest neutrino mass $m_{3}$ 
for vanishing $|M_{\tau\tau}|$ in case of inverted ordering. The red line represents $\sum m_i = 0.12\,{\rm eV}$~\cite{Zhang:2020mox}. The 
yellow line in Fig.~\ref{fig:24.24a} represents $|M_{ee}|= 0.06\,{\rm eV}$~\cite{KamLAND-Zen:2016pfg}.}
\label{fig24:tautauzero_io}
\end{figure}

We show the variation of $|M_{ee}|$, $|M_{e\mu}|$, $|M_{e\tau}|$, $|M_{\mu\mu}|$ and $|M_{\mu\tau}|$ with the lightest neutrino mass $m_{3}$ 
for vanishing $|M_{\tau\tau}|$ in Fig.~\ref{fig:24.24a}, Fig.~\ref{fig:24.24b}, Fig.~\ref{fig:24.24c}, Fig.~\ref{fig:24.24d}, and 
Fig.~\ref{fig:24.24e}, respectively. The bound on the total neutrino mass $\sum m_i < 0.12\,{\rm eV}$ is represented by the red line. 
Similarly, the upper bound of $M_{ee} = 0.06\,{\rm eV}$ is represented by the yellow line in Fig.~\ref{fig:24.24a}.

From Fig.~\ref{fig24:tautauzero_io}, it is clear that $|M_{e\mu}|$, $|M_{e\tau}|$ and $|M_{\mu\mu}|$ can vanish simultaneously with $|M_{\tau\tau}|$, however, $|M_{ee}|$ and $|M_{\mu\tau}|$ can not. 
When $|M_{\tau\tau}|$ approaches zero, the minimum value of the effective Majorana mass term is found to be 
$|M_{ee}| = 0.02\,{\rm eV}$, which is consistent with the current upper bound from EXO-200, CUORE, and KamLAND-Zen. Furthermore, this value 
of $|M_{ee}|$ falls within the cosmological upper bound on the total neutrino mass. For $M_{ee}< 0.06\,{\rm eV}$, the value of 
total neutrino mass $\sum m_i$ lies within the range $(0.10, 0.42)$ eV.

\begin{figure}[htbp]
 \begin{subfigure}{0.32\textwidth}
\includegraphics[width=5.2cm,height=4.4cm]{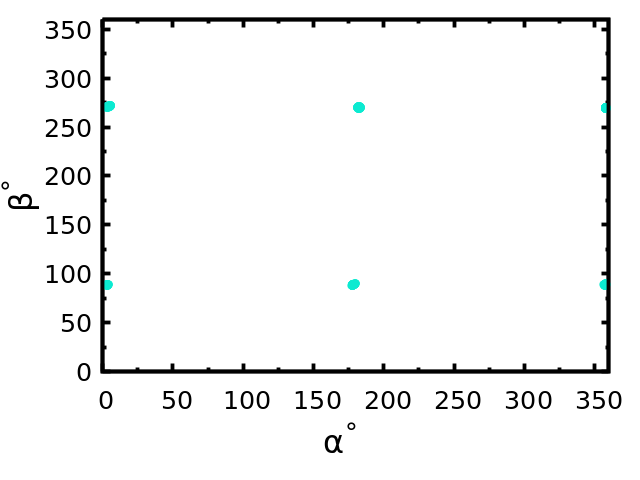}
\caption{}
\label{fig:alpbetaemutautauzero_io}
\end{subfigure}
\begin{subfigure}{0.32\textwidth}
\includegraphics[width=5.2cm,height=4.4cm]{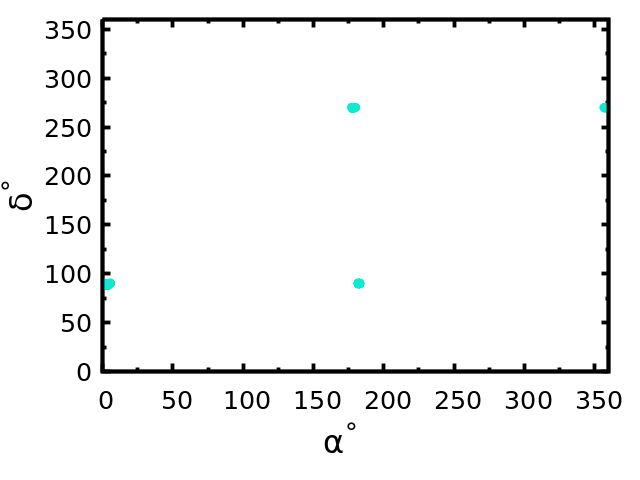}
\caption{}
\label{fig:alpdeltaemutautauzero_io}
\end{subfigure}
\begin{subfigure}{0.32\textwidth}
\includegraphics[width=5.2cm,height=4.4cm]{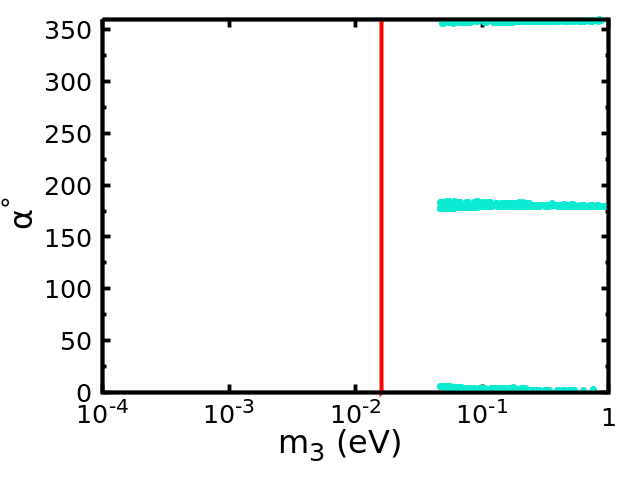}
\caption{}
\label{fig:emutautauzerom1alp_io}
\end{subfigure}
\captionsetup{justification=raggedright,singlelinecheck=false}
\caption{Correlation between $\alpha$ and $\beta$, $\alpha$ and $\delta$, and between $\alpha$ and $m_{3}$ with vanishing $|M_{e\mu}|$ and 
$|M_{\tau\tau}|$ in case of inverted ordering.}
\label{fig:emutautauzerocor_io}
\end{figure}

\begin{figure}[htbp]
 \begin{subfigure}{0.32\textwidth}
\includegraphics[width=5.2cm,height=4.4cm]{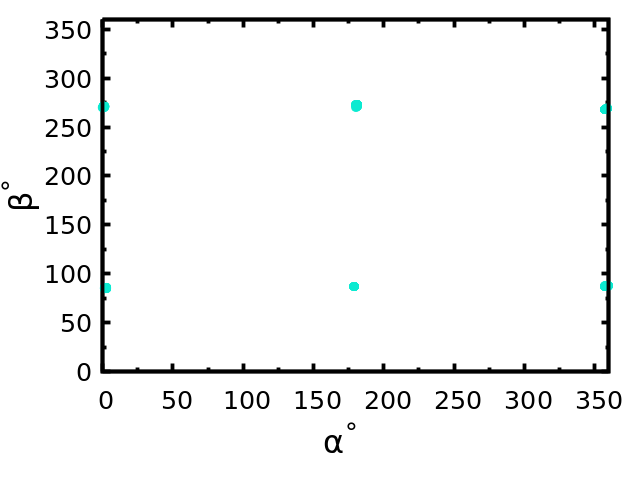}
\caption{}
\label{fig:alpbetaetautautauzero_io}
\end{subfigure}
\begin{subfigure}{0.32\textwidth}
\includegraphics[width=5.2cm,height=4.4cm]{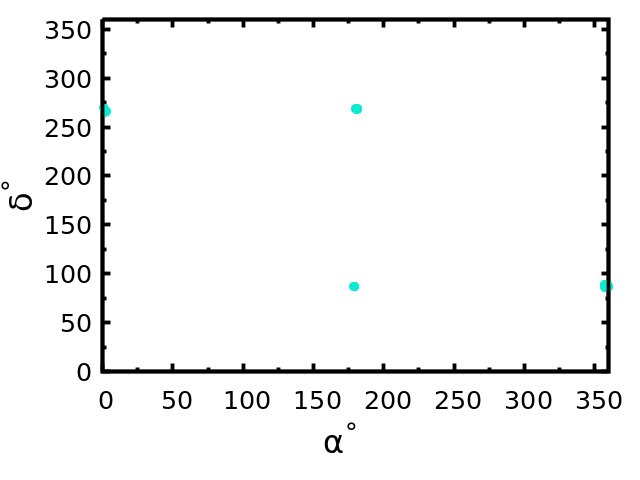}
\caption{}
\label{fig:alpdeltaetautautauzero_io}
\end{subfigure}
\begin{subfigure}{0.32\textwidth}
\includegraphics[width=5.2cm,height=4.4cm]{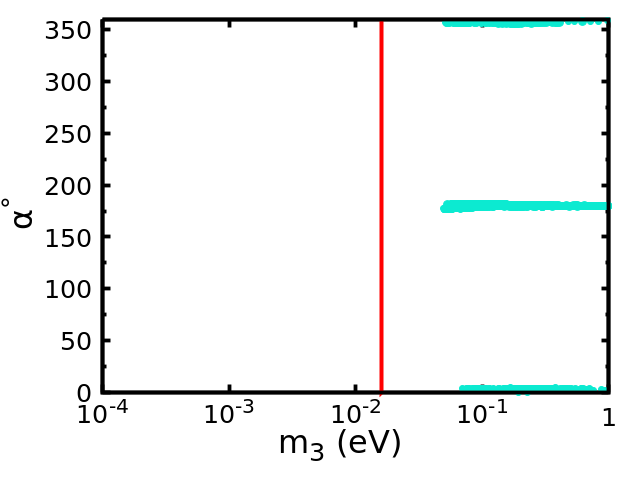}
\caption{}
\label{fig:etautautauzerom3alpha_io}
\end{subfigure}
\captionsetup{justification=raggedright,singlelinecheck=false}
\caption{Correlation between $\alpha$ and $\beta$, $\alpha$ and $\delta$, and between $\alpha$ and $m_{3}$ with vanishing $|M_{e\tau}|$ and 
$|M_{\tau\tau}|$ in case of inverted ordering.}
\label{fig:etautautauzerocor_io}
\end{figure}

\begin{figure}[h!]
 \begin{subfigure}{0.32\textwidth}
\includegraphics[width=5.2cm,height=4.4cm]{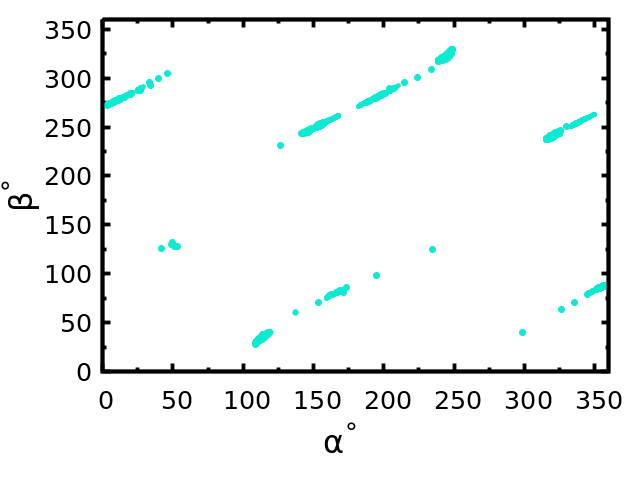}
\caption{}
\label{fig:mumutautauzeroaplbeta_io}
\end{subfigure}
\begin{subfigure}{0.32\textwidth}
\includegraphics[width=5.2cm,height=4.4cm]{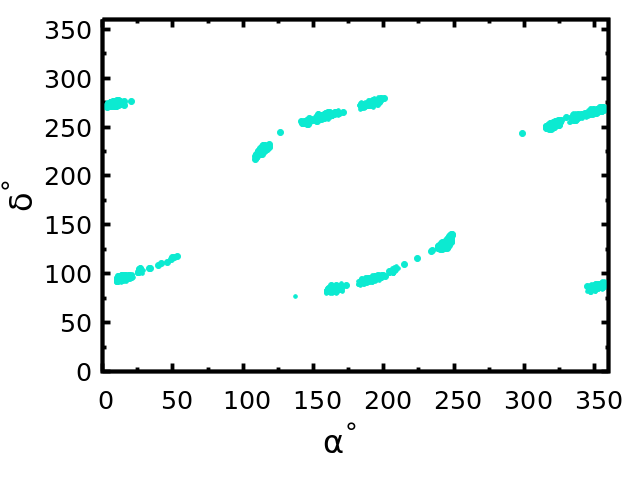}
\caption{}
\label{fig:mumutautauzeroapldelta_io}
\end{subfigure}
\begin{subfigure}{0.32\textwidth}
\includegraphics[width=5.2cm,height=4.4cm]{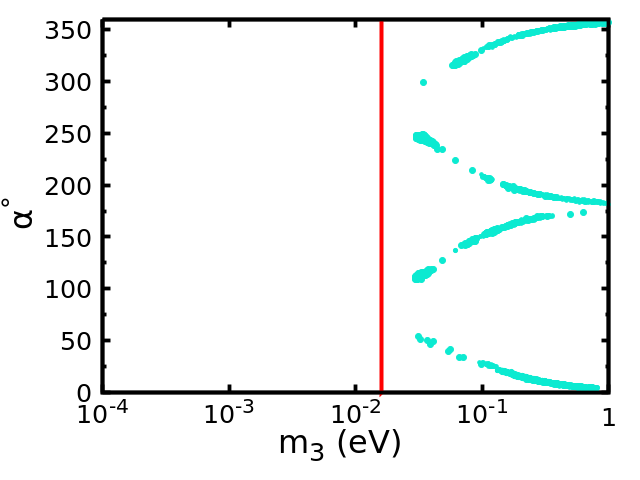}
\caption{}
\label{fig:mumutautauzeroaplham1_io}
\end{subfigure}
\captionsetup{justification=raggedright,singlelinecheck=false}
\caption{Correlation between $\alpha$ and $\beta$, $\alpha$ and $\delta$, and between $\alpha$ and $m_{3}$ with vanishing $|M_{\mu\mu}|$ and 
$|M_{\tau\tau}|$ in case of inverted ordering.}
\label{fig:mumutautauzerocor_io}
\end{figure}

We show the correlation between $\alpha$ and $\beta$, $\alpha$ and $\delta$, and between $m_{3}$ and $\alpha$ with vanishing $|M_{e\mu}|$ 
and $|M_{\tau\tau}|$ in Figs.~\ref{fig:alpbetaemutautauzero_io}, \ref{fig:alpdeltaemutautauzero_io}, \ref{fig:emutautauzerom1alp_io}, 
respectively. From Fig.~\ref{fig:emutautauzerocor_io}, it is evident that simultaneous vanishing of $|M_{e\mu}|$ and $ |M_{\tau\tau}|$ 
puts severe constraint on the CP violating phases $\alpha$, $\beta$ and $\delta$. Value of $\alpha$ is restricted to a very narrow region 
around $0^{\circ}$, $180^{\circ}$ and $360^{\circ}$. Similarly, for vanishing $|M_{e\mu}|$ and $|M_{\tau\tau}|$, value of $\beta$ and $\delta$
should lie around $90^{\circ}$ or $270^{\circ}$, respectively. We can have simultaneous vanishing of $|M_{e\mu}|$ and $|M_{\tau\tau}|$ if the lightest neutrino mass $m_3 > 4.5 \times 10^{-2}\, {\rm eV}$. We also show the correlation between $\alpha$ and $\beta$, $\alpha$ and $\delta$, and between $m_{3}$ and $\alpha$ with vanishing $|M_{e\tau}|$ 
and $|M_{\tau\tau}|$ in Figs.~\ref{fig:alpbetaetautautauzero_io}, \ref{fig:mumutautauzeroapldelta_io}, \ref{fig:etautautauzerom3alpha_io}, 
respectively. It is observed that the correlation between $\alpha$ and $\delta$ of
Fig.~\ref{fig:alpdeltaemutautauzero_io} is quite different from that of Fig.~\ref{fig:alpdeltaetautautauzero_io}. It is worth mentioning that $|M_{e\mu}|$ and $|M_{\tau\tau}|$ can vanish simultaneously only if $\theta_{23}$ lies in the lower octant, i.e, the value of $\theta_{23}$ has to be $\theta_{23} < 45^{\circ}$ to have vanishing $|M_{e\mu}|$ and $|M_{\tau\tau}|$. We
have shown it in Fig.~\ref{fig:Imalpth23emutautauzerof_io} of Appendix~\ref{app2}. Similarly, $|M_{e\tau}|$ and $|M_{\tau\tau}|$ can vanish simultaneously only if $\theta_{23}$ lies in the lower octant, i.e, the value of $\theta_{23}$ has to be $\theta_{23} < 45^{\circ}$ to have vanishing $|M_{e\tau}|$ and $|M_{\tau\tau}|$. We
have shown it in Fig.~\ref{fig:Imalpth23etautautauzerof_io} of Appendix~\ref{app2}.

We show the correlation between $\alpha$ and $\beta$, $\alpha$ and $\delta$, and between $m_{3}$ and $\alpha$ with vanishing $|M_{\mu\mu}|$ 
and $|M_{\tau\tau}|$ in Figs.~\ref{fig:mumutautauzeroaplbeta_io}, \ref{fig:mumutautauzeroapldelta_io}, \ref{fig:mumutautauzeroaplham1_io}, 
respectively. The $|M_{\mu\mu}|$ and $|M_{\tau\tau}|$ can vanish simultaneously for  $m_{3} > 2.95\times 10^{-2}$ eV. We show the correlation between $\theta_{23}$ and $\alpha$  with vanishing $|M_{\mu\mu}|$ 
and $|M_{\tau\tau}|$ in Fig.~\ref{fig:Imnalpth23mumutautauzero_io} of Appendix~\ref{app2}.  

\subsubsection{Vanishing $|M_{\mu\tau}|$:} 
\begin{figure}[htbp!]
   \begin{subfigure}{0.24\textwidth}
\includegraphics[width=\textwidth]{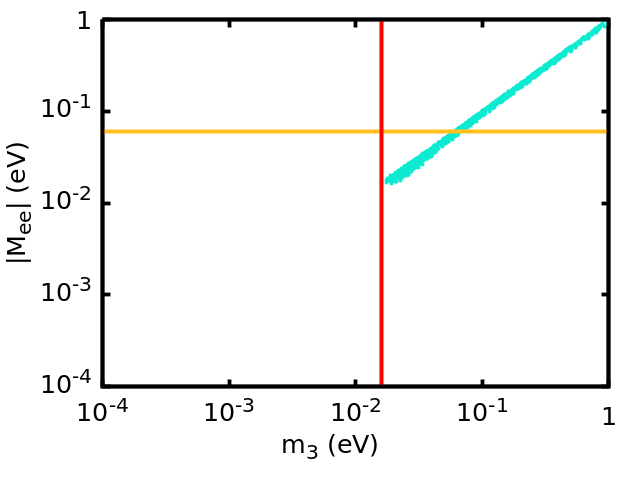}
\caption{}
\label{fig:25.25a}
\end{subfigure}
\begin{subfigure}{0.24\textwidth}
\includegraphics[width=\textwidth]{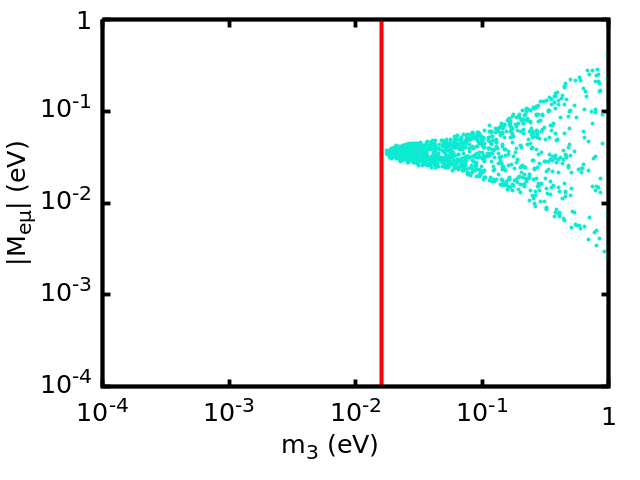}
\caption{}
\label{fig:25.25b}
\end{subfigure}
\begin{subfigure}{0.24\textwidth}
\includegraphics[width=\textwidth]{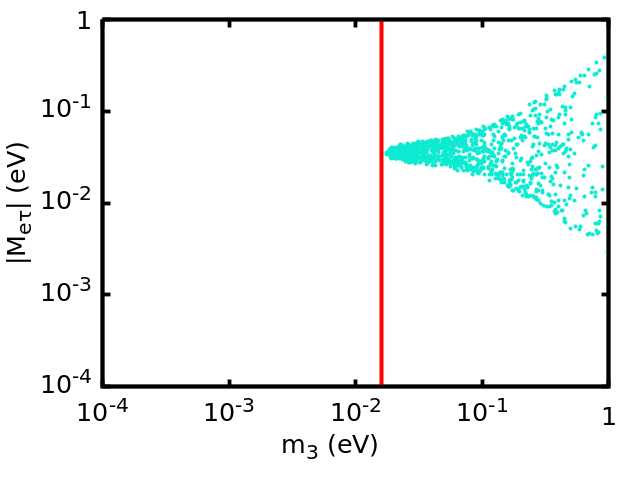}
\caption{}
\label{fig:25.25c}
\end{subfigure}
\begin{subfigure}{0.24\textwidth}
\includegraphics[width=\textwidth]{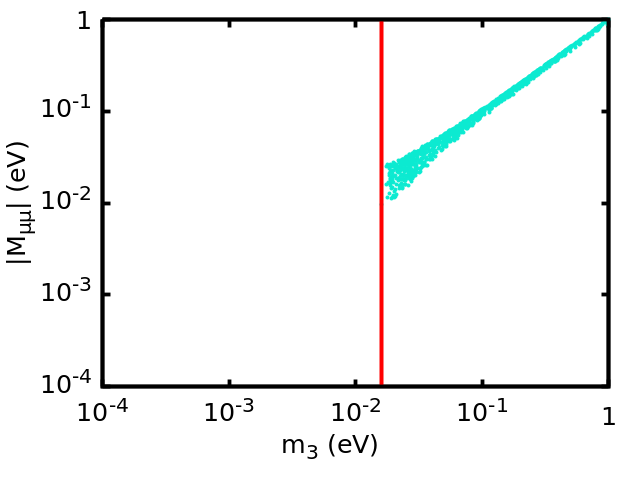}
\caption{}
\label{fig:25.25d}
\end{subfigure}
 \captionsetup{justification=raggedright,singlelinecheck=false}
 \caption{ Correlation of $|M_{ee}|$, $|M_{e\mu}|$, $|M_{e\tau}|$ and $|M_{\mu\mu}|$ with the lightest neutrino mass $m_{3}$ for vanishing 
$|M_{\mu\tau}|$ in case of inverted ordering. The red line represents $\sum m_i = 0.12\,{\rm eV}$~\cite{Zhang:2020mox}. The yellow line in 
Fig.~\ref{fig:25.25a} represents $|M_{ee}|= 0.06\,{\rm eV}$~\cite{KamLAND-Zen:2016pfg}.}
\label{fig25:mutauzero_io}
\end{figure}

Variation of $|M_{ee}|$, $|M_{e\mu}|$, $|M_{e\tau}|$, and  $|M_{\mu\mu}|$ with $m_{3}$ with vanishing $|M_{\mu\tau}|$ is shown in 
Fig.~\ref{fig:25.25a}, Fig.~\ref{fig:25.25b}, Fig.~\ref{fig:25.25c}, and Fig.~\ref{fig:25.25d}, respectively. The bound on the total neutrino 
mass $\sum m_i < 0.12\,{\rm eV}$ is represented by a red line, whereas, the bound on the effective Majorana mass $M_{ee} = 0.06\,{\rm eV}$ 
is represented by a yellow line.

From Fig.~\ref{fig25:mutauzero_io}, it is evident that we can not have vanishing $|M_{ee}|$, $|M_{e\mu}|$, $|M_{e\tau}|$, and $|M_{\mu\mu}|$ 
for vanishing $|M_{\mu\tau}|$. Hence, no two texture zero patterns are allowed for vanishing $|M_{\mu\tau}|$. It should be noted that, when 
$|M_{\mu\tau}|$ vanishes, the lowest value of $|M_{ee}|$ is found to be $0.02\,{\rm eV}$ and it lies within the allowed range 
of $|M_{ee}| < (0.06, 0.31)\,{\rm eV}$ as reported by EXO-200, CUORE, and KamLAND-Zen measurements. Additionally, the minimum value of 
$|M_{ee}|$ obtained for this scenario is consistent with the cosmological bound on the total neutrino mass. For $M_{ee}< 0.06$ eV, the 
value of $\sum m_i$ lies within the range $(0.11, 0.22)\,{\rm eV}$.

\subsubsection{Vanishing $|M_{\mu\mu}|$:} 
\begin{figure}[htbp]
 \begin{subfigure}{0.32\textwidth}
\includegraphics[width=5.2cm,height=4.4cm]{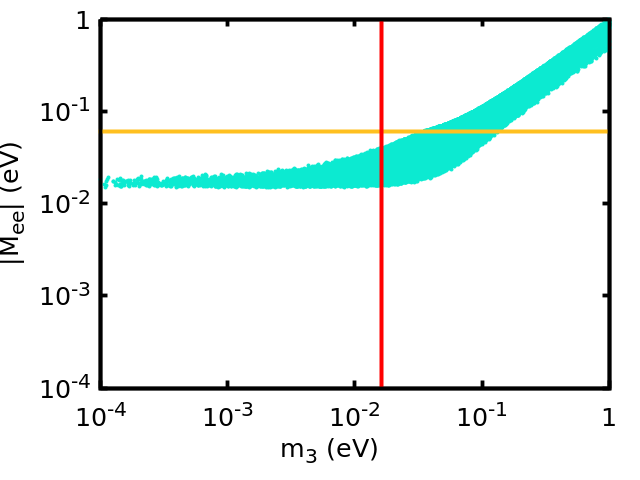}
\caption{}
\label{fig:26.26a}
\end{subfigure}
\begin{subfigure}{0.32\textwidth}
\includegraphics[width=5.2cm,height=4.4cm]{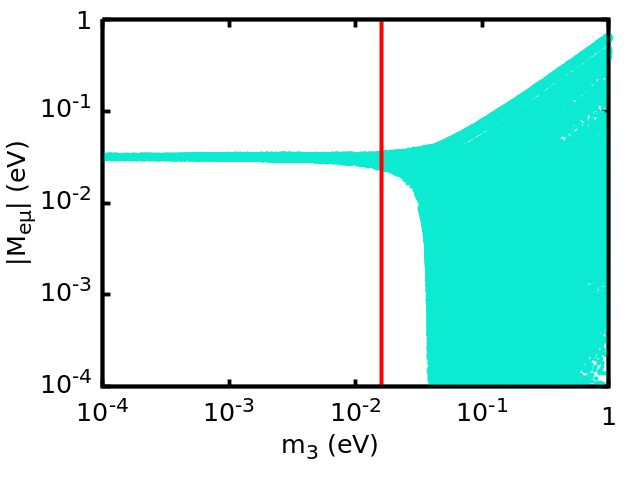}
\caption{}
\label{fig:26.26b}
\end{subfigure}
\begin{subfigure}{0.32\textwidth}
\includegraphics[width=5.2cm,height=4.4cm]{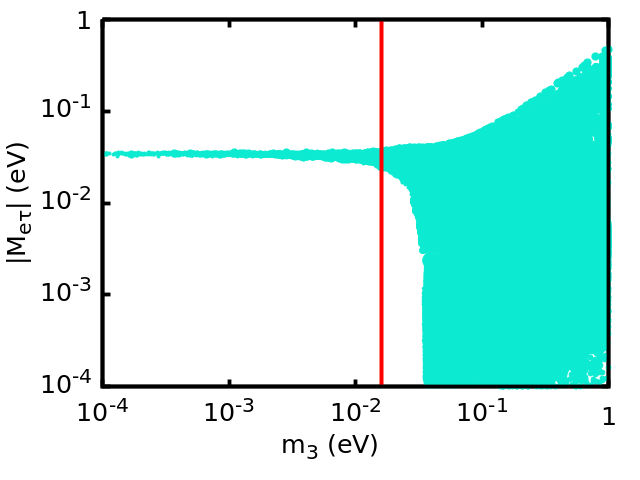}
\caption{}
\label{fig:26.26c}
\end{subfigure}
\captionsetup{justification=raggedright,singlelinecheck=false}
 \caption{ Correlation between $|M_{ee}|$ and $m_{3}$, $|M_{e\mu}|$ and $m_{3}$, and between $|M_{e\tau}|$ and $m_{3}$ with vanishing 
$|M_{\mu\mu}|$ in case of inverted ordering. The red line represents $\sum m_i = 0.12\,{\rm eV}$~\cite{Zhang:2020mox}. The yellow line in 
Fig.~\ref{fig:26.26a} represents $|M_{ee}|= 0.06\,{\rm eV}$~\cite{KamLAND-Zen:2016pfg}.}
\label{fig26:mumuzero_io}
\end{figure}

We show the variation of $|M_{ee}|$, $|M_{e\mu}|$, and $|M_{e\tau}|$ with $m_{3}$ for vanishing $|M_{\mu\mu}|$ in Fig.~\ref{fig:26.26a}, 
Fig.~\ref{fig:26.26b}, and Fig.~\ref{fig:26.26c}, respectively. The bound on the total neutrino mass $\sum m_i < 0.12\,{\rm eV}$ is 
represented by a red line and the upper bound of $M_{ee} = 0.06\,{\rm eV}$ is represented by a yellow line.

From the Fig.~\ref{fig26:mumuzero_io}, it is evident that the two texture zero configurations  
$(M_{e\mu} = 0,\,M_{\mu\mu}=0)$ and $(M_{e\tau} = 0,\, M_{\mu\mu} =0)$ are possible, however, $(M_{ee} = 0,\, M_{\mu\mu} = 0)$ is not. It 
should also be noted that for
vanishing $|M_{\mu\mu}|$, the lowest value of $|M_{ee}|$ is found to be $0.02 \,{\rm eV}$ and it lies within the allowed upper limit 
of $|M_{ee}| < (0.06, 0.31)\,{\rm eV}$ reported by EXO-200, CUORE, and KamLAND-Zen experiments. Moreover, the minimum value of $|M_{ee}|$ 
obtained for this case is consistent with the cosmological bound on the total neutrino mass. For $M_{ee}< 0.06$ eV, the value of 
$\sum m_i$ lies within the range $(0.10, 0.18)\,{\rm eV}$.

\begin{figure}[htbp]
 \begin{subfigure}{0.32\textwidth}
\includegraphics[width=5.2cm,height=4.4cm]{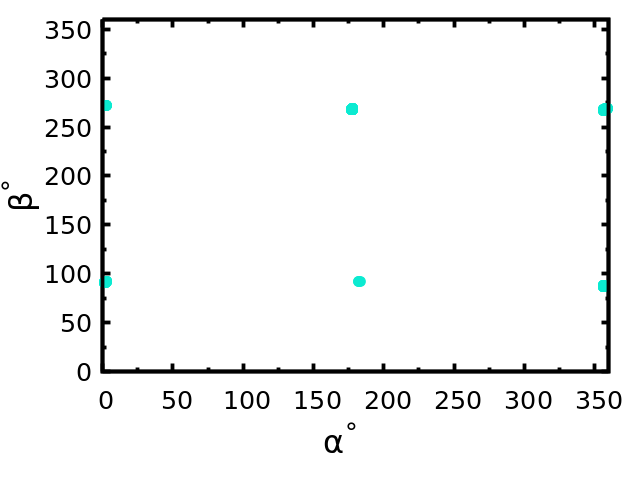}
\caption{}
\label{fig:mumuemuzeroaplbeta_io}
\end{subfigure}
\begin{subfigure}{0.32\textwidth}
\includegraphics[width=5.2cm,height=4.4cm]{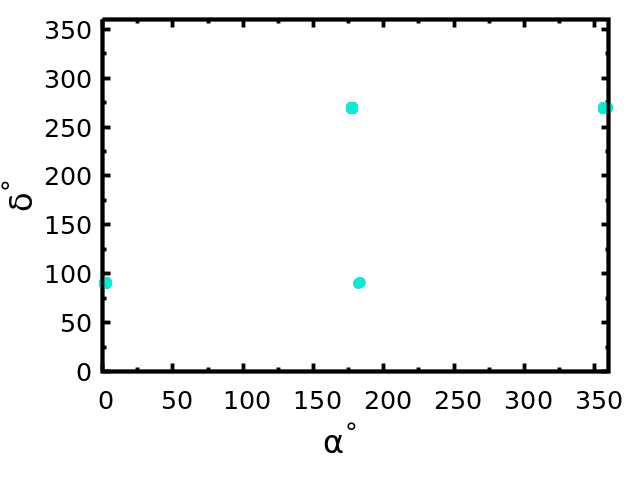}
\caption{}
\label{fig:mumuemuzeroapldelta_io}
\end{subfigure}
\begin{subfigure}{0.32\textwidth}
\includegraphics[width=5.2cm,height=4.4cm]{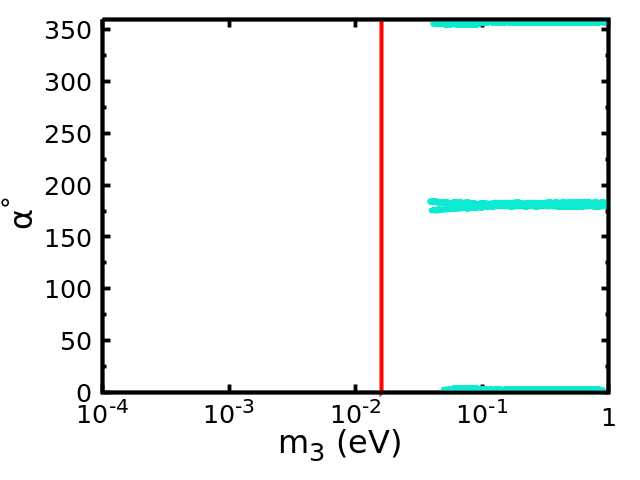}
\caption{}
\label{fig:mumuemuzeroaplham1_io}
\end{subfigure}
\captionsetup{justification=raggedright,singlelinecheck=false}
\caption{Correlation between $\alpha$ and $\beta$, $\alpha$ and $\delta$, and between $\alpha$ and $m_{3}$ with vanishing $|M_{e\mu}|$ and 
$|M_{\mu\mu}|$ in case of inverted ordering.}
\label{fig:mumuemuzero_io}
\end{figure}

\begin{figure}[h!]
 \begin{subfigure}{0.30\textwidth}
\includegraphics[width=5.2cm,height=4.4cm]{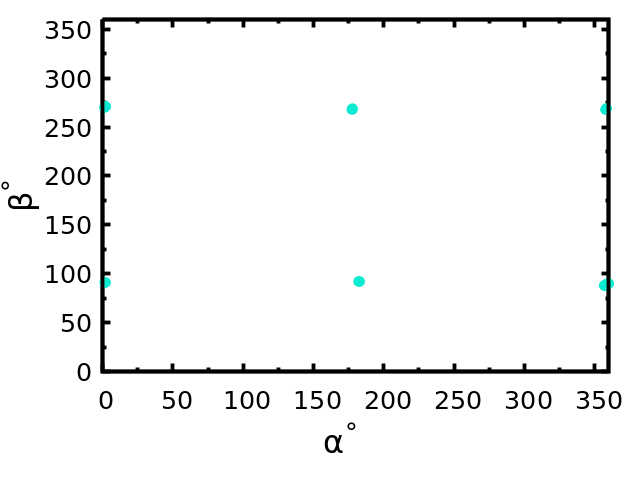}
\caption{}
\label{fig:mumuetauzeroaplbeta_io}
\end{subfigure}
\begin{subfigure}{0.30\textwidth}
\includegraphics[width=5.2cm,height=4.4cm]{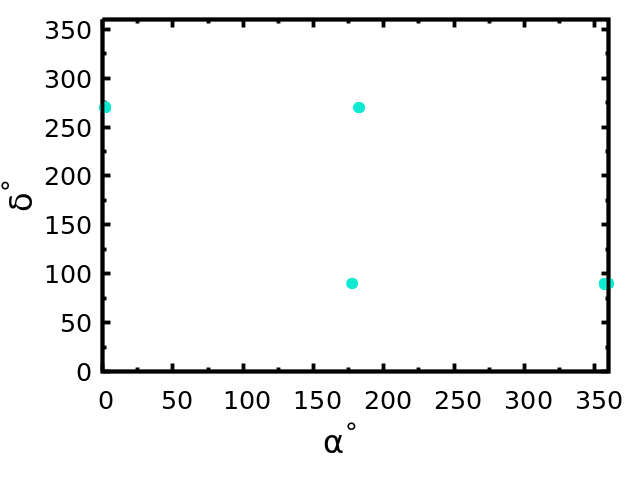}
\caption{}
\label{fig:mumuetauzeroapldelta_io}
\end{subfigure}
\begin{subfigure}{0.30\textwidth}
\includegraphics[width=5.2cm,height=4.4cm]{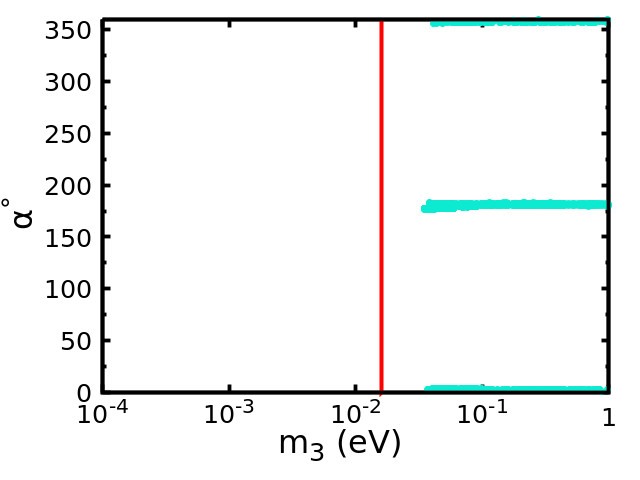}
\caption{}
\label{fig:mumuetauzeroaplham1_io}
\end{subfigure}
\captionsetup{justification=raggedright,singlelinecheck=false}
\caption{Correlation between $\alpha$ and $\beta$, $\alpha$ and $\delta$, and between $\alpha$ and $m_{3}$ with vanishing $|M_{e\tau}|$ and 
$|M_{\mu\mu}|$ in case of inverted ordering.}
\label{fig:mumuetauzero_io}
\end{figure}

We show the correlation between $\alpha$ and $\beta$, $\alpha$ and $\delta$, and between $m_{3}$ and $\alpha$ with vanishing $|M_{e\mu}|$ and 
$|M_{\mu\mu}|$ in Figs.~\ref{fig:mumuemuzeroaplbeta_io}, \ref{fig:mumuemuzeroapldelta_io}, \ref{fig:mumuemuzeroaplham1_io}, respectively. 
It is evident that we can have vanishing $|M_{e\mu}|$ and $|M_{\mu\mu}|$ for $m_{3}>3.8\times 10^{-2}\,{\rm eV}$. Value of
$\alpha$ is restricted to a very narrow region around $0^{\circ}$, $180^{\circ}$ and $360^{\circ}$. Similarly, value of $\beta$ and $\delta$ 
has to lie around $90^{\circ}$ or $270^{\circ}$ to have vanishing $|M_{e\mu}|$ and $|M_{\mu\mu}|$.
Similar correlation plots between $\alpha$ and $\beta$, $\alpha$ and $\delta$, and between $m_{3}$ and $\alpha$ with vanishing $|M_{e\tau}|$ 
and $|M_{\mu\mu}|$ are shown in Figs.~\ref{fig:mumuetauzeroaplbeta_io}, \ref{fig:mumuetauzeroapldelta_io}, \ref{fig:mumuetauzeroaplham1_io}, 
respectively. The correlation between $\alpha$ and $\beta$ and between $\alpha$ and $m_3$ of Fig.~\ref{fig:mumuemuzeroaplbeta_io} and 
Fig.~\ref{fig:mumuemuzeroaplham1_io} looks quite similar to that of Fig.~\ref{fig:mumuetauzeroaplbeta_io} and 
Fig.~\ref{fig:mumuetauzeroaplham1_io}, respectively. It is, however, observed that correlation between $\alpha$ and $\delta$ of 
Fig.~\ref{fig:mumuemuzeroapldelta_io} is quite different from that of Fig.~\ref{fig:mumuetauzeroapldelta_io}. Furthermore for inverted ordering, it is worth mentioning that $|M_{e\mu}|$ and $|M_{\mu\mu}|$ can vanish simultaneously only if $\theta_{23}$ lies in the upper octant, i.e, the value of $\theta_{23}$ has to be $\theta_{23} > 45^{\circ}$ to have vanishing $|M_{e\mu}|$ and $|M_{\mu\mu}|$. We
have shown it in Fig.~\ref{fig:Imalpth23emumumuzerof_io} of Appendix~\ref{app2}. Similarly, $|M_{e\tau}|$ and $|M_{\mu\mu}|$ can vanish simultaneously only if $\theta_{23}$ lies in the upper octant, i.e, the value of $\theta_{23}$ has to be $\theta_{23} > 45^{\circ}$ to have vanishing $|M_{e\tau}|$ and $|M_{\mu\mu}|$. We
have shown it in Fig.~\ref{fig:Imalpth23mumetauzerof_io} of Appendix~\ref{app2}.

\subsubsection{Vanishing $|M_{e\tau}|$:} 
\begin{figure}[htbp!]
   \begin{subfigure}{0.40\textwidth}
\includegraphics[width=5.6cm,height=4.9cm]{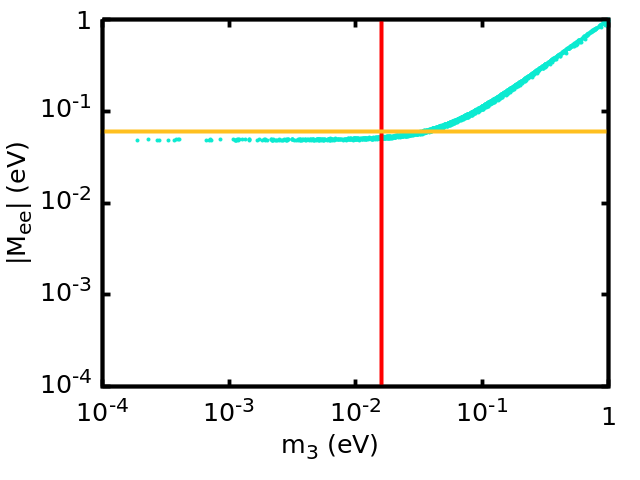}
\caption{}
\label{fig:27.27a}
\end{subfigure}
\begin{subfigure}{0.40\textwidth}
\includegraphics[width=5.6cm,height=4.9cm]{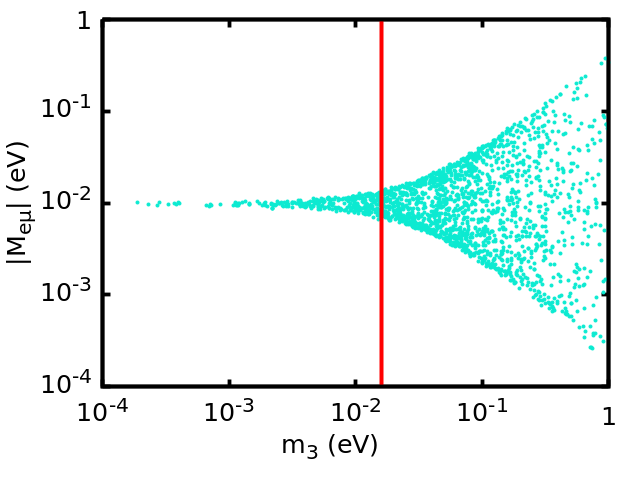}
\caption{}
\label{fig:27.27b}
\end{subfigure}
 \captionsetup{justification=raggedright,singlelinecheck=false}
 \caption{ Correlation between $|M_{ee}|$ and $m_{3}$, and $|M_{e\mu}|$ and $m_{3}$ with vanishing $|M_{e\tau}|$ in case of inverted ordering.
The red line represents $\sum m_i = 0.12\,{\rm eV}$~\cite{Zhang:2020mox}. The yellow line in Fig.~\ref{fig:27.27a} represents 
$|M_{ee}|= 0.06\,{\rm eV}$~\cite{KamLAND-Zen:2016pfg}.}
\label{fig27:etauzero_io}
\end{figure}

Variation of $|M_{ee}|$ and $|M_{e\mu}|$ with $m_{3}$ for vanishing $|M_{e\tau}|$ is shown in Fig.~\ref{fig:27.27a} and Fig.~\ref{fig:27.27b},
respectively. The bound on the total neutrino mass $\sum m_i < 0.12\,{\rm eV}$ is represented by a red line and the upper bound of 
$M_{ee} = 0.06\,{\rm eV}$ is represented by a yellow line, respectively.
 
From Fig.~\ref{fig27:etauzero_io}, it is evident that two texture zero patterns $(|M_{ee}| = 0,\,|M_{e\tau}|=0)$ and 
$(|M_{e\mu}|=0,\,|M_{e\tau}|=0)$ are not viable. It is worth mentioning that for vanishing $|M_{e\tau}|$, the effective Majorana mass 
$|M_{ee}|$ lies within the allowed upper bound $|M_{ee}| < (0.06, 0.31)\,{\rm eV}$ reported by EXO-200, CUORE, and 
KamLAND-Zen experiments. For $M_{ee}< 0.06$ eV, the value of $\sum m_i$ is found to be in the range $(0.10, 0.16)\,{\rm eV}$ eV.

\subsubsection{Vanishing $|M_{e\mu}|$:} 
\begin{figure}[htbp!]
\includegraphics[width=5.6cm,height=4.8cm]{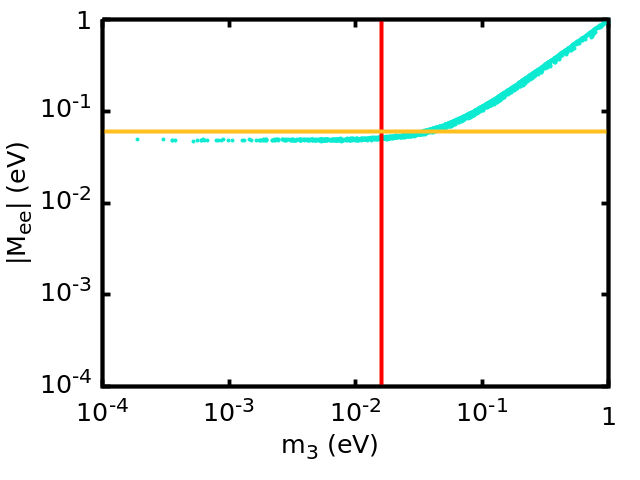}
\captionsetup{justification=raggedright,singlelinecheck=false}
 \caption{ Correlation between $|M_{ee}|$ and the lightest neutrino mass $m_{3}$ with vanishing $|M_{e\mu}|$ in case of inverted ordering. 
The red line represents $\sum m_i = 0.12\,{\rm eV}$~\cite{Zhang:2020mox}. The yellow line represents 
$|M_{ee}|= 0.06\,{\rm eV}$~\cite{KamLAND-Zen:2016pfg}.}
\label{fig27:emuzero_io}
\end{figure}

The variation of $|M_{ee}|$ with $m_{3}$ for vanishing $|M_{e\mu}|$ is shown in Fig.~\ref{fig27:emuzero_io}. The bound on the total neutrino 
mass $\sum m_i < 0.12\,{\rm eV}$ is represented by a red line. Similarly, the upper bound of $M_{ee} = 0.06\,{\rm eV}$ is represented by a 
yellow line. It is evident that $|M_{ee}|$ and $|M_{e\mu}|$ can not vanish simultaneously. However, for vanishing $|M_{e\mu}|$, the value of 
the effective Majorana mass $|M_{ee}|$ lies within the allowed upper bound $|M_{ee}| < (0.06, 0.31)\,{\rm eV}$ reported by 
EXO-200, CUORE, and KamLAND-Zen experiments. For $M_{ee} < 0.06$ eV, the value of $\sum m_i$ is found to be in the range 
$(0.10, 0.16)\,{\rm eV}$.
 
\section{Conclusion}
\label{section:4}
We have explored the detailed structure and analyze the implication of one and two texture zeros of the neutrino mass matrix for both normal 
and inverted ordering through a bottom-up approach using the current neutrino experimental data. The neutrino mass matrix elements depend on 
four unknown parameters, namely the lightest neutrino mass~$m_1~(m_3)$ and three CP violating phases $\alpha$, $\beta$ and $\delta$. 
We estimate the allowed range of each individual matrix element by varying the CP violating phases within $(0,\,2\pi)$, lightest neutrino 
mass within $(0,\,1)$ eV and all the neutrino oscillation parameters within $3\sigma$ of their central values. We have also explored 
the feasibility of one and two texture zeros in the neutrino mass matrix.  
Moreover, for each one and two texture zero cases, we have studied the correlation of all the unknown parameters for different values of the 
Dirac CP violating phase $\delta$. A brief summary of our results are as follows: 
\begin{itemize} 

\item With the latest neutrino oscillation data, the $ee$ element $|M_{ee}|$ can vanish only for normal ordering case. Value of the lightest 
neutrino mass should be $m_1 \in(2.0,\,8.0)\times 10^{-3}\,{\rm eV}$ for vanishing $|M_{ee}|$. Similarly, the Majorana CP 
violating phase $\alpha$ 
should be very close to $90^{\circ}$ or $270^{\circ}$. Vanishing of $|M_{ee}|$, however, does not put any constraint on the values of 
$\delta$ and $\beta$. 

 \item The $e\mu$ element $|M_{e\mu}|$ vanishes for normal, inverted and degenerate mass ordering. For normal ordering, the lightest 
neutrino mass should be $m_1 \ge 6.0\times 10^{-3}\,{\rm eV}$ to have vanishing $e\mu$ element. 
Depending on the value of the Dirac CP violating phase 
$\delta$, correlation between $\beta$ and $\alpha$ can change for vanishing $|M_{e\mu}|$. 
In case of normal ordering, we observe that $|M_{ee}|$ and $|M_{e\mu}|$ can vanish simultaneously for $m_1 \in (6.50,\,8.20)\times 10^{-3}\,{\rm eV}$
and $\alpha \approx 90^{\circ}$ and $270^{\circ}$.
For inverted ordering, $|M_{e\mu}|$ can vanish for 
all values of the lightest neutrino mass $m_3$. For vanishing $|M_{e\mu}|$, $\delta$ value should lie very close $90^{\circ}$ and $270^{\circ}$
once we impose $\sum m_i < 0.12\,{\rm eV}$ constraint.
Similarly, the $\alpha$ value has to lie within a very narrow region close to but not equal to $0^{\circ}$, $180^{\circ}$ and $360^{\circ}$. 

 \item $|M_{e\tau}|$ looks quite similar to $|M_{e\mu}|$ due to the approximate $\mu - \tau$ symmetry in the mass matrix. It is, however, 
worth mentioning that $|M_{e\tau}|$ and $|M_{e\mu}|$ do not vanish simultaneously. It is mainly due to the CP violating phases. We find that
the correlation between $\beta$ and $\alpha$, and between $\alpha$ and $\delta$ for vanishing $|M_{e\tau}|$ and for vanishing $|M_{e\mu}|$ are 
completely different from each other. We find that, in case of normal ordering, $|M_{ee}|$ and $|M_{e\tau}|$ vanish 
simultaneously for $m_1 \in (4.0,\, 6.0)\times 10^{-3}\,{\rm eV}$ and $\alpha \approx 90^{\circ}$ and $270^{\circ}$.

 \item The $\mu\mu$ element $|M_{\mu\mu}|$ vanishes in case of normal, inverted and degenerate mass ordering. For normal ordering, the 
lightest mass should be $m_1 \ge 5.1 \times 10^{-2}\,{\rm eV}$ for vanishing $|M_{\mu\mu}|$. However, $|M_{\mu\mu}|$ can vanish for any value
of the lightest neutrino mass $m_3$ in case of inverted ordering. It is worth mentioning that, in case of normal ordering, value of 
the atmospheric mixing angle should lie in the lower octant; more precisely $\theta_{23} < 45.3^{\circ}$ to have vanishing $|M_{\mu\mu}|$. 
There is, however, no such restriction
on $\theta_{23}$ in case of inverted ordering. In case of inverted ordering,  
it is observed that at $\delta = 0^{\circ}$, value of $\alpha$ is constrained to lie around $90^{\circ}$ or $270^{\circ}$ 
if the lightest neutrino mass $m_3 \leq 10^{-3}\,{\rm eV}$. However, value of $\alpha$ will deviate from $90^{\circ}$ or $270^{\circ}$ if the
lightest neutrino mass $m_3 \approx 4.94\times 10^{-2}\,{\rm eV}$ or larger.

\item The $\mu\tau$ element $|M_{\mu\tau}|$ can vanish in case of inverted and degenerate mass ordering. We can have vanishing 
$|M_{\mu\tau}|$ only if the lightest neutrino mass $m_3 > 10^{-2}\,{\rm eV}$. The value of the Majorana CP violating phases $\alpha$ and 
$\beta$
are restricted to lie within a very narrow range for vanishing $|M_{\mu\tau}|$ if we impose $\sum m_i < 0.12\,{\rm eV}$ constraint. 
However, without the constraint, it can vanish for any value of $\alpha$ and $\beta$ within $(0^{\circ}, 360^{\circ})$.

\item The $\tau\tau$ element $|M_{\tau\tau}|$ can vanish for normal, inverted and degenerate mass ordering. In case of normal
ordering the lightest neutrino mass has to be $m_1 > 3.4\times 10^{-2}\,{\rm eV}$ for vanishing $|M_{\tau\tau}|$. For $\delta = 0^{\circ}$,
$|M_{\tau\tau}|$ can vanish for any value of $\alpha$ and $\beta$ within $(0^{\circ}, 360^{\circ})$. However, for $\delta = 90^{\circ}$ and 
$180^{\circ}$,
$\alpha = 90^{\circ}$ and $270^{\circ}$ are not allowed for vanishing $\tau\tau$ element. Similarly, $\beta = 0^{\circ}$, $180^{\circ}$ and
$360^{\circ}$ are also not allowed for this case. 
In case of inverted ordering, we will have vanishing $|M_{\tau\tau}|$ only if the lightest neutrino mass $m_3 \ge 4.0\times 10^{-3}\,{\rm eV}$.
For vanishing $|M_{\tau\tau}|$, the Majorana CP violating phase $\alpha$ has to lie very close to $90^{\circ}$ or $270^{\circ}$ for 
$\delta = 180^{\circ}$ if $m_3 < 10^{-2}\,{\rm eV}$. However, for $m_3 > 10^{-2}\,{\rm eV}$, value of $\alpha$ will gradually deviate from 
$90^{\circ}$ or $270^{\circ}$. It is worth mentioning that value of the atmospheric mixing angle should lie in the upper octant, i.e,
$\theta_{23} > 44.3^{\circ}$ to have vanishing $|M_{\tau\tau}|$ in case of normal ordering. However, there is no such restriction on 
$\theta_{23}$ in case of inverted ordering.
 
\item In case of normal ordering, patterns $(M_{e\mu} =0,\,M_{\tau\tau} = 0)$ and $(M_{e\tau} = 0,\,M_{\tau\tau} = 0)$ are allowed for the upper octant of $\theta_{23}$, i.e, the value of $\theta_{23}$ should be $\theta_{23}> 45^{\circ}$ for these two patterns, pattern $(M_{\mu\mu} = 0,\,M_{\tau\tau} = 0)$ is allowed for $\theta_{23}$ value around $45^{\circ}$ and patterns $(M_{e\mu} =0,\,M_{\mu\mu} = 0)$ and $(M_{e\tau} = 0,\,M_{\mu\mu} = 0)$ are allowed for the lower octant of $\theta_{23}$, i.e, the value of $\theta_{23}$ should be $\theta_{23}< 45^{\circ}$ for these patterns.

\item In case of inverted ordering, patterns $(M_{e\mu} =0,\,M_{\tau\tau} = 0)$ and $(M_{e\tau} = 0,\,M_{\tau\tau} = 0)$ are allowed for the lower octant of $\theta_{23}$, i.e, for these patterns, the value of $\theta_{23}$ should be $\theta_{23} < 45^{\circ}$  and patterns $(M_{e\mu} =0,\,M_{\mu\mu} = 0)$ and $(M_{e\tau} = 0,\,M_{\mu\mu} = 0)$ are allowed for the upper octant of $\theta_{23}$, i.e, for these patterns, the value of $\theta_{23}$ should be $\theta_{23}> 45^{\circ}$. 
 
\end{itemize}

In conclusion, all six one texture zero patterns are consistent with current experimental data. We find that the $ee$ element can vanish only
for normal ordering, whereas, $e\mu$, $e\tau$, $\mu\mu$ and $\tau\tau$ entries can vanish for 
normal, inverted and degenerate mass ordering. Moreover, the $\mu\tau$ element can vanish for inverted and degenerate mass ordering.
For two texture zeros, two patterns, namely $(M_{ee} =0,\,M_{e\mu} = 0)$ and $(M_{ee} = 0,\,M_{e\tau} = 0)$ are allowed only in case of
normal ordering, whereas, $(M_{e\mu} =0,\,M_{\mu\mu} = 0)$, $(M_{e\tau} = 0,\,M_{\mu\mu} = 0)$, $(M_{\mu\mu} = 0,\,M_{\tau\tau} = 0)$,
$(M_{e\mu} =0,\,M_{\tau\tau} = 0)$ and $(M_{e\tau} = 0,\,M_{\tau\tau} = 0)$ are allowed for normal, inverted and degenerate mass ordering.
However, all the two texture zero patterns are found to be strongly correlated, i.e, a 
$1\%$ change in the value of $m_1~(m_3)$, $\alpha$, $\beta$ and $\delta$ can induce a deviation of more than $0.1\,{\rm meV}$ on the vanishing 
element. With the latest bound of $\sum m_i < 0.072\,{\rm eV}$, only three one texture zero patterns $(M_{ee} = 0)$, $(M_{e\mu} = 0)$ and
$(M_{e\tau} = 0)$, and two two-texture zero patterns $(M_{ee} =0,\,M_{e\mu} = 0)$ and $(M_{ee} = 0,\,M_{e\tau} = 0)$ remain viable for normal 
ordering. Measuring the CP-violating phases $\delta$ and $\alpha$ is crucial as they significantly affect the viability of texture zero 
patterns. Additionally, future measurements of the sign of $\Delta m_{31}^2$ and the octant of $\theta_{23}$ will also be critical in 
assessing these patterns.\\

\bigskip
\appendix
\section{}
\label{app1}

\begin{figure}[htbp]
  \begin{subfigure}{0.25\textwidth}
\includegraphics[width=\textwidth]{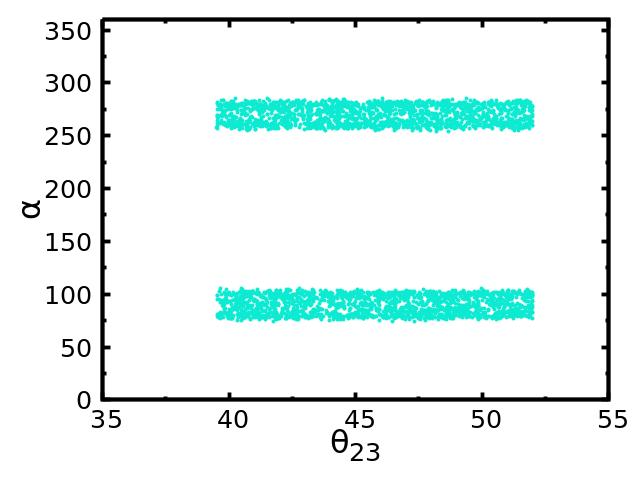}
         \subcaption{Correlation between $\theta_{23}$ and $\alpha$ with vanishing $|M_{ee}|$.}
         \label{fig:Nmalpth23meezero_no}
     \end{subfigure}\hfil
\begin{subfigure}{0.25\textwidth}
\includegraphics[width=\textwidth]{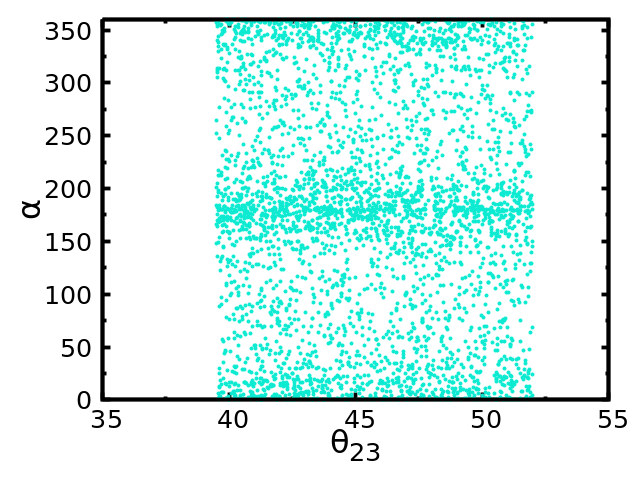}
         \subcaption{Correlation between $\theta_{23}$ and $\alpha$ with vanishing $|M_{e\mu}|$.}
         \label{fig:Nmalpth23emuzero_no}
     \end{subfigure}\hfil
 \begin{subfigure}{0.25\textwidth}
\includegraphics[width=\textwidth]{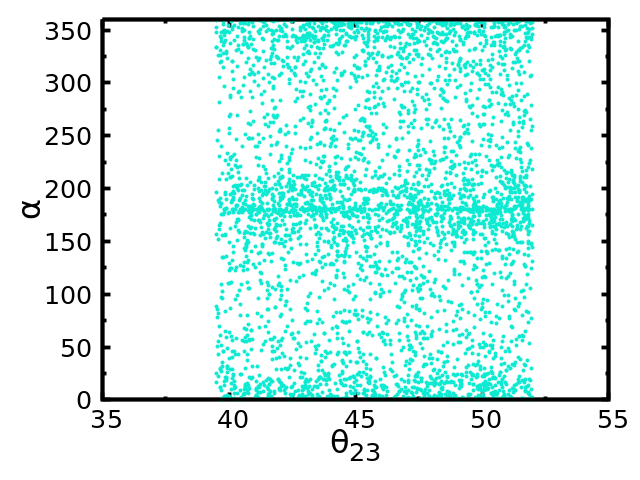}
         \subcaption{Correlation between $\theta_{23}$ and $\alpha$ with vanishing $|M_{e\tau}|$.}
         \label{fig:Nmalpth32etauzero_no}
     \end{subfigure}
\begin{subfigure}{0.25\textwidth}
\includegraphics[width=\textwidth]{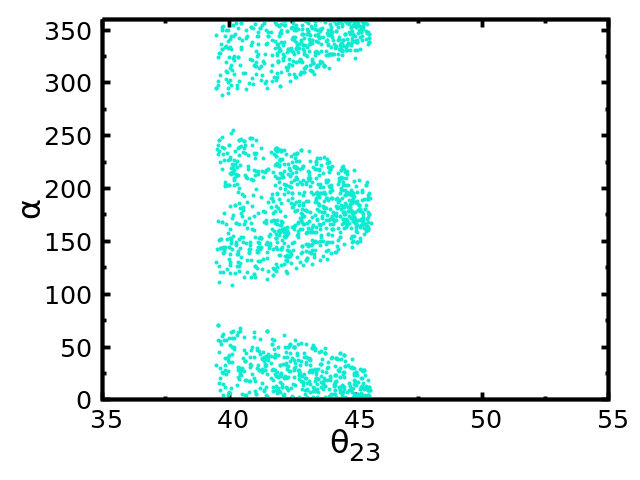}
         \subcaption{Correlation between $\theta_{23}$ and $\alpha$ with vanishing $|M_{\mu\mu}|$.}
         \label{fig:Nmalpth23mumzero_no}
     \end{subfigure} \hfil
 \begin{subfigure}{0.25\textwidth}
\includegraphics[width=\textwidth]{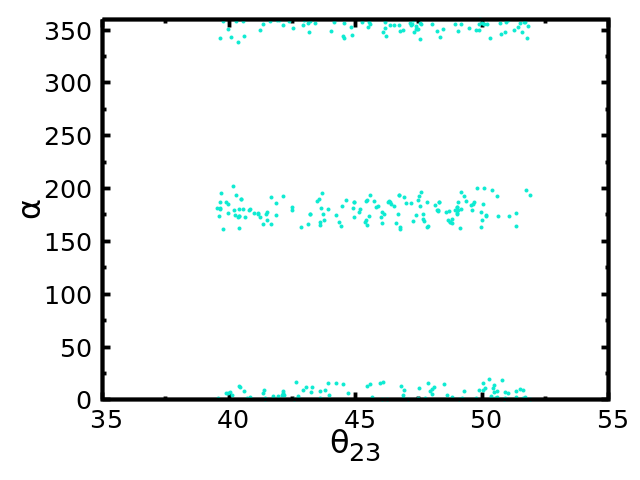}
         \subcaption{Correlation between $\theta_{23}$ and $\alpha$ with vanishing $|M_{\mu\tau}|$.}
         \label{fig:Nmalpth23mumtautauzero_no}
     \end{subfigure}\hfil
 \begin{subfigure}{0.25\textwidth}
\includegraphics[width=\textwidth]{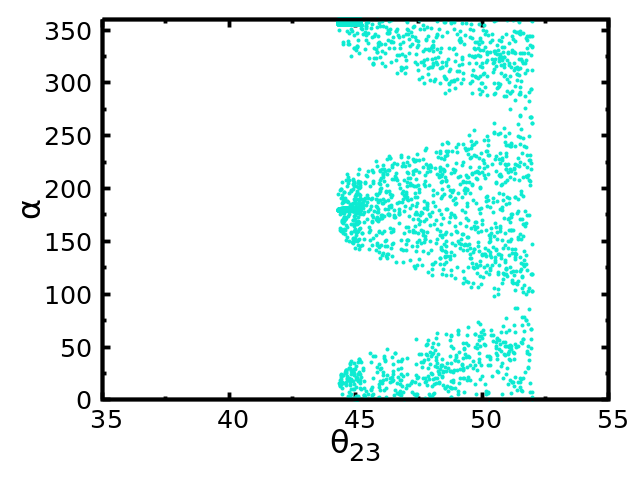}
         \subcaption{Correlation between $\theta_{23}$ and $\alpha$ with vanishing $|M_{\tau\tau}|$.}
         \label{fig:Nmalpth23tautauzero_no}
     \end{subfigure}
      \caption{Correlation between $\theta_{23}$ and $\alpha$ for one texture zero in case of normal ordering.}
      \label{Nmalpth23_no}
\end{figure}

\begin{figure}[htbp]
     \begin{subfigure}{0.25\textwidth}
\includegraphics[width=\textwidth]{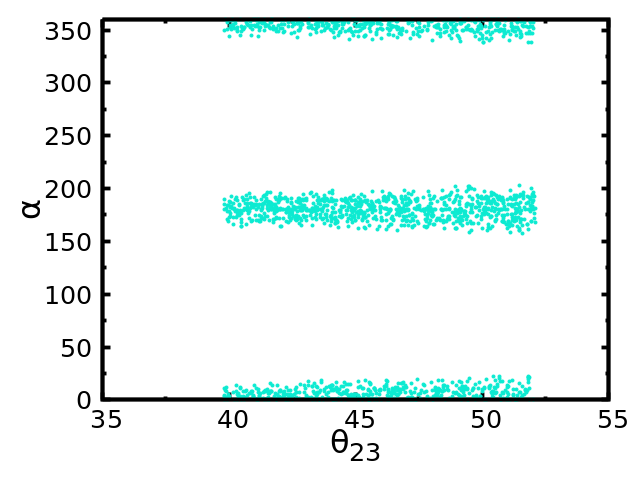}
         \subcaption{Correlation between $\theta_{23}$ and $\alpha$ with vanishing $|M_{e\mu}|$.}
         \label{fig:Imalpth23emuzero_io}
     \end{subfigure}\hspace{0.4cm}
 \begin{subfigure}{0.25\textwidth}
\includegraphics[width=\textwidth]{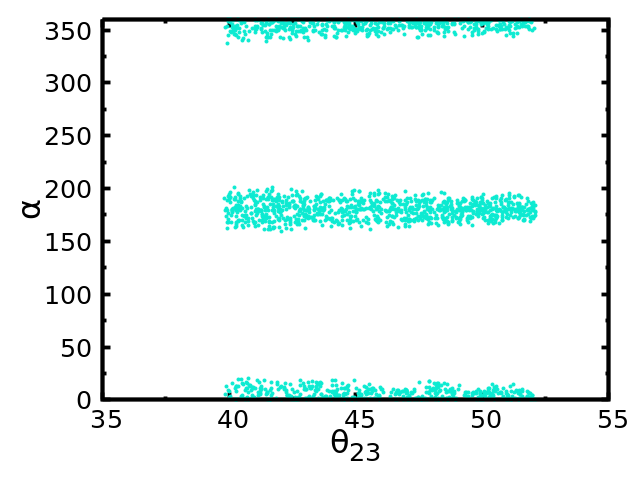}
         \subcaption{Correlation between $\theta_{23}$ and $\alpha$ with vanishing $|M_{e\tau}|$.}
         \label{fig:Imalpth23etauzero_io}
     \end{subfigure}\hspace{0.4cm}
 \begin{subfigure}{0.25\textwidth}
\includegraphics[width=\textwidth]{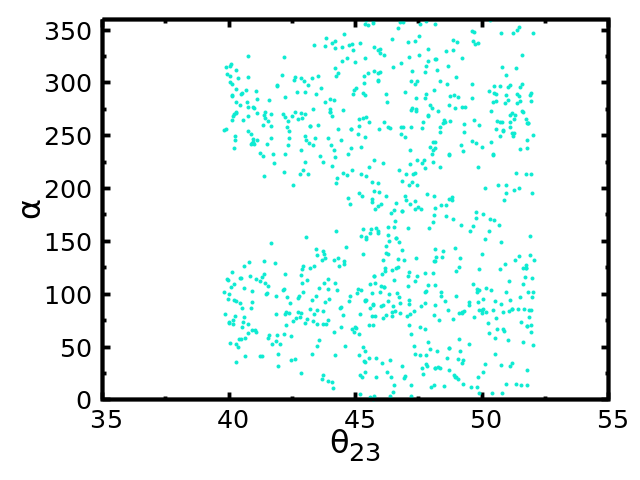}
         \subcaption{Correlation between $\theta_{23}$ and $\alpha$ with vanishing $|M_{\mu\mu}|$.}
         \label{fig:Imalpth23mumuzero_io}
     \end{subfigure}
 \begin{subfigure}{0.25\textwidth}
\includegraphics[width=\textwidth]{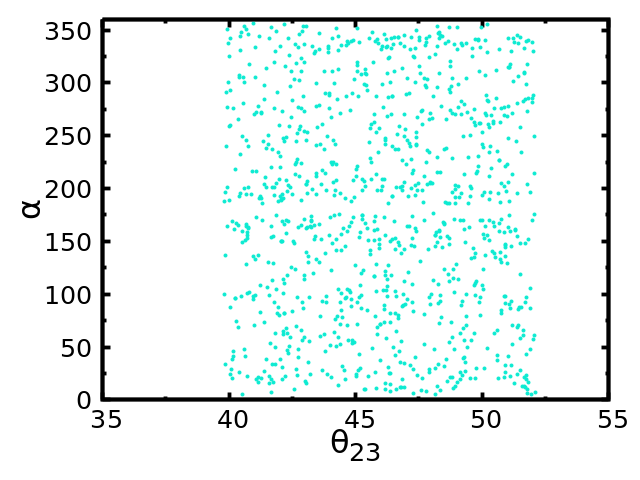}
         \subcaption{Correlation between $\theta_{23}$ and $\alpha$ with vanishing $|M_{\mu\tau}|$.}
         \label{fig:Imalpth23mutautauzero_io}
     \end{subfigure}\hspace{0.4cm}
  \begin{subfigure}{0.25\textwidth}
\includegraphics[width=\textwidth]{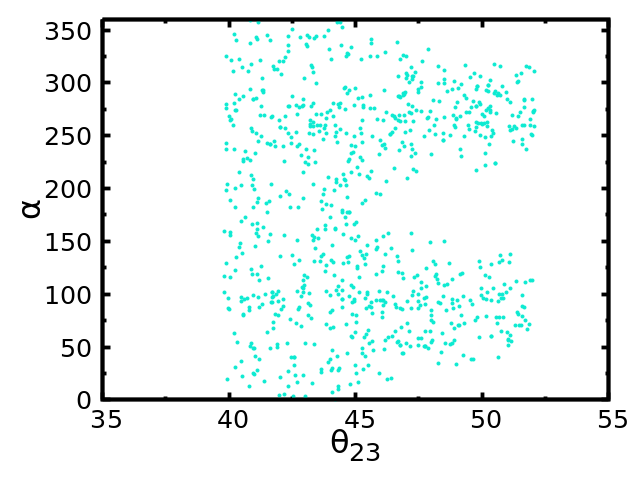}
         \subcaption{Correlation between $\theta_{23}$ and $\alpha$ with vanishing $|M_{\tau\tau}|$.}
         \label{fig:Imalpth23tautauzero_io}
     \end{subfigure}
      \caption{Correlation between $\theta_{23}$ and $\alpha$ for one texture zero in case of inverted ordering.}
      \label{Imalpth23_io}
\end{figure}

Correlation plots between $\theta_{23}$ and $\alpha$ for one texture zero are shown in Figs.~\ref{Nmalpth23_no} and ~\ref{Imalpth23_io} in case of normal and inverted orderings, respectively. 

\newpage
\section{}
\label{app2}
\begin{figure}[htbp]
     \centering
   \begin{subfigure}{0.25\textwidth}
\includegraphics[width=\textwidth]{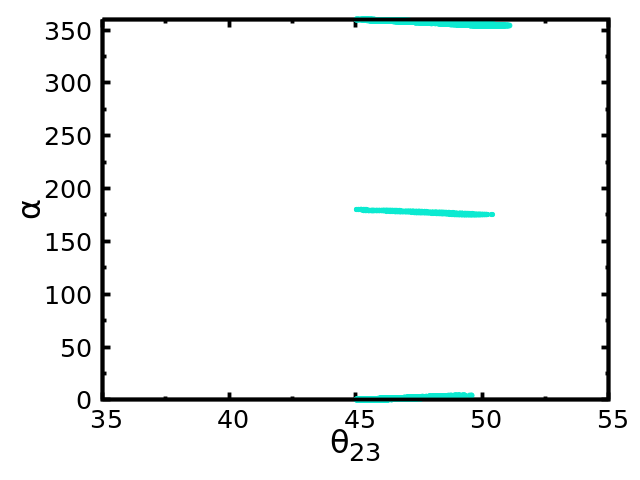}
         \subcaption{Correlation between $\theta_{23}$ and $\alpha$ with vanishing $|M_{e\mu}|$ and $|M_{\tau\,\tau}|$.}
    \label{fig:Nmalpth23emutautauzerofinal_no}
     \end{subfigure}
 \begin{subfigure}{0.25\textwidth}
\includegraphics[width=\textwidth]{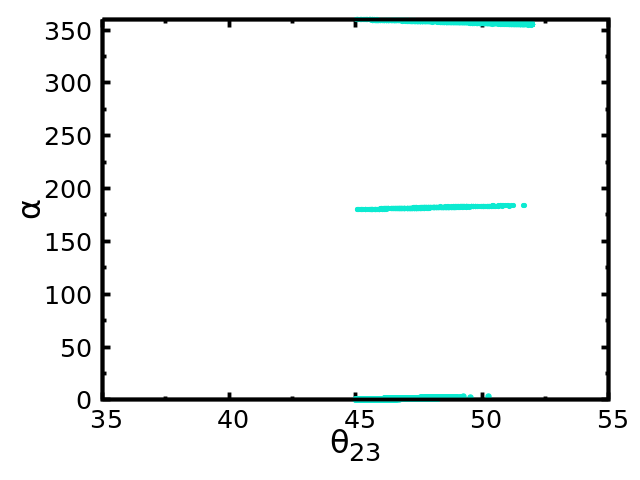}
         \subcaption{Correlation between $\theta_{23}$ and $\alpha$ with vanishing $|M_{e\tau}|$ and $|M_{\tau\,\tau}|$.}
  \label{fig:Nmalpth23etautautauzerofinal_no}
     \end{subfigure}
 \begin{subfigure}{0.24\textwidth}
\includegraphics[width=\textwidth]{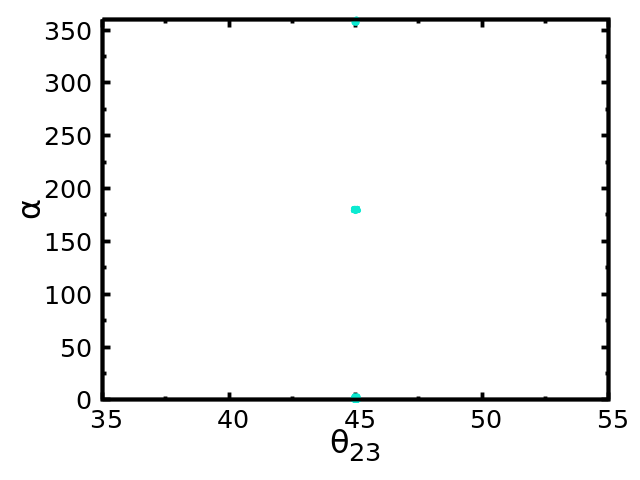}
         \subcaption{Correlation between $\theta_{23}$ and $\alpha$ with vanishing $|M_{\mu\,\mu}|$ and $|M_{\tau\,\tau}|$.}
         \label{fig:Nmalpth23mumutautauzerofinal_no}
     \end{subfigure}
    \begin{subfigure}{0.24\textwidth}
\includegraphics[width=\textwidth]{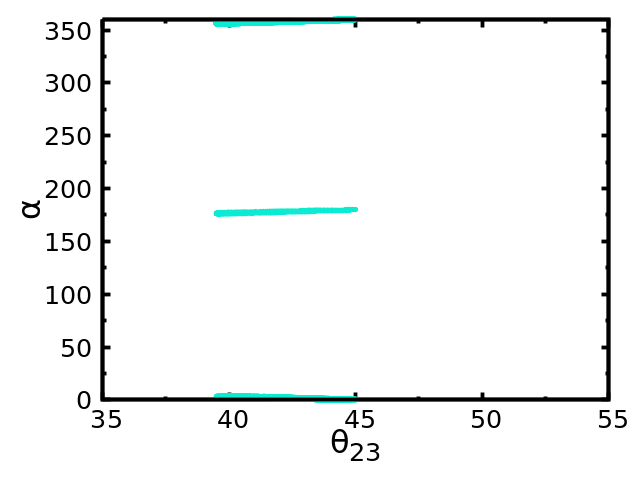}
         \subcaption{Correlation between $\theta_{23}$ and $\alpha$ with vanishing $|M_{e\mu}|$ and $|M_{\mu\,\mu}|$.}
         \label{fig:Nmalpth23mumuemuzerofinal_no}
     \end{subfigure}
\begin{subfigure}{0.25\textwidth}
\includegraphics[width=\textwidth]{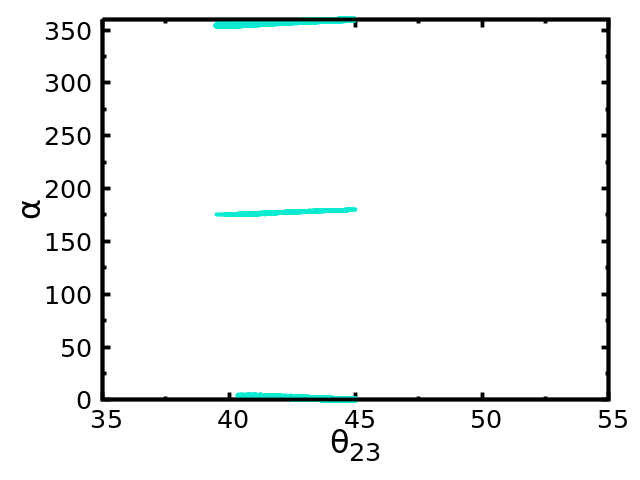}
         \subcaption{Correlation between $\theta_{23}$ and $\alpha$ with vanishing $|M_{e\tau}|$ and $|M_{\mu\,\mu}|$.}
         \label{fig:Nmalpth23mumetauzerofinal_no}
     \end{subfigure} \hspace{0.2cm}
    \begin{subfigure}{0.25\textwidth}
\includegraphics[width=\textwidth]{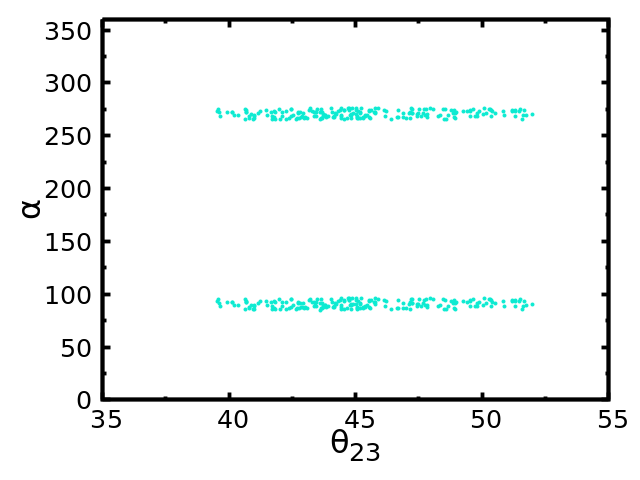}
         \subcaption{Correlation between $\theta_{23}$ and $\alpha$ with vanishing $|M_{ee}|$ and $|M_{e\tau}|$.}
        \label{fig:Nmalpth23meeetauzerofinal_no}
     \end{subfigure} \hspace{0.2cm}
 \begin{subfigure}{0.25\textwidth}
\includegraphics[width=\textwidth]{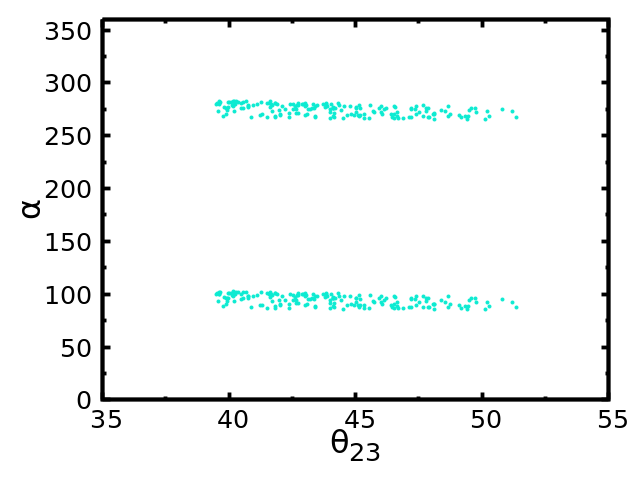}
         \subcaption{Correlation between $\theta_{23}$ and $\alpha$ with vanishing $|M_{ee}|$ and $|M_{e\mu}|$.}
         \label{fig:Nmalpth23meeemufinal_no}
     \end{subfigure}
      \caption{Correlation between $\theta_{23}$ and $\alpha$ for two texture zeros in case of normal ordering.}
      \label{Nmalpth23_two_no}
\end{figure}

\begin{figure}[h!]
    \begin{subfigure}{0.26\textwidth}
\includegraphics[width=\textwidth]{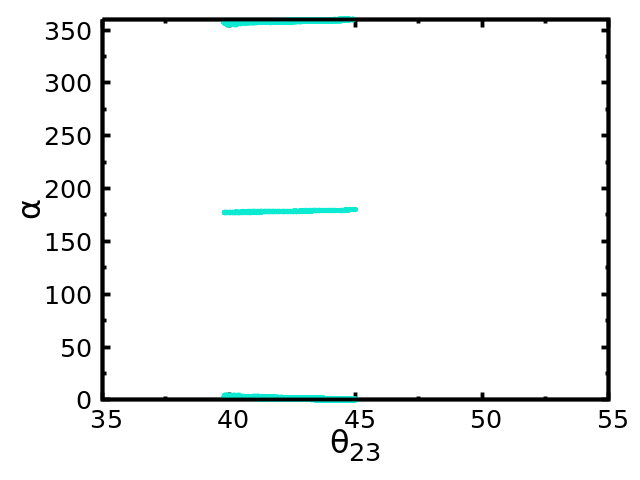}
         \subcaption{Correlation between $\theta_{23}$ and $\alpha$ with vanishing $|M_{e\mu}|$ and $|M_{\tau\,\tau}|$.}
         \label{fig:Imalpth23emutautauzerof_io}
     \end{subfigure} \hspace{0.4cm}
 \begin{subfigure}{0.26\textwidth}
\includegraphics[width=\textwidth]{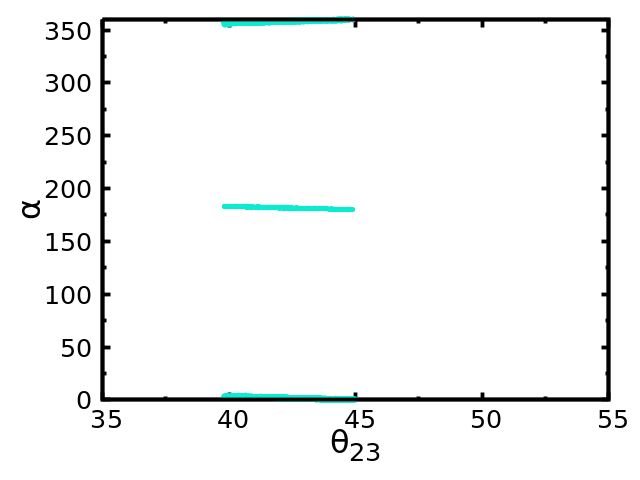}
         \subcaption{Correlation between $\theta_{23}$ and $\alpha$ with vanishing $|M_{e\tau}|$ and $|M_{\tau\,\tau}|$.}
         \label{fig:Imalpth23etautautauzerof_io}
     \end{subfigure}\hspace{0.4cm}
 \begin{subfigure}{0.26\textwidth}
\includegraphics[width=\textwidth]{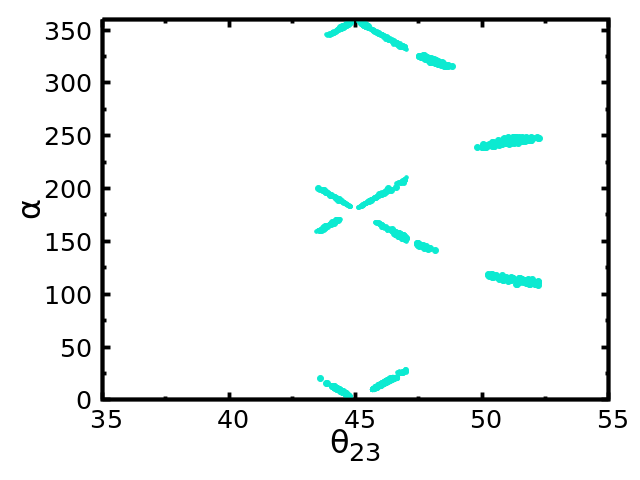}
         \subcaption{Correlation between $\theta_{23}$ and $\alpha$ with vanishing $|M_{\mu\,\mu}|$ and $|M_{\tau\,\tau}|$.}
         \label{fig:Imnalpth23mumutautauzero_io}
     \end{subfigure}
 \begin{subfigure}{0.26\textwidth}
\includegraphics[width=\textwidth]{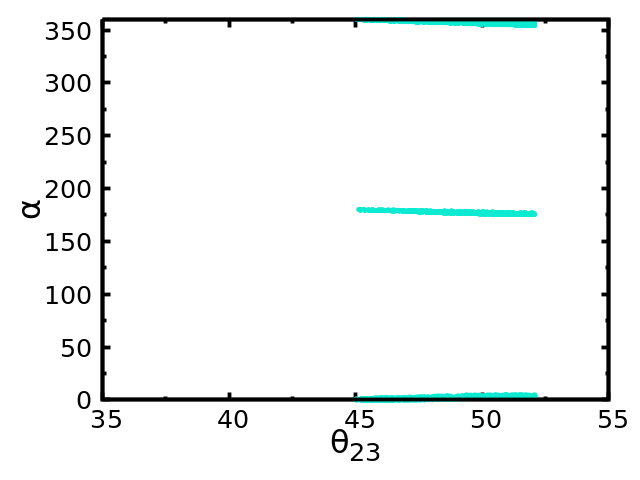}
         \subcaption{Correlation between $\theta_{23}$ and $\alpha$ with vanishing $|M_{e\mu}|$ and $|M_{\mu\,\mu}|$.}
         \label{fig:Imalpth23emumumuzerof_io}
     \end{subfigure}\hspace{0.4cm}
 \begin{subfigure}{0.26\textwidth}
\includegraphics[width=\textwidth]{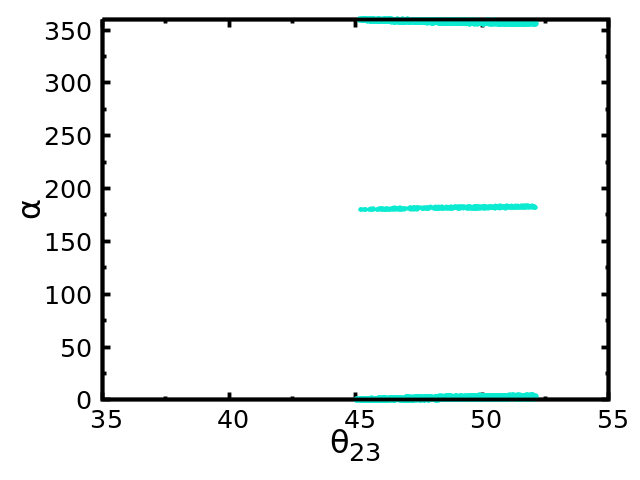}
         \subcaption{Correlation between $\theta_{23}$ and $\alpha$ with vanishing $|M_{e\tau}|$ and $|M_{\mu\,\mu}|$.}
         \label{fig:Imalpth23mumetauzerof_io}
     \end{subfigure}
           \caption{Correlation between $\theta_{23}$ and $\alpha$ for two texture zeros in case of inverted ordering.}
      \label{Imalpth23_two_io}
\end{figure}

Correlation plots between $\theta_{23}$ and $\alpha$ for two texture zero are shown in Figs.~\ref{Nmalpth23_two_no} and ~\ref{Imalpth23_two_io} in case of normal and inverted orderings, respectively. 

\end{document}